\documentclass{WileyMSP-template}
\usepackage{microtype}
\usepackage[english]{babel}
\usepackage{hyperref}
\usepackage{cleveref}
\usepackage{placeins}
\usepackage{ragged2e}
\usepackage{cite}
\usepackage[euler]{textgreek}
\begin{document}


\title{Unlocking Spin Dynamics: Spin-Orbit Coupling Driven Spin State Interconversion in Carbazole-Containing TADF Emitters}

\maketitle


\author{Annika Morgenstern$^1$$^\dagger$},
\author{Jonas Weiser$^2$$^\dagger$},
\author{Lucas Schreier$^1$},
\author{Konstantin Gabel$^{1,4}$},
\author{Tom Gabler$^3$},
\author{Alexander Ehm$^1$},
\author{Nadine Schwierz$^2$},
\author{Ulrich T. Schwarz$^4$},
\author{Kirsten Zeitler$^3$},
\author{Dietrich R.T. Zahn$^{1,5}$},
\author{Christian Wiebeler$^2$*},
\author{Georgeta Salvan$^{1,5}$*}

\begin{affiliations}

$^{1}$Semiconductor Physics, Institute of Physics, Chemnitz University of Technology, 09126 Chemnitz, Germany\\
$*$georgeta.salvan@physik.tu-chemnitz.de\\

$^{2}$Computational Biology, Institute of Physics, University of Augsburg, 86159 Augsburg, Germany  \\
$*$christian.wiebeler@uni-a.de\\

$^{3}$Organic Chemistry and Catalysis, Institute of Organic Chemistry, Leipzig University, 04103 Leipzig, Germany\\

$^{4}$Experimental Sensor Science, Institute of Physics, Chemnitz University of Technology, 09126 Chemnitz, Germany \\

$^{5}$Materials, Architecture and Integration of Nanomembranes (MAIN), Chemnitz University of Technology, 09126 Chemnitz, Germany\\
$^\dagger$ contributed equally to this work
\end{affiliations}



\begin{abstract}

The determination of transport mechanisms in organic light-emitting diodes (OLEDs) is crucial for optimizing device performance. Magnetic field measurements enable the differentiation of spin state interconversion mechanisms, but data interpretation remains challenging. Here, experimental and theoretical investigations were combined to provide a comprehensive understanding of the underlying processes.
This study systematically compares three cyanoarene-based emitters with different singlet-triplet gaps ($\Delta E_ \mathrm{ST}$) to explore factors influencing reverse intersystem crossing (RISC). The comparison of all-$^1\mathrm{H}$ and  all-$^2\mathrm{H}$ 4CzIPN isotopologues confirms that RISC is governed by spin-orbit coupling (SOC) rather than hyperfine interactions. Magnetic field-dependent measurements reveal that charge transport in OLED devices is driven by triplet-charge annihilation in 3CzClIPN and 4CzIPN, while triplet-triplet annihilation dominates for 5CzBN. Theoretical calculations further indicate that SOC-mediated RISC in 3CzClIPN and 4CzIPN can additionally occur via a $T_\mathrm{2}$ intermediate state with an activation energy distinct from $\Delta E_\mathrm{{ST}}$. A temperature-dependent analysis of the devices was conducted to quantify this activation energy and compare it with the computational findings.
These findings establish key correlations between activation energy, spin dynamics, and magnetic field effects in TADF emitters, advancing our understanding of excitonic processes in OLEDs.

\end{abstract}

  
\section{Introduction}
Organic light-emitting diodes (OLEDs) play a crucial role in optoelectronic applications, including flat-panel displays, white light sources \cite{hong2021brief, dos2024golden}, and wearable electronics \cite{yin2016efficient, kim2013soft, kwon2018weavable, song2020organic}. Beyond these applications, OLEDs are also attracting growing interest in sensor technology \cite{hauck2024perspective} and biomedicine \cite{yin2016efficient, song2020organic}.

In recent years, organic thermally activated delayed fluorescence (TADF) materials have gained significant attention due to their intrinsic ability to enable reverse intersystem crossing (RISC) from the triplet to the singlet excited states \cite{dos2024golden, cole1941dispersion, wang2024boosting}. This process allows for a theoretical internal quantum efficiency of up to $100 \ \%$ without the need for heavy atoms\cite{adachi2001nearly}. By tuning donor and acceptor units, the energy gap between the singlet and triplet states ($\Delta E_\mathrm{{ST}}$) can be adjusted. To efficiently harness triplet excitons, a small $\Delta E_\mathrm{{ST}}$
in the range of thermal energy at room temperature is required \cite{adachi2014third, nakanotani2021thermally}. If $\Delta E_\mathrm{{ST}}$ is too large, RISC becomes inefficient, reducing overall device performance \cite{uoyama2012highly, shizu2015enhanced, kotadiya2019efficient}. However, the exact mechanism behind the spin state interconversion process remains a topic of debate and strongly depends on the choice of material\cite{wang2024understanding, tanaka2020understanding}.
For exciplex-based TADF materials, where charge-transfer complexes are formed between two distinct molecules, the dominant spin state interconversion mechanism is often attributed to the $\Delta g$-mechanism or hyperfine induced spin-mixing \cite{basel2016magnetic, wang2016immense}. In contrast, these effects appear to be less significant in excitonic TADF systems. Recently, Wang \textit{et al.} \cite{wang2024understanding} proposed that spin-orbit coupling (SOC) plays a major role in facilitating the spin state interconversion process and Gibson \textit{et al.} \cite{gibson2016importance} highlighted the critical role of vibronic coupling in enabling efficient RISC in TADF materials \cite{aizawa2020kinetic}. However, determining the underlying spin state interconversion mechanism experimentally is challenging. To investigate these interactions, magnetic fields are often used to induce Zeeman splitting, thereby lifting the degeneracy of the triplet states. Depending on the dominant spin state interconversion mechanism, the broadening of the magnetic field effect curves changes accordingly. These magnetic field effects (MFEs) are generally quantified using the following equation:

\begin{equation} 
\mathrm{MFE} = \frac{MFE(B)-MFE(B=0)}{MFE(B=0)} 
\label{Eq:MFE-general} 
\end{equation}

MFEs have been exploited to gain insight into the spin dynamics of emitter materials\cite{pan2022disorder, mondal2023degradation, wang2024understanding}. The MFE response is most commonly attributed to hyperfine interactions (HFIs) \cite{mondal2023degradation, schellekens2010exploring}, the $\Delta g$- mechanism \cite{wang2016immense}, and second-order processes \cite{morgenstern2024analysis, weber2024exciplex}, such as triplet-triplet annihilation (TTA) or triplet-charge annihilation (TCA) \cite{tanaka2020understanding}. In the case of TADF materials, MFEs have also been suggested to affect the RISC process \cite{wang2024understanding, wang2016immense, basel2016magnetic}.
Each mechanism exhibits a distinct fingerprint MFE curve, aiding in identifying the dominant effect influencing charge transport\cite{wang2024understanding}.\newline 
If the device is electrically biased, the electrons and holes injected from the electrodes can recombine and form weakly bound polaron pairs as either singlets ($PP_\mathrm{1}$) or triplets ($PP_\mathrm{3}$) on neighboring molecules \cite{geng2018effect, crooker2014spectrally, wang2024understanding}. $PP_\mathrm{3}$ includes the degenerate states $PP_\mathrm{3,+}$, $PP_\mathrm{3,0}$, and $PP_\mathrm{3,-}$. If the exchange interaction energy between electrons and holes is weak, spin state interconversion is facilitated by HFI, namely intersystem crossing (ISC) ($PP_\mathrm{1} \rightarrow PP_\mathrm{3}$) and RISC ($PP_\mathrm{3} \rightarrow PP_\mathrm{1}$) \cite{wang2024understanding, ehrenfreund2012effects}. \newline
Generally, the coupling strength, originating from either HFI or SOC, can be directly linked to a characteristic magnetic field via the Zeeman energy:  

\begin{equation}  
    B_0 = \frac{E_\mathrm{coup}}{g_\mathrm{e-h} \mu_B}  
\end{equation}  

where the energy \(E_\mathrm{coup}\) is determined using the Planck relation, which can be considered as the energy of the coupling strength:  

\begin{equation}  
    E = h c \tilde{\nu}  
\end{equation}  

with \(h\) being Planck constant, \(c\) the speed of light, and \(\tilde{\nu}\) the wavenumber. Here, \(g_\mathrm{e-h}\) represents the Landé factor of an electron-hole pair, and \(\mu_B\) denotes the Bohr magneton.\newline
$E_{coup}$ is proportional to $B_0$, so the broadening of the MFE curve can be directly associated with an increased coupling strength, as the characteristic magnetic field $B_\mathrm{0}$ is increased (Eq. \ref{Eq: DL fit}, \textit{vide infra}). HFI in organic molecules is generally weak, causing a characteristic magnetic field ($\leq 10$ mT), while a broader response is expected if SOC dominates \cite{wang2024understanding, nguyen2010isotope}.\newline
In addition to the spin state interconversion analysis, the application of an external magnetic field allows for the differentiation between various charge carrier transport mechanisms, as they exhibit distinct fingerprint responses \cite{wang2024understanding, hu2019spin}.
In TCA, a $T_\mathrm{1}$ exciton collides with a charge carrier ($q$), which makes the carrier scatter and the $T_\mathrm{1}$ states quench ($T_\mathrm{1} + q \rightarrow q' + S_\mathrm{0}$) \cite{wang2024understanding, zhao2023abundant}. As Wang \textit{et al.} \cite{wang2024understanding} explained, the Zeeman splitting of the $T_\mathrm{1}$ state causes a decrease of the rate constant of the TCA process when applying an external magnetic field. This means that the external magnetic field can lower the probability for TCA, therefore enhancing the $T_\mathrm{1}$ population. Consequently, more charge carriers can contribute to the RISC process, enhancing the MEL response \cite{hu2019spin, wang2024understanding}. It is important to note that the spin state interconversion can occur only from the $T_{n,0}$ state to the excited singlet state, since intra-triplet state mixing is suppressed according to the generation of triplet states due to the applied magnetic field. \newline

The TTA process can be described as $T_\mathrm{1} + T_\mathrm{1} \rightarrow S_\mathrm{n} + S_\mathrm{0}$ and is also affected by the Zeeman splitting of the $T_\mathrm{1}$ state. Below $|\sim 20|$ mT \cite{wang2024understanding, hu2019spin, tanaka2020understanding, ogiwara2015mechanism}, the zero-field splitting is stronger than the Zeeman splitting. Consequently, the annihilation rate constant of the $T_\mathrm{1}$ state is increased, which is reflected in an increase in the MEL response. Conversely, by increasing the magnitude of the magnetic field at $|B| > 20 \ \mathrm{mT}$, the Zeeman splitting becomes stronger than the zero-field splitting. Consequently, the annihilation rate  constant decreases, and the MEL response is decreased\cite{wang2024understanding}.
\newline
Differentiating between these mechanisms in emitter materials using MFEs can be challenging, often leading to overfitting in an attempt to account for all observed effects (\textit{e.g.} TCA, TTA, RISC, ISC) \cite{wang2024understanding, hu2019spin, zhao2023abundant}. In the case of the carbazole-based molecules under study, we propose a simple model based on the combination of only two Lorentzian fitting functions, which replicates our measured data and is in line with our theoretical investigations. To reduce the impact of all materials other than the emissive carbazole-based molecules, we fabricated simple layer stack devices containing only three organic layers, namely the hole-transport layer (PEDOT:PSS), the emissive layer (carbazole-based cyanoarene molecules), and the electron-transport layer (TPBi).  
Furthermore, our advanced measurement setup, coupled with its original software, enabled a rigorous statistical analysis, enhancing the reproducibility and reliability of our results. A comprehensive description of the hard- and software components of the MFE measurement setup, along with the analysis process, is provided in the Supplementary Information (see section 1.1. Experimental Apparatus and Analysis).

\bigskip
While differentiating between the mechanisms present in emitter materials can be easier from a computational point of view, computational approaches struggle with different problems when it comes to the description of spin state interconversion processes. (Spin) dynamics in excited states are challenging to model due to the complex net of state interactions\cite{mukherjee2021modeling} and long timescales needed to simulate the RISC process. In this work, we present a workflow that allows for the discussion of spin dynamics between the potential energy surfaces (PESs) of the involved excited states without the need for explicit excited state molecular dynamics simulations. A key component of this analysis is the identification of points of contact between different PESs, namely minimum energy crossing points (MECPs) between states of different spin multiplicity and minimum energy conical intersections (MECIs) between states of the same multiplicity \cite{bearpark1994direct, maeda2010updated}. When two PESs corresponding to different electronic states intersect, the emerging energy degeneracy allows for efficient transitions between these states. Consequently, knowing the location of these intersections enables the evaluation of dynamic processes without requiring molecular dynamics simulations.
\newline
The intersection of two PESs is defined by the vectors along which the energy degeneracy is lifted, the gradient difference vector ($GDV$) and the derivative coupling vector ($DCV$) \cite{matsika2011nonadiabatic}:

\begin{equation}  
    GDV = \frac{\partial (E_1 - E_2)}{\partial q}
\end{equation}  

\begin{equation}  
    DCV = \langle \Psi_1 | \left(\frac{\partial}{\partial q} \right)| \Psi_2 \rangle = \frac{\langle \Psi_1 | \left(\frac{\partial \hat{H}}{\partial q} \right)| \Psi_2 \rangle}{E_1 - E_2} 
\end{equation} 

with the latter determining how strongly the electronic wavefunctions change with respect to a change in the nuclear coordinates $q$. This two-dimensional subspace spanned by the $GDV$ and the $DCV$ is also called the branching plane. 
\newline
For MECPs, transitions occur between states of different spin multiplicities, meaning the $DCV$ is zero as there is no direct derivative coupling between the wavefunctions. Instead, transitions proceed via SOC. In this case, the branching plane is reduced to a one-dimensional hyperline. To find the energy minimum on this hyperline, \textit{i.e.} the MECP, Bearpark \textit{et al.} proposed the gradient projection method \cite{bearpark1994direct}. Here, the $\mathrm{GDV}$ is minimized along the hyperline while the gradient of $\mathrm{E_2}$ is projected onto the remaining space orthogonal to it and minimized there.
\newline
In the case of MECI optimization, both the $GDV$ and the $DCV$ are necessary. However, as the calculation of the $DCV$ is not widely available for all quantum chemistry methods and can be costly or potentially unreliable depending on the accuracy of the method, Maeda \textit{et al.} developed an approach that does not rely on the $DCV$ \cite{maeda2010updated}. The two-dimensional branching plane is instead updated in a gradient projection framework using only the $GDV$. At each optimization step, the branching plane is represented by two orthonormal unit vectors: A unit vector $x$ parallel to the $GDV$ and a unit vector $y$ perpendicular to it that needs to be estimated through the updating method as a surrogate for the $DCV$. The method relies on the fact that even with an inaccurate initial $y$, as the optimization proceeds around the region of the conical intersection, $x$ (and thus the $GDV$) minimizes the energy difference between the electronic states and leads towards the energy minimum. This induces changes that allow the iterative algorithm to refine the estimation of $y$.

\bigskip
By combining the outlined computational and experimental approaches, we can provide a comprehensive picture of the interconversion processes in carbazole-containing TADF emitters.
\section{Results and Discussion}
\subsection{Device Fabrication}
\begin{figure}[h]
  \centering
    \includegraphics[width=0.7\textwidth]{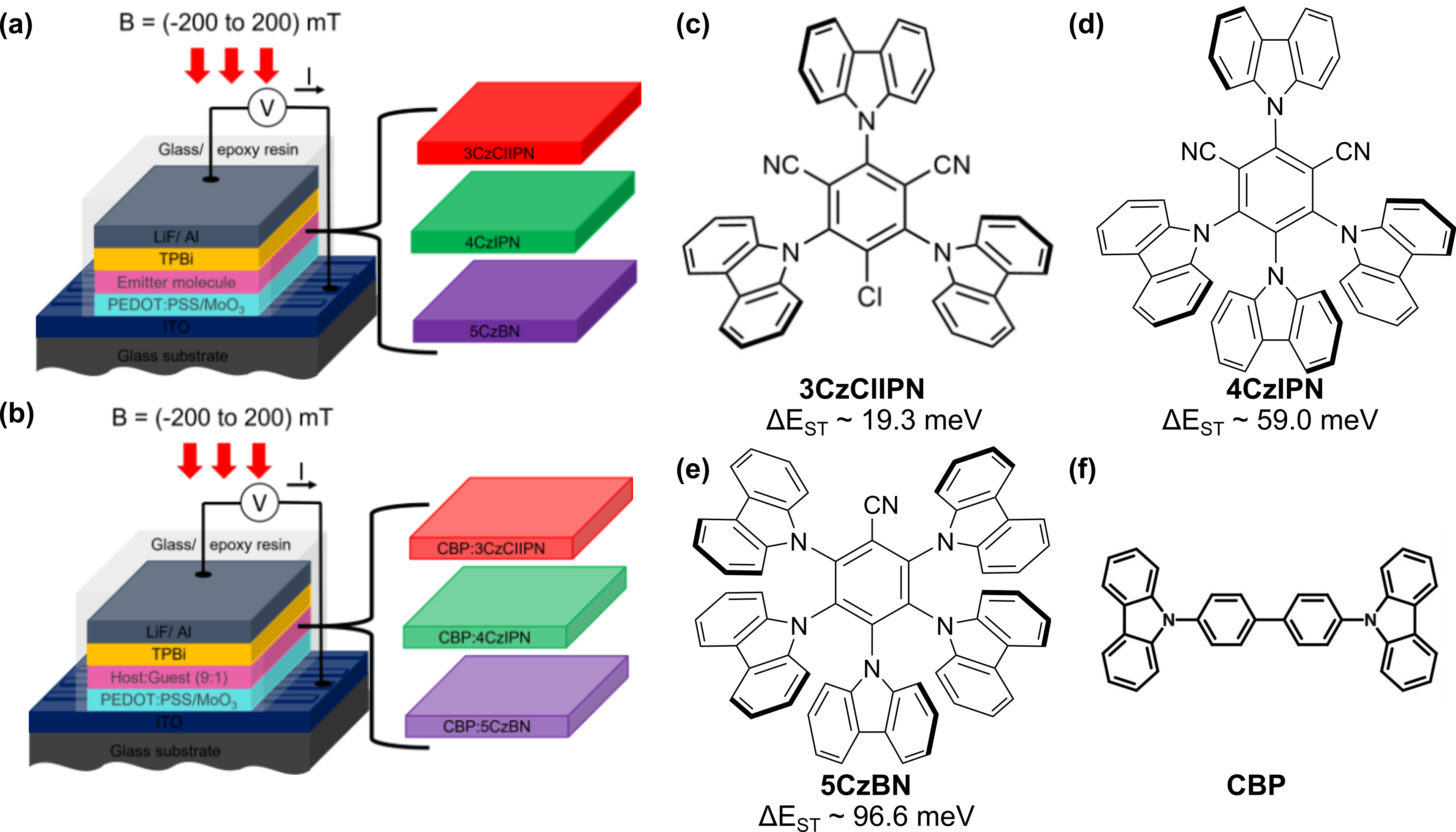} 
  \caption{Layer stacks of the devices fabricated using (a) the bare emitters, and (b) a host–guest system, along with the molecular structures of the TADF emitters in (c)–(e). The corresponding $\Delta E_{\mathrm{ST}}$, obtained from TD-DFT calculations, are also provided for (c) 3CzClIPN, (d) 4CzIPN, and (e) 5CzBN. The molecular structure of the host compound CBP is depicted in (f).}
  \label{fig:layer-stack-molecular-structure}
\end{figure}

\begin{figure}[h]
  \centering
    \includegraphics[width=0.7\textwidth]{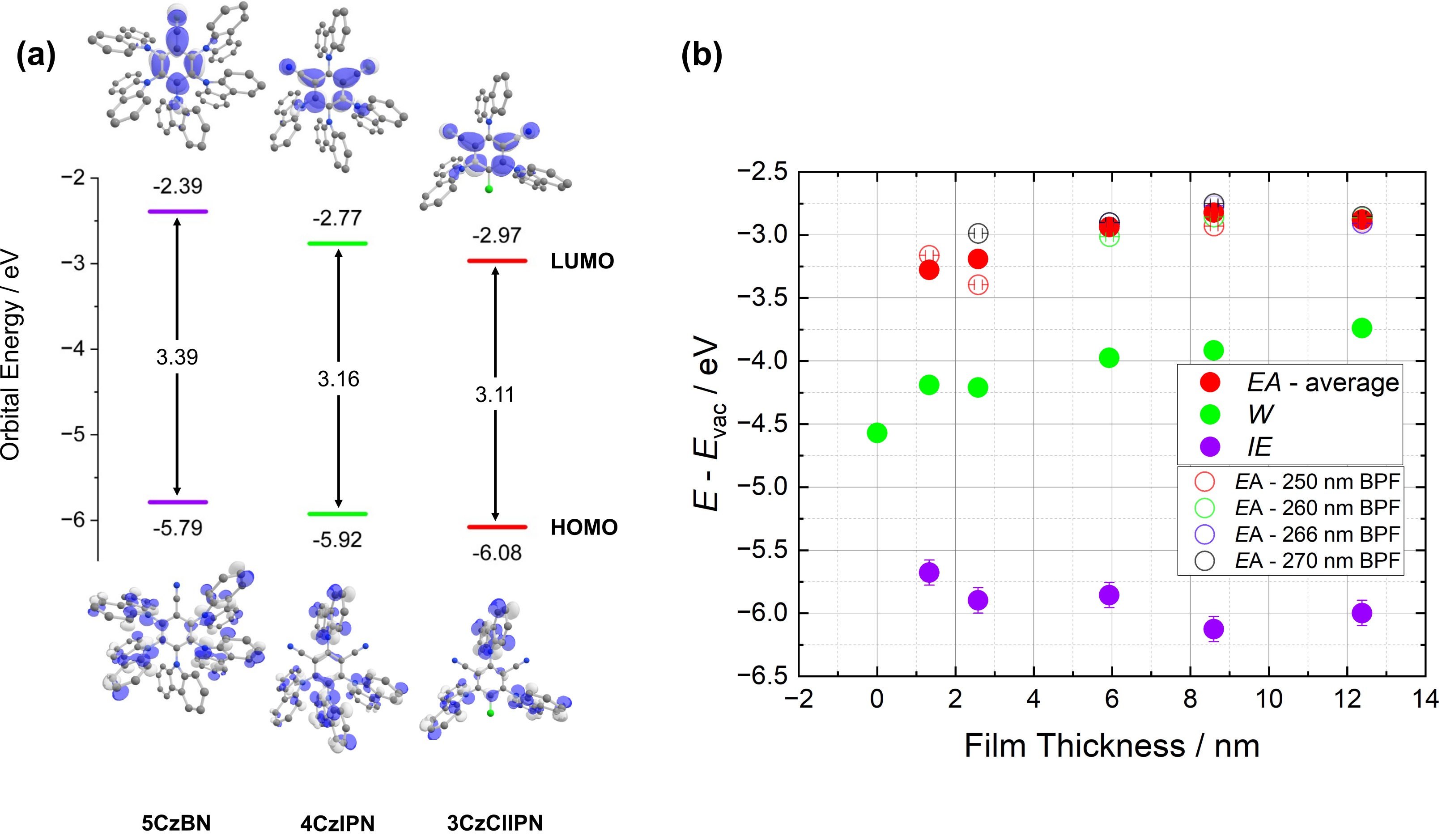} 
  \caption{(a) HOMOs and LUMOs of 5CzBN, 4CzIPN, and 3CzClIPN with corresponding absolute orbitals energies and corresponding HOMO-LUMO gaps obtained from DFT calculations. (b) Thickness-dependent energy diagram of the experimentally determined HOMO and LUMO onsets of 4CzIPN on Au, expressed by the ionization energy $IE$ and the electron affinity $EA$, as well as its work function $W$, with respect to the vacuum energy $E_\mathrm{vac}$. Multiple values of $EA$ measured with different bandpass filters (BPF) are shown to evaluate the statistical significance of the determined parameters.}
  \label{fig:HOMO-LUMO-DFT}
\end{figure}

The specific layer structure for all fabricated devices can be found in Figure \ref{fig:layer-stack-molecular-structure} (a) and (b). Throughout this study, the three cyanoarene molecules 2,4,6-Tri(9\textit{H}-carbazol-9-yl)-5-chloroisophthalonitrile (\textbf{3CzClIPN}), 1,2,3,5-Tetrakis(carbazol-9-yl)-4,6-dicyanobenzene (\textbf{4CzIPN}), and Penta-carbazolylbenzonitrile (\textbf{5CzBN}) were compared as bare molecular devices as well as embedded in a host-guest system with 4,4´-Bis(\textit{N}-carbazolyl)-1,1´-biphenyl (CBP) as host material (ratio host:guest~=~9:1). Their molecular structure is depicted in Figure \ref{fig:layer-stack-molecular-structure} (c)-(f) with the respective calculated $\Delta E_\mathrm{{ST}}$ for the TADF emitters (\textit{cf.} (c)-(e)). $\Delta E_{\mathrm{ST}}$ can be reduced by strengthening the acceptor unit(s) \cite{bassan2023visible}. In contrast, the introduction of carbazole (donor) units increases $\Delta E_\mathrm{{ST}}$. This is closely tied to the influence of donor and acceptor units on the frontier orbitals (Figure \ref{fig:HOMO-LUMO-DFT}). Since RISC in TADF molecules depends on efficient energy transfer, a small $\Delta E_{\mathrm{ST}}$ is preferred for optoelectronic applications such as OLEDs \cite{dos2024golden}.

To assess the quality of the band alignment (cf. Figure S11), the positions of the highest occupied molecular orbital (HOMO) and lowest unoccupied molecular orbital (LUMO) were calculated with DFT for all molecules. These results were then compared to values obtained through a combination of low-energy inverse photoemission (LEIPES) and ultraviolet photoemission spectroscopy (UPS) for validation.\newline
The computational results are shown in Figure \ref{fig:HOMO-LUMO-DFT} (a), illustrating a continuous increase in the band gap from 3CzClIPN to 5CzBN. Additionally, both the HOMO and LUMO positions exhibit an upward shift in energy, following this same trend.
Figure \ref{fig:HOMO-LUMO-DFT} (b) shows the experimentally determined values of the HOMO and LUMO onsets and the work function $W$ for 4CzIPN with respect to the vacuum energy $E_\mathrm{vac}$ from UPS and LEIPES measurements. The work function determined for the Au substrate is $(4.57 \pm 0.05) \ \mathrm{eV}$, which is significantly lower than typical values from literature (\textit{e.g.} $5.4$ eV \cite{uda1998work}). This is due to the presence of an adventitious carbon layer, which was not removed before the deposition of 4CzIPN to keep conditions similar to the samples that were prepared outside UHV conditions using spin-coating. With increasing organic film thickness, a progressive shift of the work function $W$ corresponding to a movement of the Fermi level towards the vacuum level can be observed. This is accompanied by an apparent opening of the HOMO-LUMO gap, which saturates near a film thickness of $8$ nm. Such effects were often observed in relation to ultra-thin films of organic semiconductors on metallic substrates. In phthalocyanine molecules, this was associated with the formation of an image charge potential, due to the mirroring of the molecular electron cloud in the conductive substrate \cite{flores2009modelling}. The energy parameters determined for the largest film thickness can be considered as the values of the bulk material. They amount to $EA = (2.88 \pm 0.03)$ eV for the electron affinity, $IE = (6.0 \pm 0.1) \ \mathrm{eV}$ for the ionization energy, and $W = (3.74 \pm 0.05) \ \mathrm{eV}$. This adds up to a transport bandgap of $(3.12 \pm 0.13) \ \mathrm{eV}$. The uncertainty of $EA$ is given here by the standard deviation of the values determined from measurements under different conditions, \textit{i.e.} using different bandpass filters (BPFs), as proposed by Sugie \textit{et al.} \cite{sugie2023dependence}. The measured Fermi level being closer to the LUMO onset than to the HOMO onset suggests an n-type character of the organic film. \newline
The obtained values for the electron affinity and ionization energy can be associated to the LUMO and HOMO position, respectively. The obtained values are in very good agreement with the DFT results.\newline
A more detailed description of the UPS, XPS, and LEIPES experiments and their analysis can be found in the Supporting Information (cf. section 1.1.4 and Figures S8-S10) .

\subsection{Optoelectronic Characterization}
\begin{figure}[h]
\centering
  \includegraphics[width=0.7\textwidth]{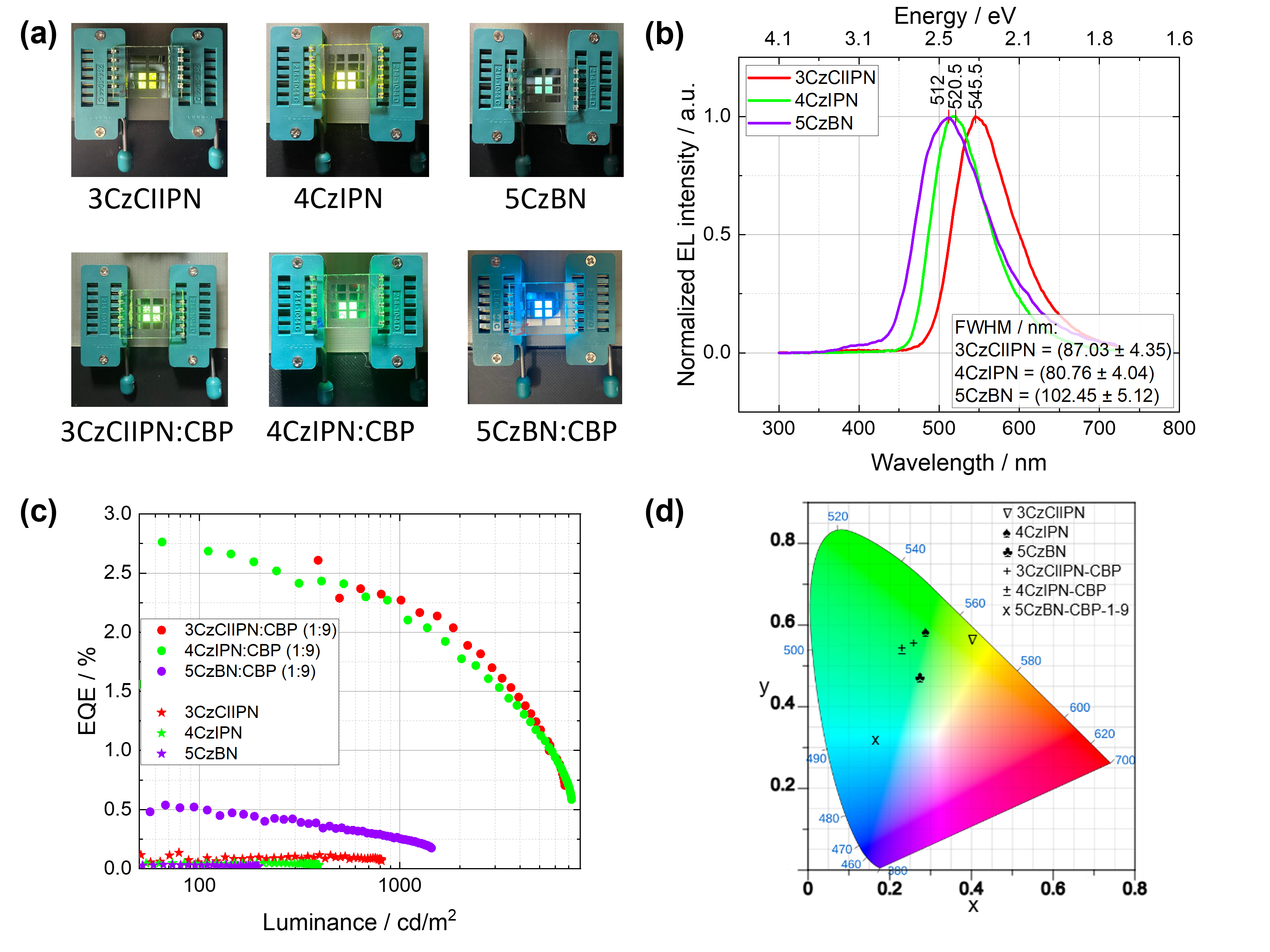}
  \caption{(a) Images of the devices fabricated in this work with four of the 8 pixels emitting light, (b) shows the EL response for the bare molecules depicted in (a). (c) EQE for the simple layer stack with the bare molecules compared to the host-guest system, containing CBP as a host material and showing a strong enhancement. (d) CIE chromaticity diagram extracted from the EL response (\textit{cf.} Figure \ref{fig:optoelectronic-characterization} (b)) }
  \label{fig:optoelectronic-characterization}
\end{figure}

\begin{figure}[h]
\centering
  \includegraphics[width=0.7\textwidth]{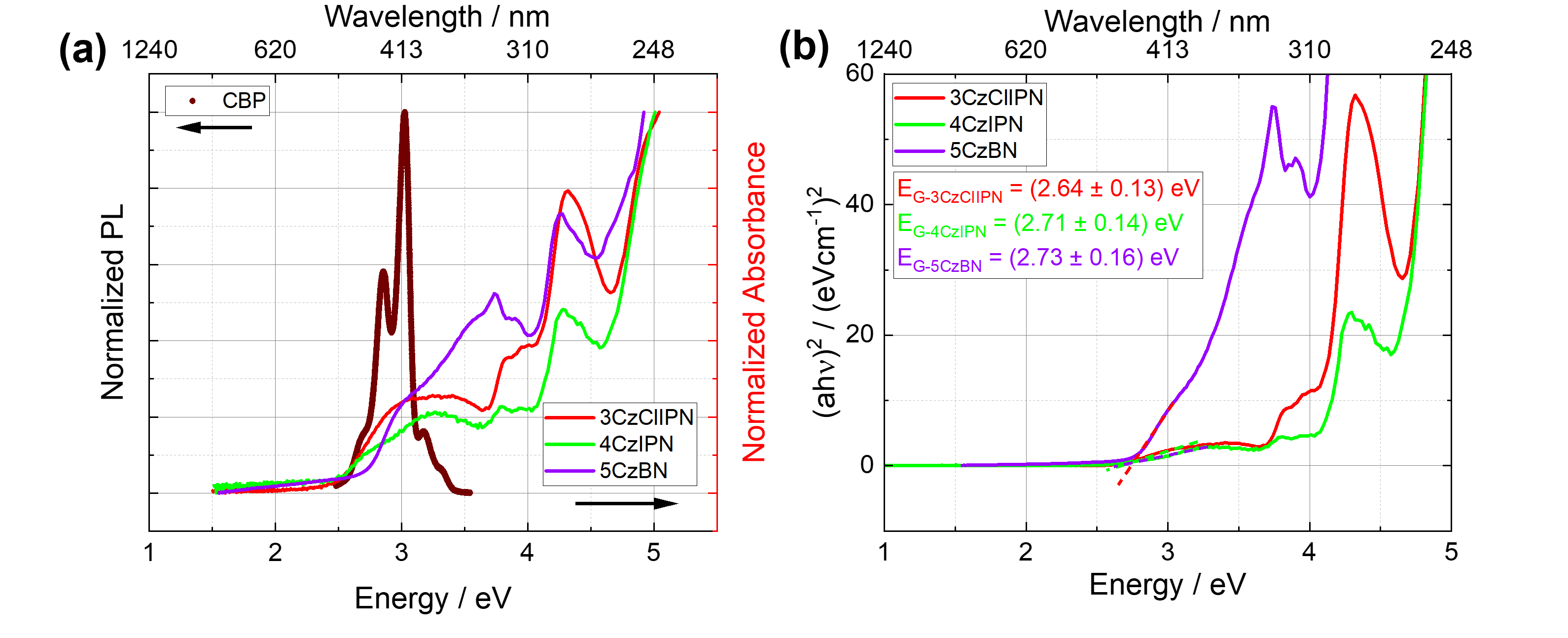}
  \caption{(a) Absorbance spectra for the emitter molecules and PL spectra for the CBP host. The absorbance of all emitter materials overlaps with the PL of CBP indicating an efficient FRET. (b) Tauc plot with energy band gap for the bare emitter materials. $\Delta E_\mathrm{{opt}}$ slightly increases with increasing number of carbazole units, which is well in line with the observed blue shift in the PL and EL response (\textit{vide supra}).}
  \label{fig:absorption}
\end{figure}

The difference in emitted color is visible to the naked eye when comparing the light emission of the investigated bare molecular devices (\textit{cf.} Figure \ref{fig:optoelectronic-characterization} (a)). As can be seen in Figure \ref{fig:optoelectronic-characterization} (b), the electroluminescence (EL) response shows a blue shift as the number of carbazole units increases. This can be explained by an increase in the HOMO-LUMO gap from 3CzClIPN to 5CzBN as shown in Figure \ref{fig:HOMO-LUMO-DFT}. Since the HOMO is localized on the carbazole donor units and the LUMO on the acceptor unit, changes in the donor or acceptor composition directly affects the energy gap. The trend from 3CzClIPN to 5CzBN shows that successively adding carbazole donor units leads to a steady increase in the energy of the HOMO due to the increasing electron-donating effect of the electron-rich carbazole moiety. Conversely, starting from 5CzBN and going to 3CzClIPN, introducing electron-withdrawing groups such as CN and Cl to the acceptor unit lowers the energy of the LUMO. Notably, while the relative energy increase of the HOMOs remains similar across the series from 3CzClIPN to 5CzBN, the relative energy decrease observed for the LUMOs differs more significantly. Adding a chlorine atom (from 4CzIPN to 3CzClIPN) has a much weaker effect than adding a cyano group (from 4CzIPN to 5CzBN), resulting in a comparatively large HOMO-LUMO gap for 5CzBN while the gap for 4CzIPN and 3CzClIPN remains similar. This trend in the HOMO-LUMO gaps is closely related to the respective $\Delta E_\mathrm{{ST}}$ of the molecules as the first singlet and triplet excited states are characterized by HOMO to LUMO transitions.
\newline
The photoluminescence (PL) (\textit{cf.} Figure S12 (d) in the Supporting Information) and the EL response exhibit a strong correlation, suggesting that the molecules behave similarly under optical and electrical excitation.
The corresponding CIE (\textit{Commission internationale de l'éclairage}) chromaticity diagram is given in Figure \ref{fig:optoelectronic-characterization} (d) with the CIE values summarized in Table S2. 
\newline
In the host-guest system, a distinct blue shift in the EL response is observed (\textit{cf.} Figure S12 (c)). This shift is attributed to environmental changes induced by CBP, preventing aggregation-induced quenching, which can also lead to a blue shift in the EL response \cite{stavrou2020photophysics}. The reduced full width at half maximum values (FWHM) demonstrate the higher selectivity along with the more efficient light emission as interactions between the emitter molecules are decreased, which can be attributed to a modification of the energy transfer processes, \textit{e.g.} Förster (FRET) and Dexter energy transfer (DET) \cite{stavrou2020photophysics, karunathilaka2020suppression}. FRET mainly depends on the overlap of the PL spectrum of the host molecule and the absorption spectrum of the guest molecule \cite{wang2024understanding}. As shown in Figure \ref{fig:absorption} (a), all TADF emitters exhibit a strong overlap with the PL spectrum of CBP, indicating efficient FRET between the host and guest molecules. This is corroborated by simulated absorption and PL spectra (see Figure S17). From the UV-vis absorption spectra, we also determined the optical band gap ($\Delta E_\mathrm{{opt}}$) using the Tauc plot method assuming a direct band gap \cite{viezbicke2015evaluation}. The exact experimental values of $\Delta E_\mathrm{{opt}}$ for the three molecules are presented in Figure \ref{fig:absorption} (b). These values correlate with the increasing HOMO-LUMO energy gap, and with the observed blue shift in the PL and EL responses correlated with the increasing number of carbazole units in the cyanoarene molecules. 

Considering the external quantum efficiency (EQE) results in Figure \ref{fig:optoelectronic-characterization} (c), the introduction of the host-guest system significantly enhances the EQE compared to the device consisting only of bare molecules. An increase in the EQE was detected when using a material with a smaller $\Delta E_\mathrm{{ST}}$, as the RISC process is more efficient. The $\mathrm{EQE_{max}}$ for 3CzClIPN and 4CzIPN is very similar, while 5CzBN shows a much lower EQE related to the larger $\Delta E_\mathrm{{ST}}$. The exact values for the optoelectronic characterization are summarized in Table S1. As mentioned, the increased EQE when comparing the host-guest system to the bare molecular systems can be attributed to better charge carrier transport according to an improved energy transfer, mostly originating from FRET (\textit{cf.} Figure \ref{fig:absorption}). \newline Nevertheless, as the devices are not optimized, the total EQE values remain quite low. Due to the higher emitter concentration in the bare molecular devices, a concentration quenching effect is assumed. The lack of separation of emitter and transport functions in such layers favors exciton quenching \cite{zhang2019suppressing}. The relatively high energy barriers (cf. Figure S11) and the strong efficiency roll-off could be compensated in further investigations by improving the band alignment using transport and blocking layers \cite{nakanotani2013promising}. However, the lower EQE positively impacts the MFE amplitude, as will be discussed in detail below. 

\begin{figure}
    \centering
    \includegraphics[width=0.3\linewidth]{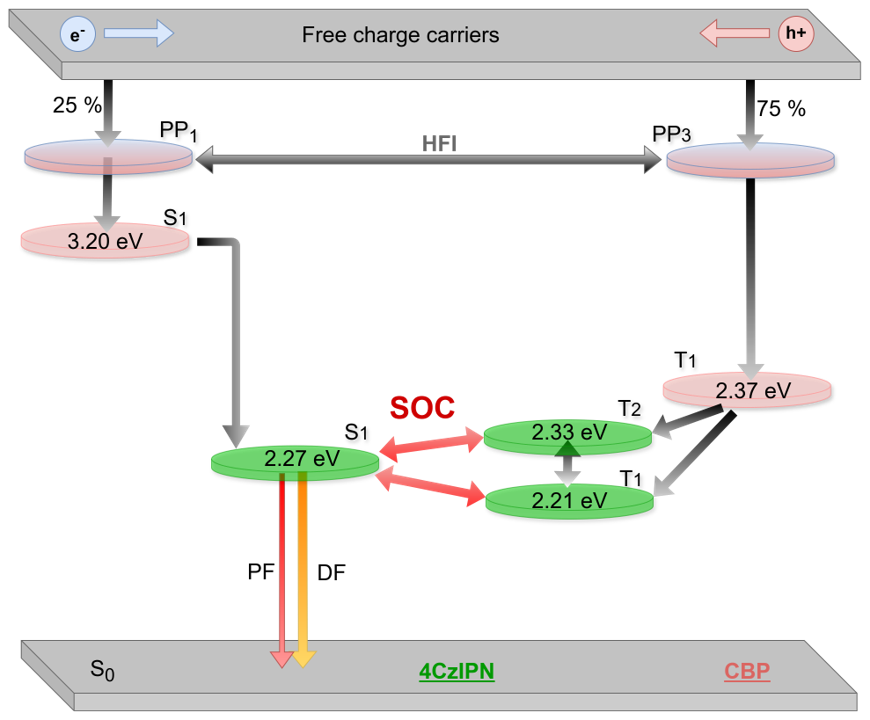}
    \caption{Jablonski diagram of 4CzIPN based on TD-DFT calculations. Depicted is the energy alignment of 4CzIPN relative to the CBP matrix, with a focus on its application in a host-guest system.}
    \label{fig:Jablonski 4CzIPN}
\end{figure}

Examining the calculated relative energy differences between the triplet energy levels of CBP and the excitonic TADF emitters reveals a difference of  $0.22$ eV for 3CzClIPN (\textit{cf.} Figure S18 (a)), $0.16$ eV for 4CzIPN (\textit{cf.} Figure \ref{fig:Jablonski 4CzIPN}), and $0.0$2 eV for 5CzBN (\textit{cf.} Figure S18 (b)). Since DET mainly depends on the alignment of the host and guest triplet states \cite{kim2014study}, it should be most efficient for the combination of CBP and 5CzBN. However, the optoelectronic properties, particularly the EQE, remain significantly higher for the other two TADF emitter molecules. This indicates that DET plays a less critical role in these devices compared to the intrinsic properties of the TADF emitter.

\subsection{Characterization of Magnetic Field Effects}
\begin{figure}[h]
\centering
  \includegraphics[width=1\textwidth]{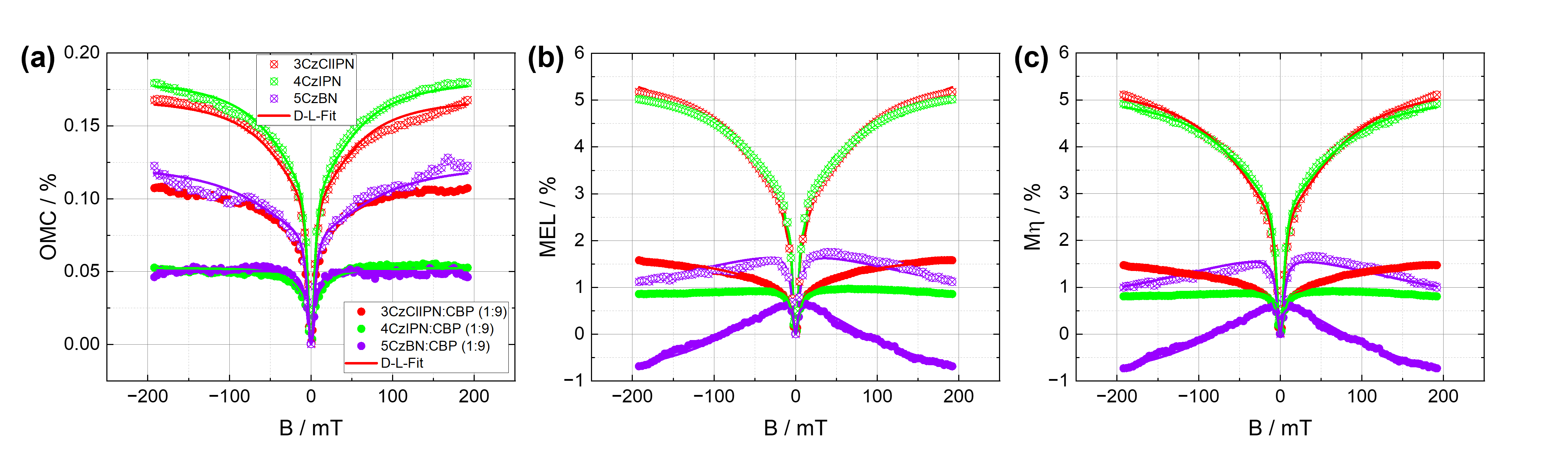}
  \caption{OMC (a), MEL (b), and $M_\mathrm{\eta}$ (c) detected at a constant current density of $175 \ \mathrm{mA/cm^2}$ for the bare molecules and the host-guest system containing CBP as host matrix.}
  \label{fig:MEL-OMC-Mn}
\end{figure}

As discussed in section 2.2, the introduction of a host-guest system using CBP as the host matrix dramatically increased EQE (\textit{cf.} Figure \ref{fig:optoelectronic-characterization}). In Figure \ref{fig:MEL-OMC-Mn}, the organic magnetoconductance (OMC), magnetoelectroluminescence (MEL), and magnetoefficiency ($M_\mathrm{\eta}$) responses of the bare molecular devices are compared to their respective host-guest equivalents as an example at a constant current density of $175\ \mathrm{mA/cm^2}$. Since the OMC is less sensitive to radiative recombination and the MEL is affected by both radiative and non-radiative recombination, we propose the magneto-efficiency (\(M_\mathrm{\eta}\)) as the most reliable metric. The relationship between the EL and the current ($I$) in OLEDs can generally be described by the formula $\mathrm{EL} = \eta I/e$ \cite{zhang2009low}, with $\eta$ being the EL quantum efficiency and $e$ the elementary charge. If $\eta$ is B-dependent, the following relationship exists \cite{pan2019extraordinary}:

\begin{equation}
    M_\mathrm{\eta} = \mathrm{MEL - OMC}
    \label{Eq: Magnetoefficiency}
\end{equation}

The OMC signal was found to be quite low, which is in line with previous observations \cite{wang2016immense, wu2022identifying}. Interestingly, the opposite trend was observed when compared to the EQE analysis: The magnetic field effects are significantly quenched as device efficiency is improved. Niedermeier \textit{et al.} \cite{niedermeier2008enhancement} demonstrated that an electrical conditioning procedure could enhance the MFE response, which was associated with introducing trap states. As device efficiency increases, charge transport is also improved. Consequently, the interaction time of charge carriers with the external magnetic field is reduced, leading to a suppression of the MFE. Additionally, concentration quenching effects need to be considered\cite{kim2017concentration}, since the emitter concentration decreases by introducing a host-guest system, lowering the probability of interactions between the TADF emitters and the magnetic field. 

\begin{figure}[h]
\centering
  \includegraphics[width=1\textwidth]{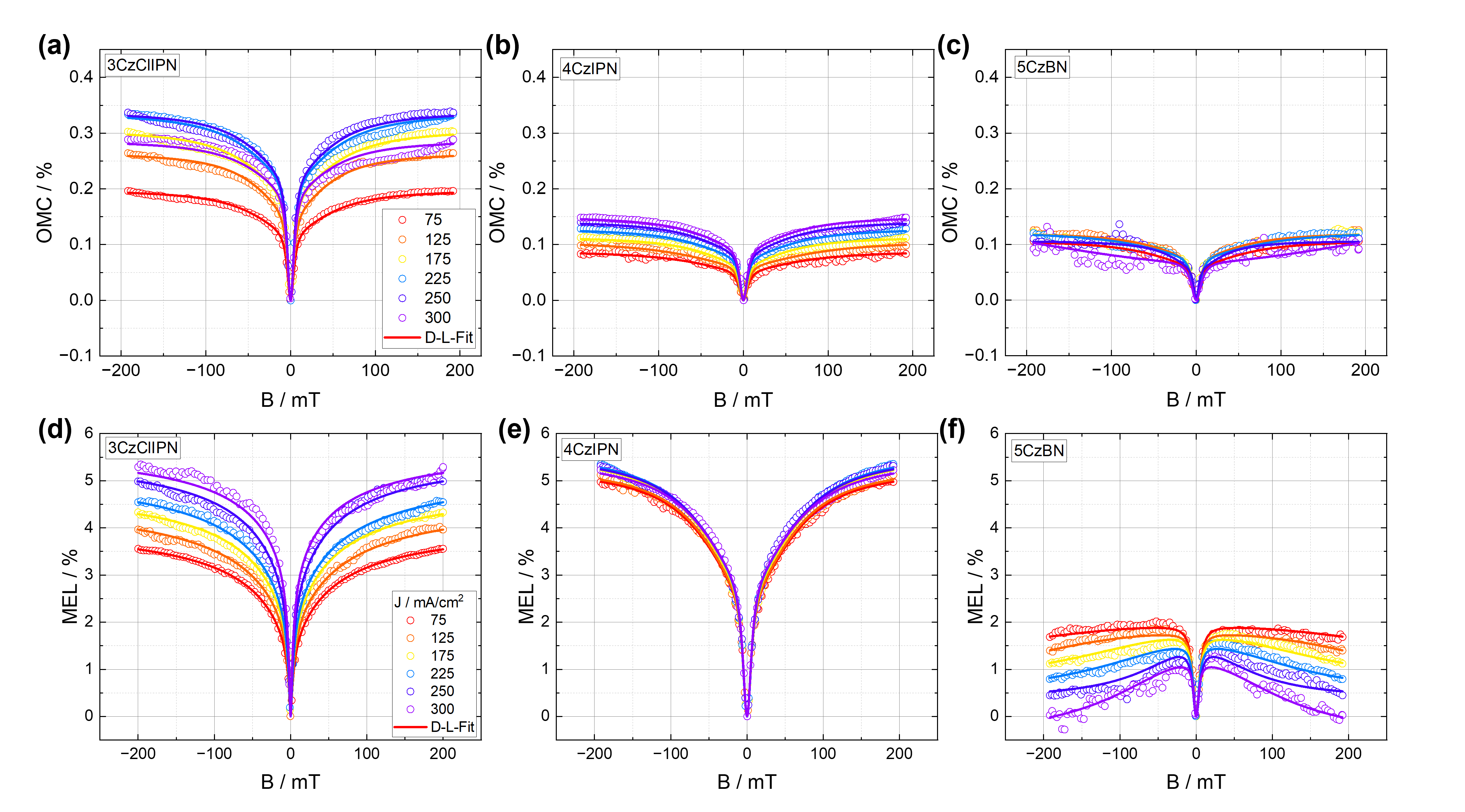}
  \caption{OMC (a)-(c) and MEL (d)-(f) detected at current densities from $75 \ \mathrm{mA/cm^2}$ up to $300 \ \mathrm{mA/cm^2}$  for the bare molecules 3CzClIPN (a) and (d), 4CzIPN (b) and (e), and 5CzBN (c) and (f).}
  \label{fig:MFE-I-dependency}
\end{figure}

One can further observe in both cases, for the bare emitter devices and their host-guest systems, that $M_\mathrm{\eta}$ is quenched upon increasing $\Delta E_\mathrm{{ST}}$. The MEL response of 5CzBN exhibits a markedly different lineshape compared to those of the other two molecules.
For the following analysis, we will focus on the bare molecular systems, as their measured MFE signals are stronger and hence easier to analyze. To gain deeper insights, the OMC and MEL responses were measured at various current densities, as demonstrated in Figure \ref{fig:MFE-I-dependency} for the bare molecular devices. From left to right, the molecules exhibit an increasing $\Delta E_\mathrm{ST}$, which reduces the likelihood of RISC \cite{aizawa2020kinetic}. As expected, the MFE decreases accordingly, emphasizing its sensitivity to the RISC process. Interestingly however, the lineshape does not replicate the characteristic fingerprint curves of the RISC process, which would typically correspond to a negative MEL response \cite{wu2022identifying, wang2024understanding}. \newline

\begin{figure}[h]
\centering
  \includegraphics[width=0.7\textwidth]{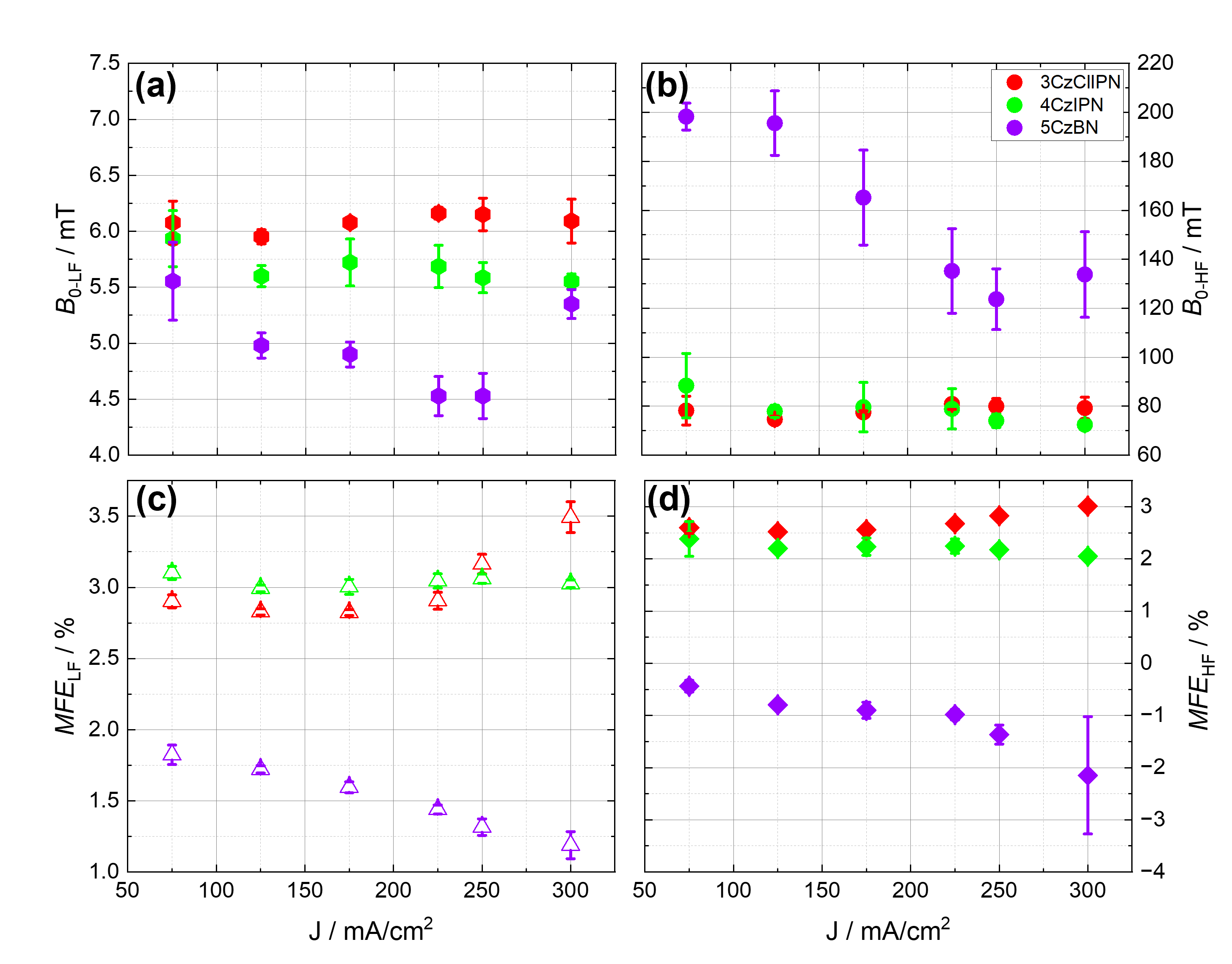}
  \caption{Fit analysis of the characteristic values  obtained from the $M_\mathrm{\eta}$ curves in dependence of the current density for the three bare emitter molecular devices based on 3CzClIPN, 4CzIPN, and 5CzBN, respectively. The parameters obtained from the Double-Lorentzian fit defined in Eq. \ref{Eq: DL fit} include the characteristic magnetic fields for the low- and high-field regimes, $B_\mathrm{0-LF}$ and $B_\mathrm{0-HF}$. Additionally, the amplitudes corresponding to these effects are represented by $MFE_\mathrm{{LF}}$ and $MFE_\mathrm{{HF}}$, respectively. }
  \label{fig:Magnetoefficiency-ramp-data}
\end{figure}

For a more detailed analysis, the curves depicted in Figure \ref{fig:MFE-I-dependency} were fitted by a combination of two Lorentzian functions for the $M_\mathrm{\eta}$ response:

\begin{equation}
    M_\mathrm{\eta_{fit}}(\%) = {\frac{MFE_\mathrm{LF} B^2}{B^2 + B_\mathrm{0-LF}^2} + \frac{MFE_\mathrm{HF} B^2}{B^2 + B_\mathrm{0-HF}^2}}
    \label{Eq: DL fit}
\end{equation}

with $MFE_\mathrm{{LF}}$ and $MFE_\mathrm{{HF}}$ corresponding to the amplitudes and $B_\mathrm{{0-{LF}}}$ and $B_\mathrm{{0-{HF}}}$ corresponding to the broadening of the curves (half width at half maximum - HWHM) for the low-field (LF) and high-field (HF), respectively. The two terms determine the major effects corresponding to the LF ($< |20| \ \mathrm{mT}$), and HF effect regime ($>20 \ \mathrm{mT}$). The broadening $B_\mathrm{0}$ of the $M_\mathrm{\eta}$ response is also determined as the characteristic magnetic field. A comparison of several models (cf. Figure S13 and Table S4) showed that Eq. (\ref{Eq: DL fit}) best replicates the $M_\mathrm{\eta}$ response. The error bars shown in the diagrams are derived from the standard deviation of ten separate measurements.
Upon examination of the data presented in Figure \ref{fig:Magnetoefficiency-ramp-data}, a decrease in the $M_\mathrm{\eta}$ response is observed with increasing $\Delta E_\mathrm{ST}$, which is well replicated in the $MFE_\mathrm{{LF}}$ variable in Figure \ref{fig:Magnetoefficiency-ramp-data} (c). 

Interestingly, the $M_\mathrm{\eta}$ response for 4CzIPN is slightly higher than that of 3CzClIPN, particularly at lower current densities (\textit{cf.} Figure \ref{fig:Magnetoefficiency-ramp-data} (c)). A comparison of the MFE results with the J-V and Luminance-V responses (\textit{cf.} Supporting Information Figure S12 (a) and (b)) reveals that the variations in turn-on voltage (determined at $10 \ \mathrm{cd/m^2}$) and operational regimes are well reflected in the MFE responses, causing the different onset of the increasing $MFE_\mathrm{{LF}}$. The high turn-on voltage observed for 3CzClIPN is attributed to a relatively large energy barrier, as evidenced by the band alignment (cf. Figure S11). $B_\mathrm{{0-{LF}}}$ exhibits values below $10$ mT for all three emitter molecules, which was previously assigned to hyperfine-driven interaction between polaron pairs \cite{nguyen2012isotope, wang2024understanding}. A decrease was observed with increasing $\Delta 
E_\mathrm{ST}$, suggesting an enhanced HFI as the acceptor strength increases and the number of carbazole units decreases. Additionally, a slight increase with rising current density was noted. We therefore attribute the detected low-field effect, with ${B_\mathrm{0-LF}} \sim 5-6$ mT, to hyperfine-induced spin state interconversion of the polaron pair species in 3CzClIPN and 4CzIPN. \newline 

For 5CzBN, a pronounced decrease in the broadening of $B_\mathrm{{0-LF}}$ was observed. Since the characteristic magnetic field is directly proportional to HFI, the decrease in HFI with increasing current density results in a significant drop in efficiency, as singlet polarons can no longer be harvested. This trend is further supported by the negative correlation of $MFE_\mathrm{{LF}}$, as shown in Figure \ref{fig:Magnetoefficiency-ramp-data} (c).
For 3CzClIPN and 4CzIPN, the amplitude of the effect increases with increasing current density (\textit{cf.} $MFE_\mathrm{{LF}}$ and $MFE_\mathrm{{HF}}$), revealing a positive low- and high-MFE. The $MFE_\mathrm{{HF}}$ behavior observed for 3CzClIPN and 4CzIPN is a clear indicator for triplet- charge annihilation (TCA)\cite{tanaka2020understanding, wu2022identifying}. The values extracted for ${B_\mathrm{0-HF}}$ are in the range of $80$ mT (\textit{cf.} Figure S14 (a)), which are well in line with the characteristic magnetic field determined from spin-orbit coupling (SOC) obtained from TD-DFT calculations (\textit{cf.} Table S5).
By increasing $\Delta E_\mathrm{ST}$, the efficiency of the RISC dramatically decreases. For 5CzBN, which is exhibiting the highest $\Delta E_\mathrm{ST}$, even a slight negative trend for the $MFE_\mathrm{{LF/HF}}$ upon applied current density was observed (\textit{cf.} Figure \ref{fig:Magnetoefficiency-ramp-data} (c) and (d)). \newline
The results obtained for 5CzBN differ significantly from those observed for 3CzClIPN and 4CzIPN, especially for the MEL response (\textit{cf.} Figure \ref{fig:MFE-I-dependency} (f)), which is further reflected in $MFE_\mathrm{HF}$ exhibiting negative values. Those are clear indicators of a dominant triplet-triplet-annihilation (TTA) mechanism.
\newline
Comparing our analysis to previous ones (\textit{e.g.} \cite{wang2024understanding, zhao2023abundant}), a noticeable discrepancy arises. Their analysis employs a different fitting function, which separately determines the characteristic field of the RISC and TCA processes for the MEL response. However, from our perspective, this fitting function overfits the data. \newline
Furthermore, under an applied magnetic field, the TCA process is quenched, thereby enhancing the RISC process. This implies that TCA and RISC cannot be considered independent processes when TCA is present, as MFE measurements do not allow for their differentiation. This finding is further supported by the observed trend with increasing $\Delta E_\mathrm{ST}$, which must stem from changes in the efficiency of the RISC process. In the case of 5CzBN, the characteristic field for the HF effect is significantly higher compared to 3CzClIPN and 4CzIPN. Given the quenched MEL response at increased current density and the calculated characteristic magnetic field for SOC, the TCA process appears to play a minor role in this material system. For a deeper analysis of the model fit, we refer to the Supporting Information, especially to Figure S13 and Table S3 and S4. \newline

\subsection{Evaluation of Spin Dynamics}
\begin{figure}[h] 
\centering
\includegraphics[width=0.7\textwidth]{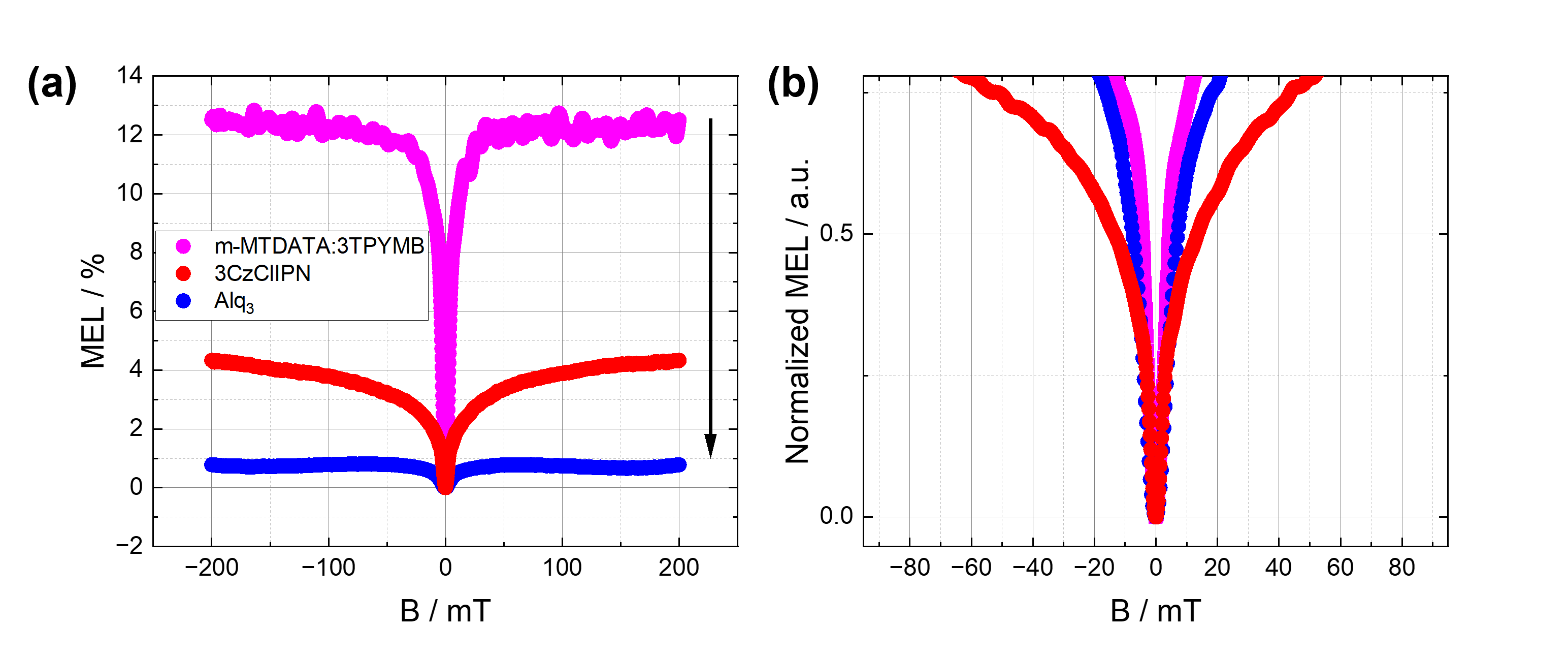}
  \caption{MEL detected at a constant current density of $175 \mathrm{mA/cm^2}$ for 3CzlIPN (a) compared to the reference device with m-MTDATA:3TPYMB as a well-known exciplex system with dominant $\Delta g$ and $\mathrm{Alq_3}$ as a non-TADF emitter with the main spin state interconversion mechanism known to be hyperfine interaction. (b) Normalized MEL response for the materials shown in (a) for better visibility of the difference in broadening.}
  \label{fig:MFE-mol-comparison-and-ref-device}
\end{figure}

To unravel the spin state interconversion process between the excited triplet and singlet excitonic states, we first fabricated reference devices with well-established properties. We examined $\mathrm{Alq_3}$ as a non-TADF emitter \cite{kalinowski2003magnetic} and a well-known exciplex-forming TADF system consisting of m-MTDATA:3TPYMB at a donor-to-acceptor concentration ratio of 1:4 \cite{wang2016immense}. In Figure \ref{fig:MFE-mol-comparison-and-ref-device} (a), the MEL response for the reference devices is compared to the one obtained for 3CzClIPN at a constant current density of $J_\mathrm{OLED} = 175\ \mathrm{mA/cm^2}$. The absolute MEL at the highest applied magnetic field ($B_\mathrm{max} = 200$ mT) differs. $\mathrm{Alq_3}$ exhibits a lower absolute MEL response compared to m-MTDATA:3TPYMB and 3CzClIPN while the broadening of the curve is much more narrow compared to 3CzClIPN, as can be observed in the normalized plot in Figure \ref{fig:MFE-mol-comparison-and-ref-device} (b). The main spin state interconversion mechanism was previously attributed to hyperfine-induced spin-mixing of polaron pairs \cite{schellekens2010exploring, nguyen2012isotope}. The MEL lineshape observed for m-MTDATA:3TPYMB can be mainly attributed to the $\Delta g$ mechanism as the dominant spin state interconversion process\cite{wang2016immense, wang2024understanding} while the broadening of the curve remains, as can be observed in the normalized plot in Figure \ref{fig:MFE-mol-comparison-and-ref-device} (b), similar to $\mathrm{Alq_3}$. According to the spin state interconversion properties, the small $\Delta E_\mathrm{{ST}}$ for exciplex materials, and consequently the efficient RISC process, they are well known to exhibit a strong MEL response \cite{wang2016immense, basel2016magnetic, lei2016ultralarge}. The small RISC activation energy, below the thermal energy at room temperature ($ \sim 25$ meV), was identified as the primary cause of the pronounced magnetic field effects \cite{ling2015large}.

\begin{figure}[h]
\centering
  \includegraphics[width=0.7\textwidth]{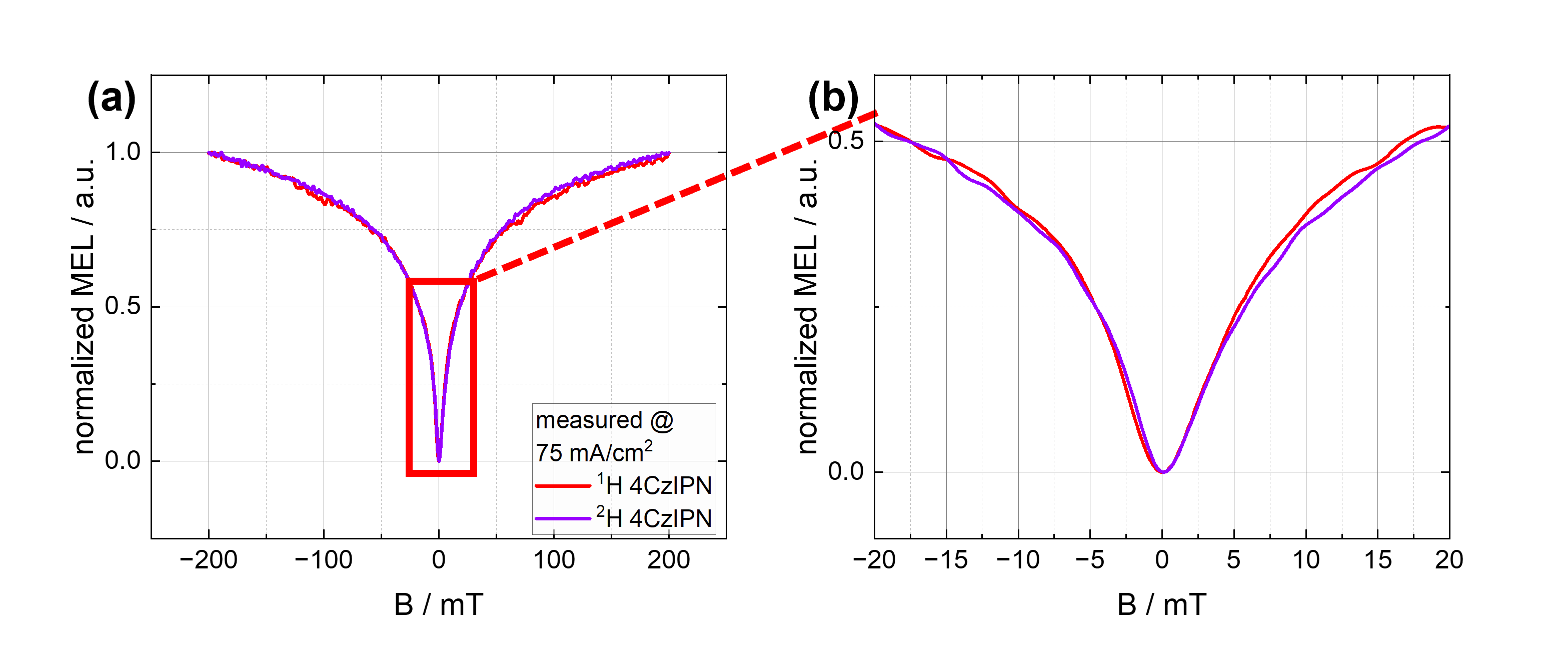}
  \caption{(a) represents the normalized MEL response for the $^1\mathrm{H}$ and $^2\mathrm{H}$ isotopologues of 4CzIPN at a constant current density of $75 \ \mathrm{mA/cm^2}$, (b) shows the inset for the data obtained in (a) in the range from -20 to 20 mT. No difference in the broadening of the curves or the lineshape could be observed.}
  \label{fig:MFE-deuteration}
\end{figure}

In addition to the difference in the maximum MFE value, the broadening of the curve is different for 3CzClIPN. Even at high MFEs ($\sim 200$ mT), a saturation cannot be observed. Hence, the observed MFE response for the excitonic TADF emitter 3CzClIPN cannot be attributed to the already-mentioned mechanisms. The structural distinctions between the systems can partly explain this difference. In exciplex materials, the donor and acceptor units are located on separate molecules, leading to out-of-phase behavior in their Larmor frequencies under an external magnetic field \cite{schellekens2010exploring, wang2016immense}. Such behavior may disrupt the precision of spin alignment between $PP_\mathrm{1}$ and $PP_\mathrm{3}$, enhancing the MFE \cite{zhang2015magnetic, mondal2023degradation}. $\mathrm{Alq_3}$ is not a TADF emitter, excluding the possibility of RISC and hence, of populating the $S_\mathrm{1}$ state via the excited triplet states.
\newline
In contrast, excitonic TADF emitters are characterized by donor and acceptor units located on the same molecule. As a result, $\Delta g$ -induced RISC can be ruled out as the dominant spin state interconversion mechanism for this system \cite{wang2024understanding}. However, hyperfine-induced spin-mixing was previously reported in excitonic TADF emitters, as discussed above. 
\newline
The key question remains whether the spin state interconversion for the excitonic bound electron-hole pairs (e-h) is driven by HFI or SOC. To address this, we compared the MEL response of all -$^1\mathrm{H}$ and all -$^2\mathrm{H}$ 4CzIPN isotopologues. Since the mass of deuterium is approximately twice that of a proton, HFI is expected to weaken significantly, leading to a narrower MEL response, as previously observed in studies on other molecular systems \cite{nguyen2012isotope}.  
Interestingly, our results deviate from these prior findings, as no noticeable difference in curve broadening is observed (\textit{cf.} Figure \ref{fig:MFE-deuteration}). Similar results were recently published by Liu \textit{et al.} \cite{liu2020isotope}, where a lack of an isotopic effect was found for TADF-OLEDs based on B3PYMPM at room temperature. 
Since variations in HFI are expected to influence the broadening of the MEL curve, the absence of such an effect suggests that HFI is not the dominant mechanism governing the MEL lineshape. We therefore propose that the observed lineshape is primarily determined by SOC, as reflected in the characteristic field $B_\mathrm{0-HF}$ (\textit{vide supra}). It is important to note that the interconversion of polaron pair (PP) spin states remains hyperfine-driven, as evidenced by the low-field feature $B_\mathrm{0-LF}$ (\textit{cf.} Figure~\ref{fig:Magnetoefficiency-ramp-data}~(a)). While a detailed investigation of PP spin dynamics is beyond the scope of this work and has been addressed in previous studies, it is worth emphasizing that SOC-induced spin state interconversion within PP species is highly unlikely. This is because SOC is strongly dependent on the e-h pair separation, whereas HFI is not (cf. ref. \cite{hu2009magnetic}, Figure~5(a) of that work). Assuming HFI remains relatively constant with decreasing e-h distance, SOC increases significantly under such conditions. Consequently, for tightly bound excitonic e-h pairs, where SOC is orders of magnitude stronger than HFI, even subtle isotope effects on PP spin-mixing are likely masked by the dominant SOC-driven spin state interconversion processes occurring within the excitonic species.

\begin{figure}[h]
\centering
  \includegraphics[width=1\textwidth]{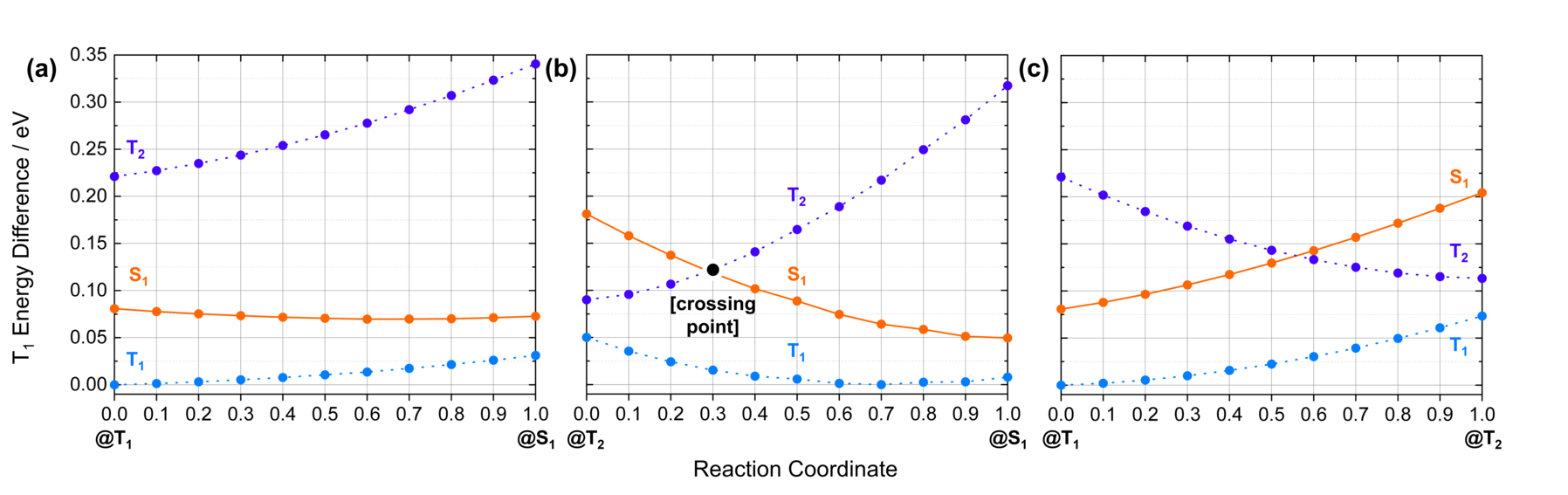}
  \caption{Excited state energies at LIPs between selected electronic state minima of 4CzIPN. (a) shows the $T_\mathrm{1} \rightarrow S_\mathrm{1}$ LIP, (b) the $T_\mathrm{2} \rightarrow S_\mathrm{1}$ LIP, and (c) the $T_\mathrm{1} \rightarrow T_\mathrm{2}$ LIP. Dotted lines correspond to triplet PES, while continuous lines represent singlet state PES. All energies are in relation to the energy of the $T_\mathrm{1}$ state in its minimum.}
  \label{fig:1D-LIPs}
\end{figure}

\bigskip
Computational investigations support this hypothesis: The calculated SOC and hyperfine coupling (HFC) for 4CzIPN in $T_\mathrm{1}$ equilibrium amount to $1.51 \times 10^{-1} \ \mathrm{cm}^{-1}$ and $5.00 \times 10^{-4} \ \mathrm{cm}^{-1}$, respectively, indicating a much stronger interaction via SOC as it is three orders of magnitude larger than the HFC which mediates the HFI.
\newline
To further explore the SOC-mediated RISC mechanism, linear interpolated pathways (LIPs) between the equilibrium geometries of the involved excited states in 4CzIPN were constructed (Figure \ref{fig:1D-LIPs}). LIPs model the transition between two equilibrium states by interpolating their geometries along a reaction coordinate. This provides qualitative insights into the energy profiles along the path of interconversion and allows for the identification of relaxation pathways.
\newline

\begin{figure}[h]
\centering
  \includegraphics[width=0.5\textwidth]{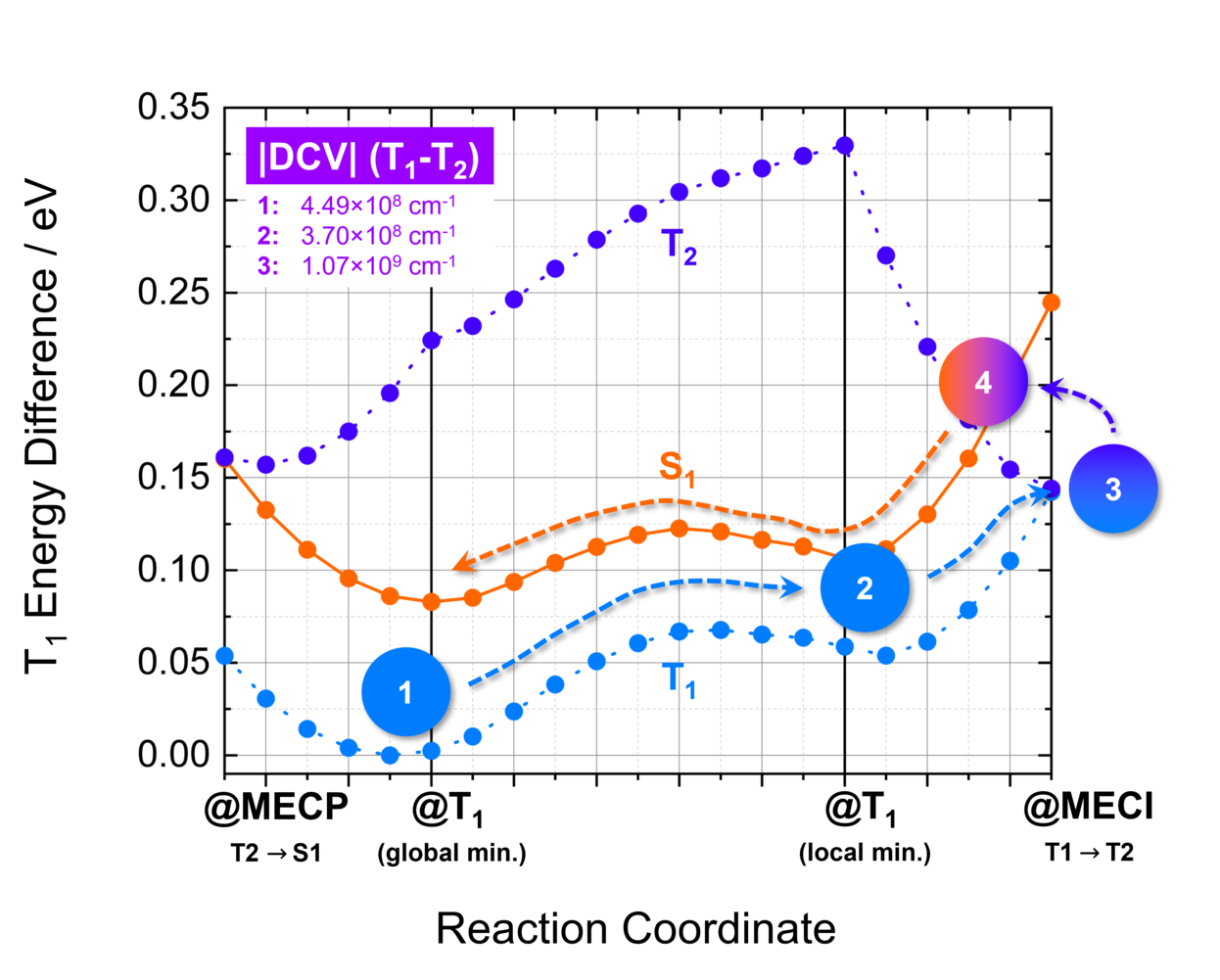}
  \caption{LIP between the $T_\mathrm{2}-S_\mathrm{1}$ MECP, the global and local $T_\mathrm{1}$ minima, and the $T_\mathrm{1}-T_\mathrm{2}$ MECI. A $T_\mathrm{2}$-mediated RISC pathway that passes a MECI and a MECP is indicated by arrows. The table inset lists the derivative coupling at the MECI (3) as well as at the local (2) and global (1) $T_\mathrm{1}$ minima. All energies are in relation to the energy of the $T_\mathrm{1}$ state in its minimum.}
  \label{fig:2D-LIP}
\end{figure}

Upon electrical excitation, the RISC mechanism is typically described as a spin state interconversion between the $T_\mathrm{1}$ and $S_\mathrm{1}$ states. However, the relatively large energy gap between $T_\mathrm{1}$ and $S_\mathrm{1}$ ($81$ meV, Figure \ref{fig:1D-LIPs} (a), left) suggests that, despite the strong SOC, this direct RISC mechanism is not as straightforward as initially thought unless additional processes are involved. Additionally, no direct crossings between these states can be observed, further reducing the possibility of efficient RISC directly from the $T_\mathrm{1}$ state. Conversely, a crossing between the $T_\mathrm{2}$ and $S_\mathrm{1}$ states is observed along the $T_\mathrm{2}-S_\mathrm{1}$ relaxation pathway (Figure \ref{fig:1D-LIPs} (b)) with a SOC of $1.05 \times 10^{-1} \ \mathrm{cm}^{-1}$ at the pictured crossing point. This suggests that the higher triplet state $T_\mathrm{2}$ could serve as an intermediate in a possible additional RISC path. 
\newline
This would first require an upconversion from the $T_\mathrm{1}$ to the $T_\mathrm{2}$ state before the RISC could occur. Although the $T_\mathrm{1}$ and $T_\mathrm{2}$ states along the $T_\mathrm{1}-T_\mathrm{2}$ pathway do not directly intersect (Figure \ref{fig:1D-LIPs} (c)), a minimum energy conical intersection (MECI) between the $T_\mathrm{1}$ and $T_\mathrm{2}$ states could provide a potential route for upconversion as a prerequisite for the subsequent RISC to $S_\mathrm{1}$. However, while PESs of different spin multiplicities cross along the whole one-dimensional hyperline, a true energy degeneracy between states of the same multiplicity is only achieved at a single point on the branching plane. This means that a one-dimensional cutout of the PESs along one coordinate (such as a LIP) can pick up a singlet-triplet crossing with a higher likelihood than a triplet-triplet crossing. Therefore, a MECI between the $T_\mathrm{1}$ and $T_\mathrm{2}$ states might still exist even though it is not directly represented in the corresponding LIP. Consequently, a MECI search between the $T_\mathrm{1}$ and $T_\mathrm{2}$ states was performed using the updating branching plane gradient projection method by Maeda \textit{et al.} \cite{maeda2010updated}. In addition to the MECI search, a minimum energy corssing point (MECP) search between the $T_\mathrm{2}$ and $S_\mathrm{1}$ states was conducted with the default gradient projection method by Bearpark \textit{et al.} \cite{bearpark1994direct}. The $T_\mathrm{2}-S_\mathrm{1}$ pathway does already show a crossing between these states, but it might not yet represent a minimum along the corresponding hyperline. A second MECP search between the $T_\mathrm{1}$ and $S_\mathrm{1}$ states was also performed to ensure that no such intersection was overlooked in the LIP analysis. The combined insights of the MECI and MECP searches allow for a more detailed understanding of the interconversion mechanism, revealing key points that would otherwise be overlooked using just LIPs. Both a MECI between the $T_\mathrm{1}$ and $T_\mathrm{2}$ states and a MECP between the $T_\mathrm{2}$ and $S_\mathrm{1}$ states could be identified while the $T_\mathrm{1}$ and $S_\mathrm{1}$ states remained non-intersecting. Upon closer examination of the two identified crossing points, it became clear that the MECI is geometrically similar to a local minimum of the $T_\mathrm{1}$ state (Root mean square deviation (RMSD) of $0.334 \ \mathrm{\r{A}}$ from the local minimum) while the MECP is in closer spatial proximity to the global minimum (RMSD of $0.017 \ \mathrm{\r{A}}$ from the global minimum). Consequently, if the $T_\mathrm{1}-T_\mathrm{2}$ upconversion via the MECI occurs first, as mentioned above, a transition from the global to the local $T_\mathrm{1}$ minimum will precede it. With this, a LIP spanning three distinct regions was constructed to provide a better overview of the overall interconversion mechanism (Figure \ref{fig:2D-LIP}): The transition from the MECP to the global $T_\mathrm{1}$ minimum, from the global to the local $T_\mathrm{1}$ minimum, and from the local $T_\mathrm{1}$ minimum to the MECI. An additional two-dimensional LIP directly illustrating the region between the MECI and the MECP, provided for improved visualization of these points, is shown in Figure S19 of the Supporting Information.
\newline

Following the path laid out in Figure \ref{fig:2D-LIP} and starting from the $T_\mathrm{1}$ minimum (marked as 1 in Figure \ref{fig:2D-LIP}), a small energy barrier must be overcome first to move from the global minimum to the local minimum closer to the MECI (2). Once the $T_\mathrm{2}$ state is populated via the MECI (3), an additional nearby crossing can be used to transition to the $S_\mathrm{1}$ state (4). There, the molecule must again overcome a small energy barrier to reach the $S_\mathrm{1}$ minimum. While the barrier separating global and local minima in the $S_\mathrm{1}$ state is below $25$ meV and therefore thermally accessible, the corresponding barrier in the $T_\mathrm{1}$ state exceeds that limit, at least along the LIP reaction coordinate. However, global and local minima in both states are also connected by the displacement of a thermally active molecular normal mode ($14.6 \ \mathrm{cm}^{-1}$ in the the $T_\mathrm{1}$ state and $13.2 \ \mathrm{cm}^{-1}$ in the $S_\mathrm{1}$ state, respectively) which can facilitate this transition. A similar connection was also suggested by Noda \textit{et al.}\cite{noda2019T2intermediate}.
\newline
Furthermore, large derivative couplings between $T_\mathrm{1}$ and $T_\mathrm{2}$ of a similar magnitude to the coupling at the MECI can be found at selected investigated points on the $T_\mathrm{1}$ PES (see table inset in Figure \ref{fig:2D-LIP}). This introduces inter-triplet state-mixing and promotes a general equilibrium between the two triplet states, meaning that reaching a MECI might not be strictly required for the inter-triplet state-mixing to take place. The described pathway could only be one of many involving a $T_\mathrm{2}$ intermediate, although leveraging the MECI would be especially efficient due to the energy degeneracy. This additional observation includes cases where no MECI is present or where PES crossings are not as clearly associated with different minima.
\newline
This analysis shows that the RISC interconversion process in 4CzIPN is more complex than initially expected and that future discussions of TADF emitter materials require more nuance, as $ \Delta E_\mathrm{{ST}}$ is not necessarily directly related or proportional to the RISC activation energy but can instead be attributed to two different potential interconversion pathways.
\newline
Similar considerations can also be made for 3CzClIPN and 5CzBN, with the respective results for SOC and HFC presented in Table \ref{tab:coupling-comparison} and the LIPs shown in the Supporting Information (Figure S20 and S21). Notably, if the $T_\mathrm{2}$ state is involved in the interconversion process, the relevant SOC occurs at the $T_\mathrm{2}-S_\mathrm{1}$ MECP. Since this point bears no special significance in the HFI regime, the corresponding coupling constants were calculated in $T_\mathrm{2}$ equilibrium instead for comparison. Due to the complex excited state character of the $T_\mathrm{2}$ state of 5CzBN, no equilibrium geometry or $T_\mathrm{2}-S_\mathrm{1}$ MECP could be converged. Instead, the unoptimized crossing found in LIP analysis was taken as the closest approximation for the calculation of the SOC. This is indicated by a dagger in Table \ref{tab:coupling-comparison} and Table \ref{tab:activation-energies}.
\newline

\begin{table}[htbp]
\centering
\caption{Spin-orbit- and hyperfine coupling for 3CzClIPN, 4CzIPN, and 5CzBN in the respective electronic states. The SOC values extracted from the experimental data are shown for comparison.}
\label{tab:coupling-comparison}
\begin{tabular}{@{}l|lll|ll@{}}
    \hline
    molecule & SOC electronic state & $\mathrm{SOC_{comp}}$ / $\mathrm{cm}^{-1}$ & $\mathrm{SOC_{exp}}$ / $\mathrm{cm}^{-1}$ & HFC electronic state & $\mathrm{HFC_{comp}}$ / $\mathrm{cm}^{-1}$ \\
    \hline
    3CzClIPN  & $T_\mathrm{1}$ & $1.21 \times 10^{-1}$ & - & $T_\mathrm{1}$ & $5.66 \times 10^{-4}$  \\
              & $T_\mathrm{2}-S_\mathrm{1}$ MECP & $1.37 \times 10^{-1}$ & $(1.46 \pm 0.04) \times 10^{-1}$ & $T_\mathrm{2}$ & $5.46 \times 10^{-4}$  \\
    \hline
    4CzIPN  & $T_\mathrm{1}$ & $1.51 \times 10^{-1}$ & - & $T_\mathrm{1}$ & $5.00 \times 10^{-4}$  \\
            & $T_\mathrm{2}-S_\mathrm{1}$ MECP & $1.95 \times 10^{-1}$ & $(1.46 \pm 0.10) \times 10^{-1}$ & $T_\mathrm{2}$ & $4.96 \times 10^{-4}$  \\
    \hline
    5CzBN  & $T_\mathrm{1}$ & $6.90 \times 10^{-1}$ & - & $T_\mathrm{1}$ & $5.02 \times 10^{-4}$  \\
           & $T_\mathrm{2}-S_\mathrm{1}$ $\mathrm{MECP^{\dagger}}$ & $3.07 \times 10^{-1}$ & $(2.96 \pm 0.61) \times 10^{-1}$ & $T_\mathrm{2}$ & - \\
    \hline
\end{tabular}
\end{table}

These findings suggest that SOC remains the primary mediator of spin-pair state interconversion across all three molecules. The upward trend in SOC from 3CzClIPN to 5CzBN can be attributed to a corresponding decrease in the torsional angles separating donor and acceptor moieties. This relationship was observed and elucidated for other TADF emitters as well \cite{monka2022understanding, kaminski2024balancing}. In the absence of additional interactions, a perpendicular orientation of $90^\circ$ between donor and acceptor units is usually sterically favored. Such an arrangement results in strong HOMO–LUMO separation, leading to a pronounced charge transfer (CT) character in the first excited singlet and triplet states. However, excessive spatial separation reduces orbital overlap, thereby weakening SOC. Moreover, direct SOC is significantly diminished between CT states involving the same orbitals and consequently showing similar CT character. In 3CzClIPN, the carbazole donor units are positioned too far apart to interact significantly, resulting in an average torsional angle of $82.6^\circ$, and leading to CT characters of $0.86$ and $0.85$ for the respective $S_\mathrm{1}$ and $T_\mathrm{1}$ transitions (A value of $1$ would signify a perfectly charge separated state and a value of $0$ a local excitation \cite{plasser2020theodore}). In contrast, when carbazole groups are placed in closer proximity, as they are in 4CzIPN and 5CzBN, steric repulsion and stabilizing $\mathrm{\pi}$ interactions drive the average torsional angles down to $66.7^\circ$ and $62.6^\circ$, respectively. This reduced angle lowers CT character but enhances SOC (CT values are provided in Table S6 in the Supporting Information). The effect becomes more pronounced with an increasing number of carbazole units. A substantial contribution to the large SOC enhancement in 5CzBN relative to the other two cyanoarene derivatives is the discrepancy in CT character between the $S_\mathrm{1}$ and $T_\mathrm{1}$ transitions ($0.81$ and $0.64$, respectively). 
\newline
The data further shows that in 3CzClIPN, spin state interconversion can also occur via a $T_\mathrm{2}$ intermediate, given the high similarity in excited state energies, crossing points, and coupling constants compared to 4CzIPN. However, the extent to which both pathways are involved in the RISC process is likely to be different for each system and cannot be quantified without further investigations.
\newline
In contrast, 5CzBN exhibits significantly stronger SOC but maintains a large $\Delta E_\mathrm{{ST}}$, generally indicative of inefficient RISC between these states. However, in the absence of a converged $T_\mathrm{2}$ equilibrium geometry or a confirmed $T_\mathrm{2}-S_\mathrm{1}$ MECP, a definitive conclusion cannot yet be drawn, as the $T_\mathrm{2}$-mediated mechanism discussed earlier is currently not computationally accessible. Additionally, as previously noted, TTA likely plays a dominant role in the spin state interconversion process of 5CzBN as well, which will be the subject of further computational investigations.
\newline
To further assess the feasibility of the RISC process, a quantitative approach complementing this qualitative analysis can be helpful. To this end, we determined the RISC activation energy using experimental and computational methods.

\subsection{Determination and Interpretation of Activation Energies in TADF Emitters}
\begin{figure}
\centering
  \includegraphics[width=0.7\textwidth]{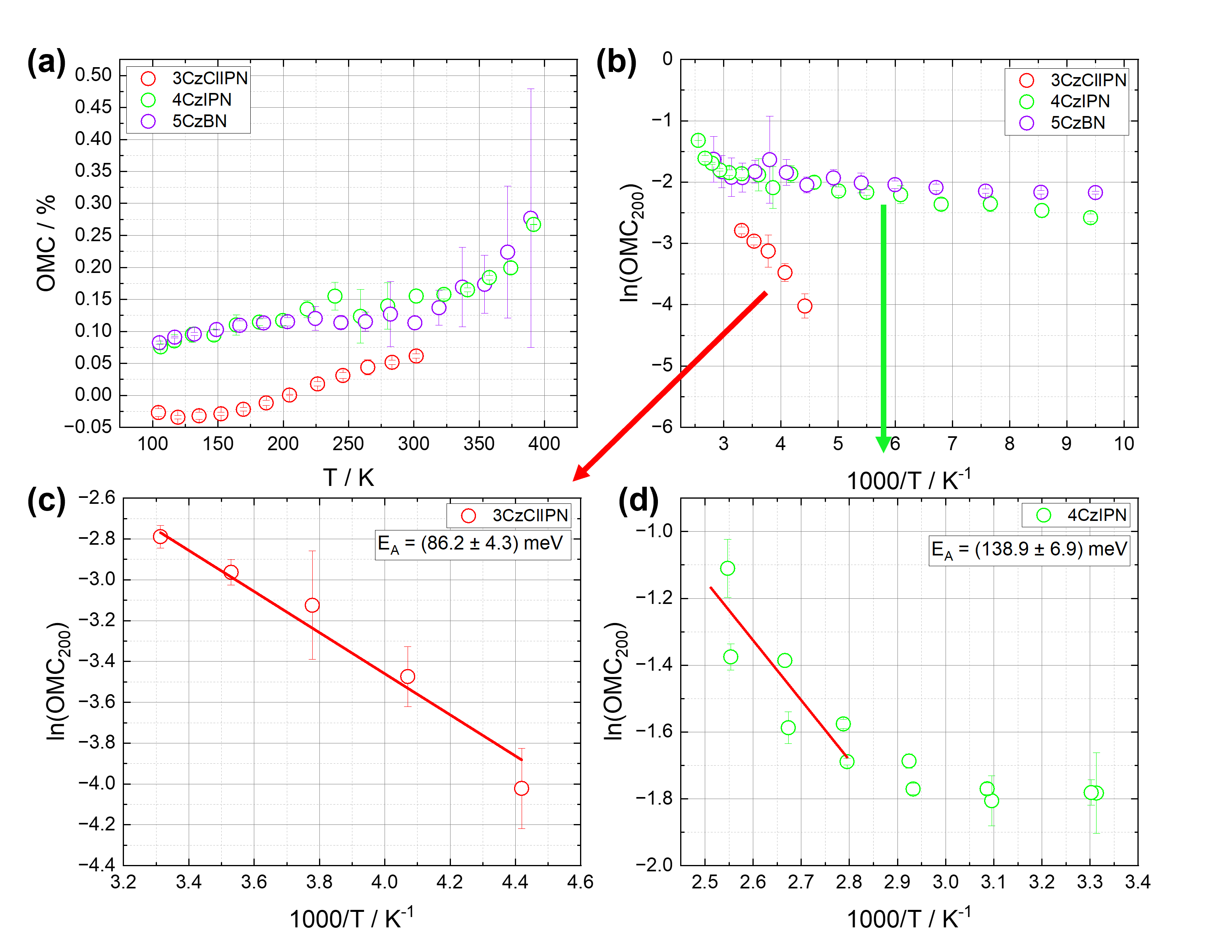}
  \caption{(a) and (b) show the temperature dependence observed for the OMC signal plotted over temperature (a) and the ln(OMC) over 1000/T (b) for 3CzClIPN, 4CzIPN and 5CzBN, respectively. (c) and (d) show the respective Arrhenius plots with the values extracted from (b) for 3CzClIPN and 4CzIPN, respectively. The activation energy was determined by the strongest onset in the OMC response.}
  \label{fig:Temperature dependency}
\end{figure}

Temperature-dependent MFE measurements were conducted to determine the activation energy associated with RISC for various TADF emitters. The corresponding data for the OMC response at a constant current density of $J_\mathrm{OLED} = 175 \ \mathrm{mA/cm^2}$ is shown in Figure \ref{fig:Temperature dependency}. The temperature dependence was fitted using the following expression:
\begin{equation}
    \mathrm{MFE} \propto exp(-\frac{E_\mathrm{act}}{k_\mathrm{B}T})
\end{equation}
where $E_\mathrm{act}$ denotes the activation energy, $k_\mathrm{B}$ the Boltzman constant, and $T$ the temperature\cite{ling2015large}. The complete MFE responses for both OMC and MEL are presented in Figure S15. 
A distinct difference between the temperature-dependent trends of the OMC and MEL responses was observed. Activation energies derived from the MEL response consistently fall below $\Delta E_\mathrm{ST}$, contradicting Marcus theory \cite{aizawa2020kinetic, serdiuk2021vibrationally}. Despite this, previous studies by Liu \textit{et al.} \cite{liu2020isotope} and Basel \textit{et al.} \cite{basel2016magnetic} extracted RISC activation energies from MEL data, reporting values significantly lower than the thermal energy at room temperature. Given that the reorganization energy is expected to be on the order of several tens of meV \cite{aizawa2020kinetic}, these discrepancies likely result from competing loss mechanisms.
\newline
In this work, we extracted the activation energy from the temperature-dependent OMC response (\textit{cf.} Figure \ref{fig:Temperature dependency}). For 3CzClIPN, a sign change in the OMC response was observed, with negative values appearing at low temperatures (\textit{cf.} Figure S15 (a)). This behavior has previously been attributed to the onset of bipolar transport being temperature-dependent, where at low temperatures only one type of charge carrier dominates the charge transport \cite{bloom2008temperature}. From the Arrhenius plots in Figure \ref{fig:Temperature dependency} (c) and (d), the activation energy was determined to be $(86.2 \pm 4.3)$ meV for 3CzClIPN and $(138.9 \pm 6.9)$ meV for 4CzIPN (\textit{cf.} Table \ref{tab:activation-energies}). It can further be observed from panels (c) and (d) that the onset of a significant OMC response shifts to higher temperatures with increasing activation energy.
\newline
Interestingly, no activation energy could be extracted for 5CzBN. As discussed previously, its MFE response is predominantly governed by hyperfine-induced spin-mixing of polaron pairs (visible in the LF effect regime) and TTA (dominating the HF effect regime). Therefore, the temperature dependence does not reflect the RISC process alone. Moreover, the activation energy is likely too high to be reliably measured within the accessible temperature range. Upon increasing the temperature, structural degradation, most likely within the emissive layer, was observed, compromising the measurement reliability. This is reflected in the large error bars in Figure \ref{fig:Temperature dependency} (a). Additionally, a poor signal-to-noise ratio, as illustrated in Figure S16, further hindered reliable analysis.

Computationally, the energy barrier that must be overcome to drive the RISC process via the $T_\mathrm{2}$-mediated pathway can be described as the energy difference between $T_\mathrm{1}$ equilibrium and the $T_\mathrm{2}-S_\mathrm{1}$ MECP \cite{aizawa2020kinetic}. For the direct $T_\mathrm{1}-S_\mathrm{1}$ pathway, an activation energy can be estimated from the following Marcus Theory-like expression incorporating $\Delta E_\mathrm{{ST}}$ \cite{serdiuk2021vibrationally}:

\begin{equation}
    E_{act} = \frac{(\Delta E_\mathrm{{ST}} + \lambda)^2}{4 \lambda}
\end{equation}

where $\lambda$ is the reorganization energy. The results for both pathways are summarized in Table \ref{tab:activation-energies}.

\begin{table}
\centering
 \caption{Experimental and computational activation energies for 3CzClIPN, 4CzIPN, and 5CzBN.}
  \begin{tabular}[htbp]{@{}lll@{}}
    \hline
    Molecule & $E_\mathrm{act}$ (exp.) / meV & $E_\mathrm{act}$ (comp.) / meV \\
    \hline
    3CzClIPN  & $(86.2 \pm 4.31)$ & 24.6 ($T_\mathrm{1}-S_\mathrm{1}$) \\
      & & 101.4 ($T_\mathrm{1}-T_\mathrm{2}-S_\mathrm{1}$) \\
    \hline
    4CzIPN  & $(138.9 \pm 6.9)$ & 74.2 ($T_\mathrm{1}-S_\mathrm{1}$) \\
      & & 160.0 ($T_\mathrm{1}-T_\mathrm{2}-S_\mathrm{1}$) \\
    \hline
    5CzBN  & - & 99.6 ($T_\mathrm{1}-S_\mathrm{1}$) \\
      & & 127.9 $(T_\mathrm{1}-T_\mathrm{2}-S_\mathrm{1})^{\dagger}$ \\
    \hline
    \label{tab:activation-energies}
  \end{tabular}
\end{table}

Even though the computationally and experimentally determined activation energies seem to align better with the idea of a dominant $T_\mathrm{2}$-mediated pathway, a noticeable discrepancy between both sets of values is apparent. From our perspective, this difference can be attributed to processes such as non-radiative recombination and the need to overcome energy barriers resulting from the simplified layer stack. Additionally, as discussed above, the MFE responses are influenced not only by RISC but also by triplet-related processes. Our computational modeling does not include the influence of these effects on the activation energy of the RISC, limiting direct comparability. Conversely, the activation energy derived from MFE measurements reflects a superposition of various effects, making it impossible to isolate the RISC activation energy alone. In the cases of 3CzClIPN and 4CzIPN, the observed activation energy likely consists of contributions from both TCA and RISC, as well as additional energy required to overcome interfacial energy barriers due to imperfections in the layer stack and energy level alignment. 
\newline
Consequently, no decisive conclusion regarding the dominance of one RISC pathway over the other should be drawn from this comparison alone.

\section{Conclusion}
Throughout this study, we compared three cyanoarene-type molecules with different singlet-triplet energy gaps ($\Delta E_\mathrm{{ST}}$). By combining experimental and computational approaches, we found that the interconversion between triplet and singlet excited states of excitonically bound electron–hole pairs is predominantly driven by spin–orbit coupling rather than hyperfine interactions.
\newline
Furthermore, the reverse intersystem crossing process in these molecules can also occur via the $T_\mathrm{2}$ state in addition to the direct $T_\mathrm{1}-S_\mathrm{1}$ pathway. This indicates that it is not directly influenced by the $\Delta E_\mathrm{{ST}}$ between $T_\mathrm{1}$ and $S_\mathrm{1}$ states, but instead by the respective activation energies of the different pathways and the dynamics associated with the spin state interconversion mechanism.
\newline
A comparison between 3CzClIPN, 4CzIPN, and 5CzBN revealed significant differences in charge transport properties. Charge transport in 5CzBN was primarily dominated by triplet-triplet annihilation, resulting in an inefficient RISC process. Triplet-charge annihilation is the primary mechanism in the other two molecules, contributing to the RISC process under applied magnetic field. These findings demonstrate that the application of an external magnetic field is a powerful and non-destructive tool to investigate charge transport properties. 
\newline
Using a Double-Lorentzian model to fit the magnetic field effect response, we successfully replicated the measured data without overfitting. This approach also revealed that distinguishing between the reverse intersystem crossing and triplet-charge annihilation processes is not possible in the presence of a magnetic field, and that they are inherently linked.
\newline
This work unravels the spin state interconversion and the dominant charge transport properties of carbazole-containing cyanoarene TADF emitters and paves the way for the development of efficient spintronic devices.
\newline
The dynamics of triplet-triplet annihilation as a competitive spin state interconversion mechanism in 5CzBN still demand a broader computational investigation to complement our initial experimental findings.  Given the vast diversity within the class of cyanoarene molecules, our findings could be compared to other molecules within this toolbox \cite{speckmeier2018toolbox} to identify broader trends, especially the influence of different donor groups, \textit{e.g.} diphenyamine-containing cyanoarene molecules.  

\FloatBarrier
\section{Experimental Section}
\subsection*{Materials}
As substrates, we used pre-patterned ITO glass substrates provided by Ossila. Poly-(3,4-ethylendioxythiophene)-poly-(styrene sulfonate) (PEDOT:PSS) was purchased from Heraeus (Clevios™—PEDOT:PSS). Additionally, Molybdenum(VI) oxide powder ($\mathrm{MoO_3}$) and lithium fluoride (LiF) were purchased from Sigma-Aldrich. 2,4,6-Tri(9\textit{H}-carbazol-9-yl)-5-chloroisophthalonitrile (\textbf{3CzClIPN}), and Penta-carbazolylbenzonitrile (\textbf{5CzBN}) were synthesized by Tom Gabler. 2,4,5,6-Tetra-9\textit{H}-carbazol-9-yl (\textbf{4CzIPN}) was purchased from Angene. 4,4'-Bis(9\textit{H}-carbazol-9-yl)biphenyl (\textbf{CBP}) was purchased from TCI chemicals. 2,2',2''-(1,3,5-Benzinetriyl)-tris(1-phenyl-1-\textit{H}-benzimidazole) (TPBi) was purchased from Ossila. All materials were used without further purification. The PEDOT:PSS solution was diluted with isopropanol (IPA) at a ratio of $1:0.04 \ \mathrm{w} \%$ . $\mathrm{MoO_3}$ was mixed with PEDOT:PSS at a ratio of $0.02:1$. Prior to this, the $\mathrm{MoO_3}$ powder was dissolved in ammonium hydroxide ($\mathrm{NH_3OH}$) ($0.25 \ \mathrm{g/ml}$). The initial solutions for 4CzIPN, 3CzClIPN, and 5CzBN as well as for CBP were prepared to be $20 \ \mathrm{mg/ml}$ in toluol. The two solutions were mixed afterward at a mass ratio of $9:1$ for CBP:TADF emitter. Prior to device fabrication, all solutions were treated in an ultrasonic bath at $80 \ ^\circ\mathrm{C}$ for at least $30$ min and filtered through a nylon syringe with a pore size of $0.22 \ \mathrm{\mu m}$. The remaining materials were deposited via thermal evaporation.

\subsection*{Experimental Details}
The TADF devices were fabricated with the following structure (\textit{cf.} Figure \ref{fig:layer-stack-molecular-structure} (a) and (b): ITO ($115 \ \mathrm{nm}$, purchased by Ossila, pre-patterned cathode (eight pixels)/ PEDOT:PSS + $\mathrm{MoO_3}$ ($35 \ \mathrm{nm}$)/ pristine TADF emitter ($\approx 30 \mathrm{nm}$)/ TPBi ($20 \ \mathrm{nm}$)/ LiF ($0.8\  \mathrm{nm}$)/ Al ($110 \ \mathrm{nm}$). The thickness of the spin-coated layer was measured using profilometry, with an associated thickness variation of \(8\%\). Before the PEDOT:PSS + $\mathrm{MoO_3}$ was deposited, the samples were cleaned by alkalic acid + DI water and IPA in an ultrasonic bath for $30$ min, respectively. Afterward, the samples were dried with nitrogen. The surface was further activated by a UV-Ozon treatment with a custom-built setup (after ref. \cite{weber2023cost}) for $35$ min. All solutions were spin-coated at 3000 rpm, (acceleration $300 \ \mathrm{rpm/s}$) for $30$ s and post-annealed at $80$ $^\circ$C for 30 min. TPBi was evaporated at constant rates of 1 $\mathrm{\r{A}}$/s.  LiF and Al were evaporated at a constant rate of 0.1 $\mathrm{\r{A}}$/s and 5 $\mathrm{\r{A}}$/s, respectively. The deposition via thermal evaporation in the vacuum chamber was performed at a pressure of $10^{-6} \ \mathrm{mbar}$. The devices were encapsulated with epoxy resin and a glass slide, followed by post-treatment under a UV lamp for 15 minutes.\newline
For a detailed description of the measurement setup, we refer to the Supplementary Information, section 1.1. The hardware and software are explained in detail, and the software Python code can be found on GitHub: \url{https://github.com/semiconductor-physics/Organic-MFE-Measurement}.
The PL spectrum was recorded by a Jasco CD spectrometer, where the excitation wavelength was chosen to be $\mathrm{\lambda_{ex}}$ = $405 \ \mathrm{nm}$ for 3CzClIPN and 4CzIPN, while the excitation wavelength was chosen to be $\mathrm{\lambda_{ex}}$ = $370 \ \mathrm{nm}$ for 5CzBN. The UV-vis absorption spectra were measured using a Cary 60 UV-vis spectrometer. The baseline was determined using quartz glass and subtracted from the spectra of interest. To perform photometric characterization of the fabricated OLEDs, a Gigahertz-Optik UMBK-150 integrating sphere (sphere diameter of $150 \ \mathrm{mm}$) was utilized to detect all the emitted light. The electrical control and measurement of the OLEDs were carried out using a Keithley 2636B SourceMeter. The photocurrent output by the integrating sphere was converted into a voltage via an amplifier and detected by an Agilent 34411A multimeter. The entire measurement setup was controlled automatically using a LabVIEW program, recording the I-V curve of the OLEDs and measuring light current, optical power, and EQE as a function of voltage.\newline
The ionization energy $IE$ and electron affinity $EA$ of 4CzIPN, corresponding to its HOMO and LUMO onsets, as well as the work function $W$ of 4CzIPN were determined experimentally by the combination of ultraviolet photoemission spectroscopy (UPS) and low-energy inverse photoemission spectroscopy (LEIPES). Both experiments were combined in the same analysis chamber of an ultra-high vacuum (UHV) setup with a base pressure of $1 \times 10^{-10}$ mbar. The techniques were applied to thin films of 4CzIPN thermally evaporated on a Si substrate with a $100$ nm thick Au layer on top in a connected preparation chamber. The substrates were prepared by successive sonication in acetone, ethanol and DI water before loading them into the UHV chamber. Thermal evaporation was carried out successively in five steps to achieve five different film thicknesses to identify possible interfacial effects of energy band alignment. The evaporation rate was controlled using a quartz crystal microbalance (QCM) and resulting thickness values and chemical purity of the organic films were confirmed using X-ray photoemission spectroscopy (XPS). For UPS, the He I emission line from a He-discharge lamp was used. XPS was carried out using the unmonochromated AlK$\alpha$ line from a XR50 x-ray source by SPECS. The photoemission signal in both cases was detected using a PHOIBOS 150 hemispherical analyzer in combination with a 1D delay-line detector, also by SPECS. For LEIPES a custom-built setup, designed after \cite{yoshida2015principle}, was used, including a custom-built electron source designed after \cite{erdman1982low} with a BaO cathode by Kimball Physics, and a photodetection system using the R2078 photomultiplier by HAMAMATSU. LEIPES measurements were carried out in the isochromat mode, using a variety of Bragg filters purchased from Edmund Optics and Laser Components Germany as optical bandpass. The central wavelengths of the used filters were $250$ nm, $260$ nm, $266$ nm, and $270$ nm, all with a FWHM of $10$ nm. The spectral resolution of the LEIPES setup was determined to be $0.3$ eV via the spectral broadening of the Fermi edge of a clean polycrystalline Ag substrate. The same procedure resulted in a spectral resolution of $110$ meV for UPS. Measurement of the FWHM of the Ag3d core levels yielded a spectral resolution of $1.3$ eV for XPS. 

\bigskip
The computational details can be found in the Supporting Information.
\bigskip

\FloatBarrier

\bigskip
\textbf{Supporting Information} \par 
Supporting Information is provided below.

\bigskip
\textbf{Acknowledgements} \par 
Annika Morgenstern and Jonas Weiser contributed equally to this work.
\newline
The authors would like to acknowledge Jörn Langenickel for the possibility to use the optoelectronic measurement setup. We extend our thanks to Prof. Michael Mehring for the possibility to use the UV-vis spectrometer. We further thank Prof. Carsten Deibel for the possibility to use the evaporation chamber.
\newline
Annika Morgenstern and Georgeta Salvan would like to thank SAB for funding this research under the project number 100649391 (ReSIDA-H2). The authors gratefully acknowledge funding by the Deutsche Forschungsgemeinschaft (DFG, German Research Foundation) through DFG-TRR 386-B05/B06 (514664767) as well as the resources on the LiCCA HPC cluster of the University of Augsburg, co-funded by the DFG – Project-ID 499211671.\newline
Dietrich R.T. Zahn and Alexander Ehm would like to thank the DFG Research Unit FOR5387 for funding.


\newpage
\subsection{Table of Contents}
This study explores overlapping charge transport mechanisms in TADF materials and identifies spin-orbit coupling as the dominant mechanism for triplet-to-singlet transition in carbazole-based cyanoarene emitters. Computational investigations reveal a second spin-orbit-coupling-mediated reverse intersystem crossing pathway, significantly enhancing the understanding of charge transport dynamics and spin state interconversion in intermolecular TADF emitter systems.
\begin{figure}
    \centering
    \includegraphics[width=0.7\linewidth]{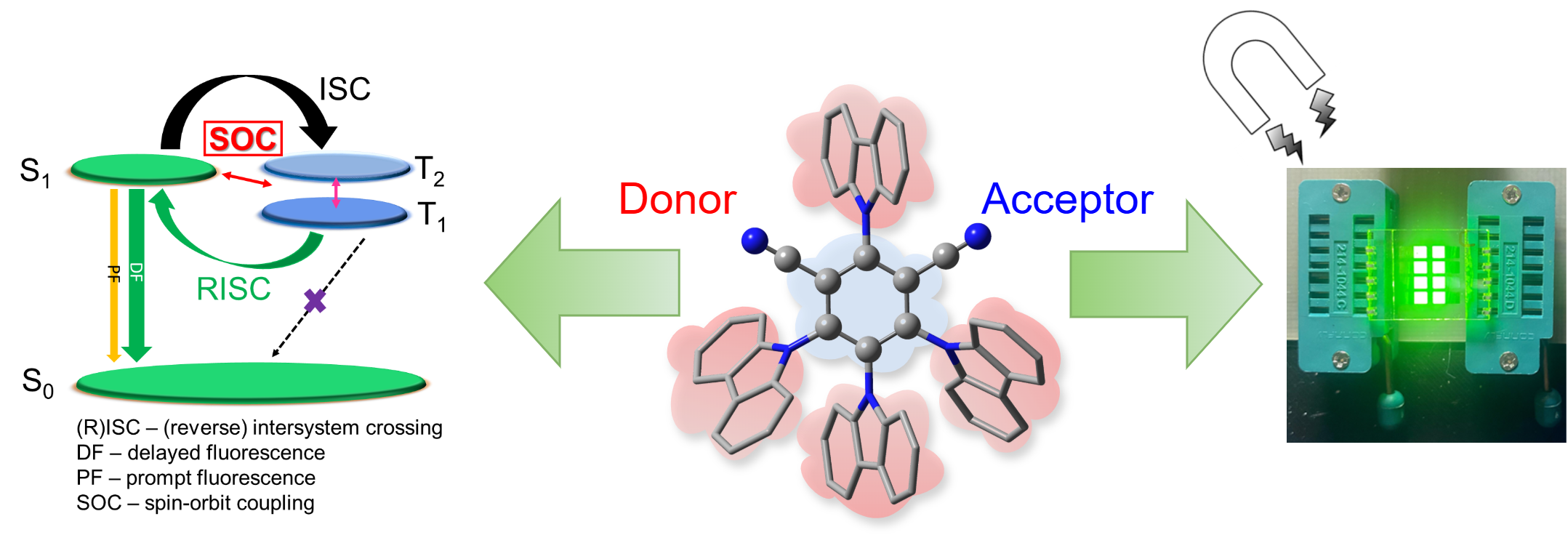}
    \caption{Table of Contents (ToC) image}
    \label{fig:toc}
\end{figure}

\newpage

\newcommand{\supplementarysection}{
  \setcounter{figure}{0}
  \setcounter{table}{0}
  \setcounter{section}{1}
  \renewcommand{\thefigure}{S\arabic{figure}}
  \renewcommand{\thetable}{S\arabic{table}}
  \section*{Supplementary Information}
  \addcontentsline{toc}{section}{Supplementary Information}
}
\newpage
\supplementarysection
\subsection{Experimental Apparatus and Analysis}
\subsubsection{Experimental Setup}
\begin{figure}
    \centering
    \includegraphics[width=0.8\linewidth]{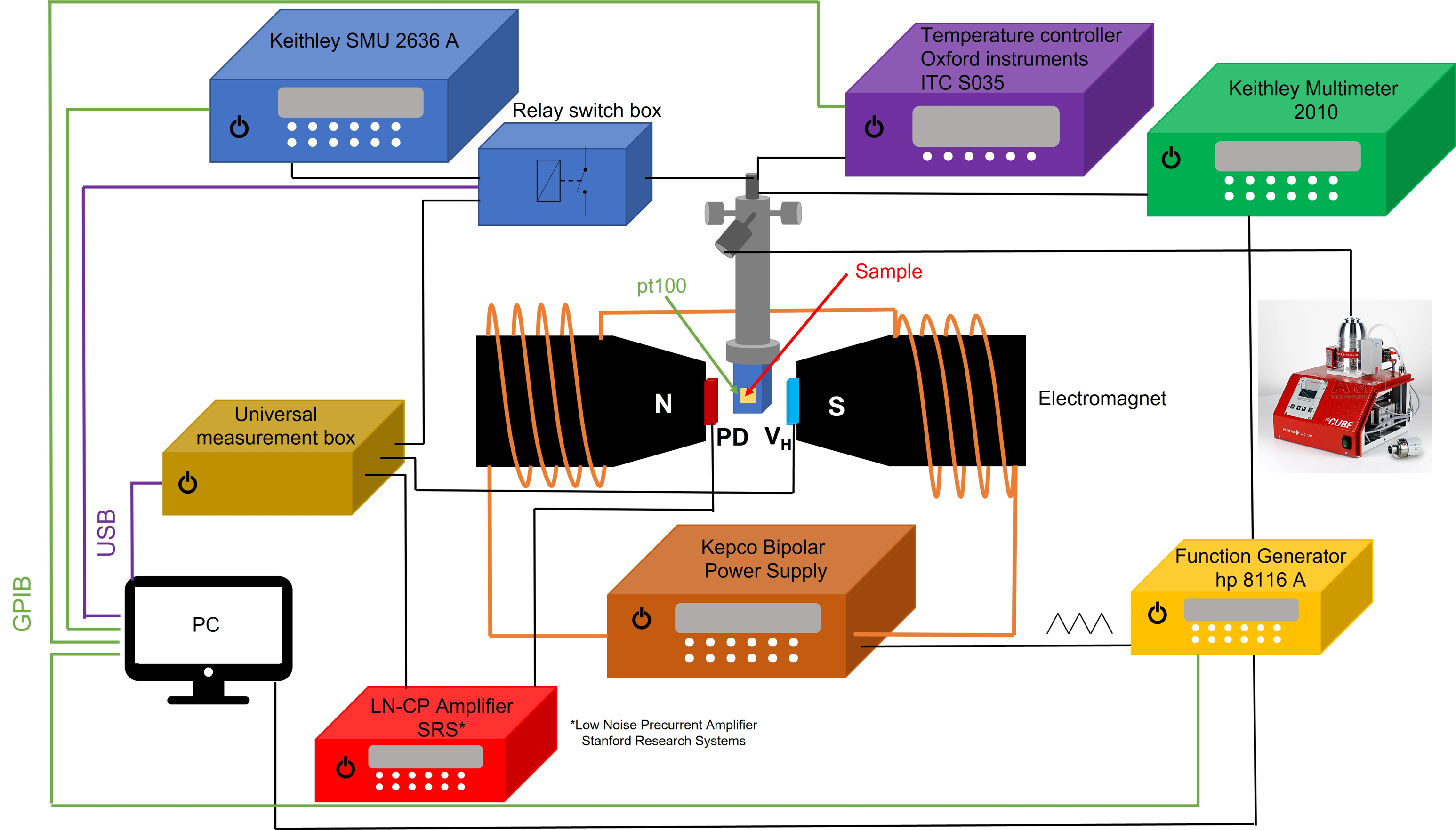}
    \caption{Schematic illustration of the MFE measurement setup for measuring the OMC and MEL response of the OLED device. The cryostat can be exchanged by a room temperature aperture for easier sample exchange.}
    \label{fig:setup-schematic}
\end{figure}

\begin{figure}
    \centering
    \includegraphics[width=0.6\linewidth]{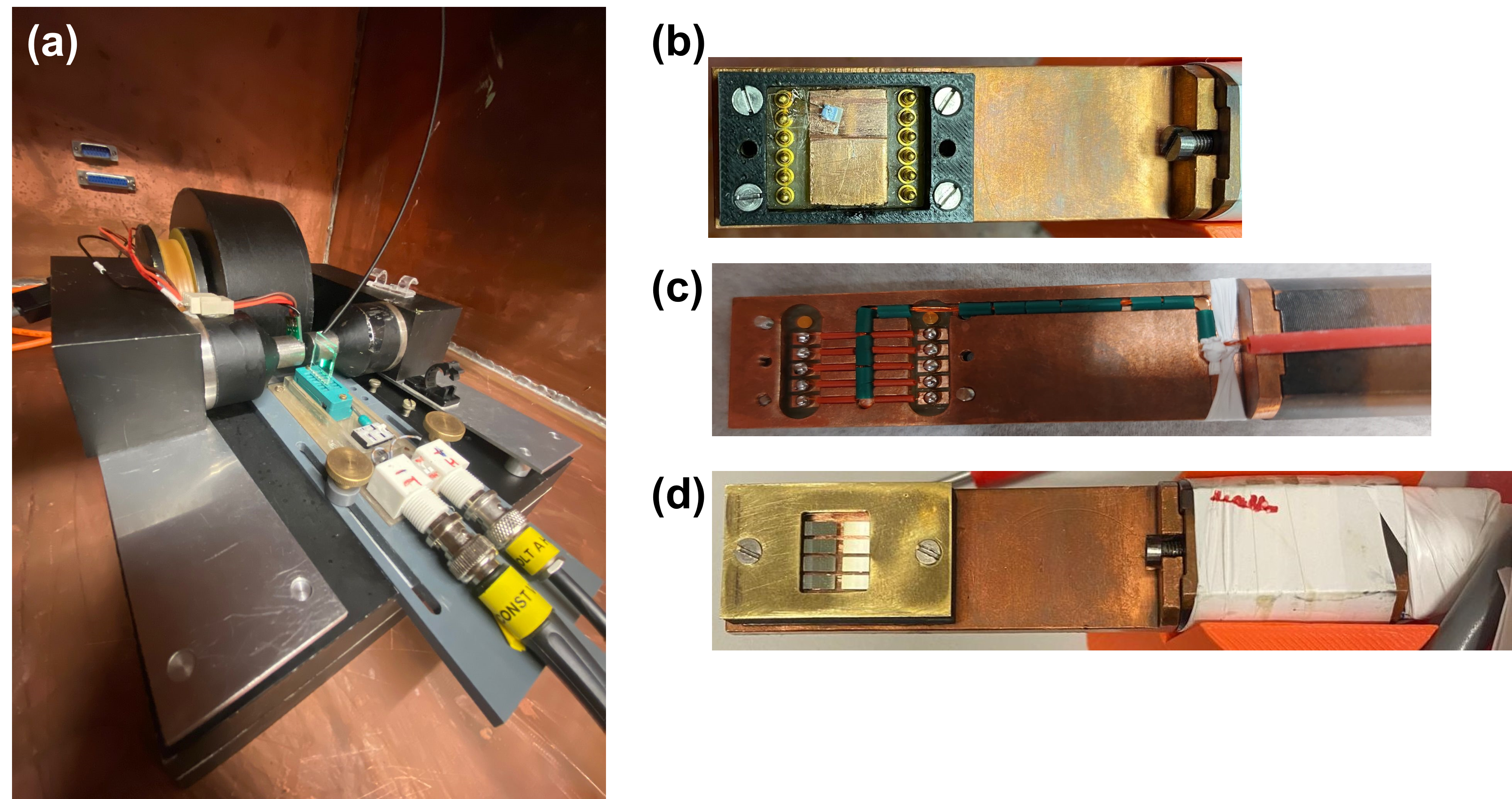}
    \caption{(a) RT measurement setup containing a DIP switch-controlled sample holder to enable measuring of all pixels (at one OLED side), (b) sample holder for inside the cryostat containing the contact pins and an additional pt100 at the sample position, (c) electrical connections made on the backside of the sample holder, (d) sample holder with a mounted OLED substrate and the top plate to ensure electrical contact and increased thermal conductivity. }
    \label{fig:RT-setup-cryo-sample-holder}
\end{figure}

\begin{figure}
    \centering
    \includegraphics[width=0.9\linewidth]{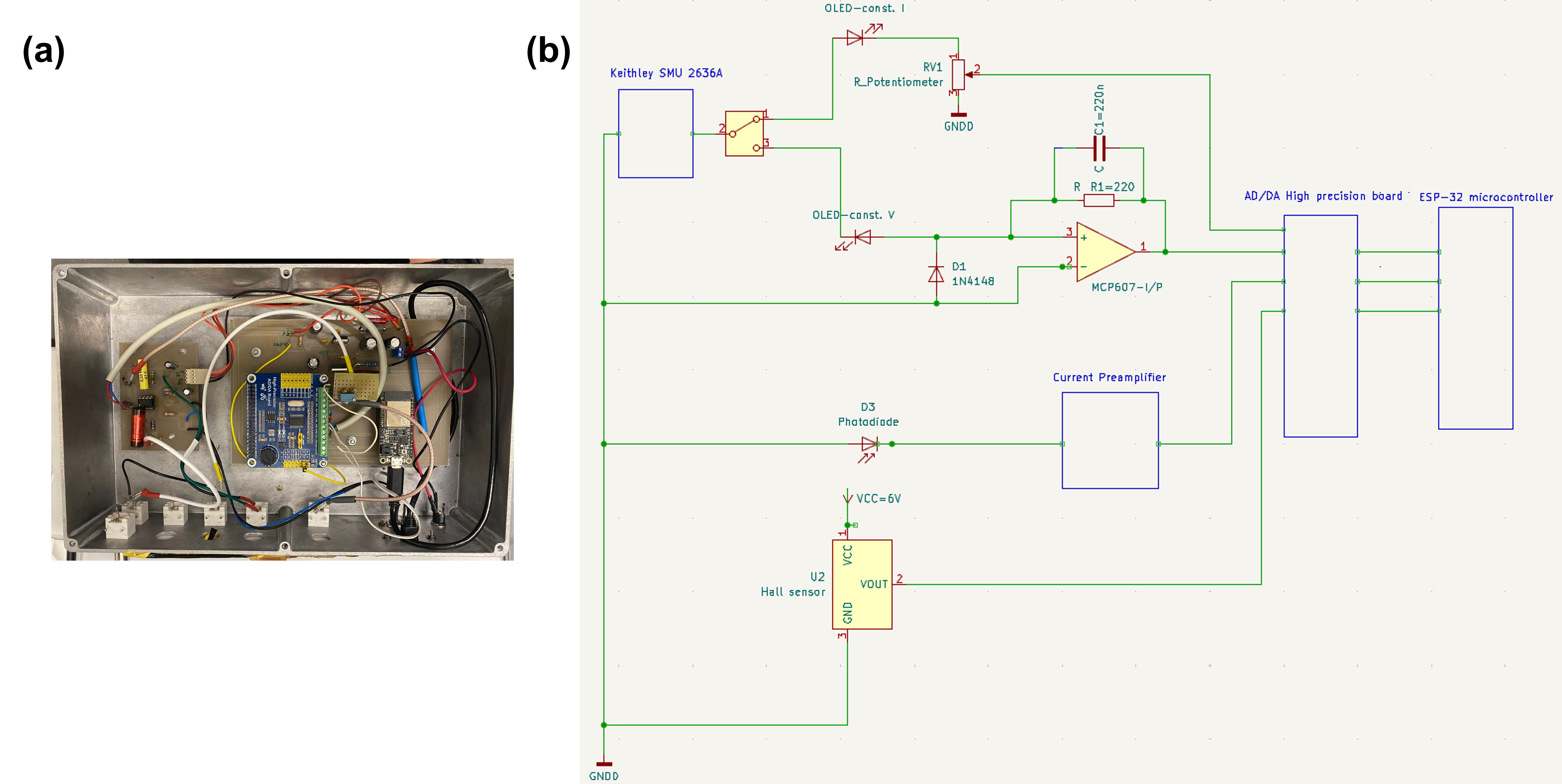}
    \caption{(a) Universal measurement box containing the electrical circuit shown in (b). The main part consists of an AD/DA converter (24 bit ADC ADS1256 from Texas Instruments in a 11010 high-precision AD/DA board from waveshare) and an ESP-32 microcontroller.}
    \label{fig:ADC-box}
\end{figure}

A schematic illustration of the measurement setup is shown in Figure \ref{fig:setup-schematic}. The device, housed in a cryostat (Oxford Instruments), can be operated in either constant current or constant voltage mode. The applied voltage or current is generated by a Keithley SMU 2636A. A custom-built relay switch box, controlled by an Arduino Nano, allows all pixels on the OLED structure to be addressed.
For making electrical contact with the OLED device, a specially designed sample holder was constructed, with the electrical connections detailed in Figure \ref{fig:RT-setup-cryo-sample-holder} (b)-(d). An additional measurement setup was utilized for room temperature (RT) measurements to enable easier sample exchange, as depicted in Figure \ref{fig:RT-setup-cryo-sample-holder} (a). The fundamental operation remains the same. The RT system can be placed within an electrically and magnetically shielded enclosure made of PCB material to minimize noise.
Figure \ref{fig:RT-setup-cryo-sample-holder} (b) illustrates the spring pins used to connect the sample electrodes to the electrical circuit. A pt100 temperature sensor was integrated to measure the temperature directly at the sample position, with readings taken via a Keithley 2010 multimeter connected to a computer via a GPIB interface. To improve thermal conductivity between the circuit board and the sample, an additional copper plate was installed in the middle of the sample holder. The backside of the sample holder, containing the copper wires, is shown in Figure \ref{fig:RT-setup-cryo-sample-holder} (c). Figure \ref{fig:RT-setup-cryo-sample-holder} (d) displays the mounted sample secured with the top plate to ensure reliable contact by pressing the sample tightly against the pins.
A function generator was employed to generate a triangular signal, which was applied to a Kepco Bipolar Power Supply to produce an alternating magnetic field. Throughout this study, the amplitude was maintained at 10 V, with a frequency of 0.1 Hz. Hence, the magnetic field is swept with 99.84 mT/s. The magnetic field was measured using a Hall sensor (CYSJ2A-T ChenYang) positioned directly on the magnet’s pole shoe.
The emitted light from the sample was detected by a photodiode (Hamamatsu S1227-1010BQ) placed directly in front of the cryostat. The signal was amplified using a Low Noise Precurrent Amplifier (LNPCA) from Stanford Research Systems, employing a low-pass filter set at 6 dB roll-off and 100 Hz cut-off frequency throughout the measurements. The LNPCA was operated in low noise mode, with sensitivity adjusted to match the light intensity of the sample. In constant voltage mode, the OLED signal was enhanced using a transimpedance amplifier, while in constant current mode, a potentiometer was employed for signal amplification (\textit{cf.} Figure \ref{fig:ADC-box} (b)).
Currents from the photodiode and OLED were converted to voltage signals and either attenuated or amplified analogously to match the input voltage range of the analog-to-digital converter (24-bit ADC ADS1256 from Texas Instruments in a 11010 high-precision AD/DA board from waveshare). The signal and reference were differentially connected to the ADC inputs via BNC connectors. The conversion was performed sequentially. The converted data was then read by an ESP-32 microcontroller, stored as floating-point numbers, and transmitted via USB to a computer. The data is retrieved by the Python host and further processed on the PC.
Figure \ref{fig:ADC-box} shows the custom-made data acquisition system (DAQ) containing the transimpedance amplifier for constant voltage measurements or potentiometer for constant current measurements, the ADC and the microcontroller. Figure \ref{fig:ADC-box} (b) provides a schematic of the electrical circuit, including the transimpedance amplifier or potentiometer for the OLED signal input, while the Hall voltage was directly read out by the ADC. The sample rate was kept constant for all measurements at 866 Hz. 

\subsubsection{Measurement Software}
\begin{figure}
    \centering
    \includegraphics[width=0.9\linewidth]{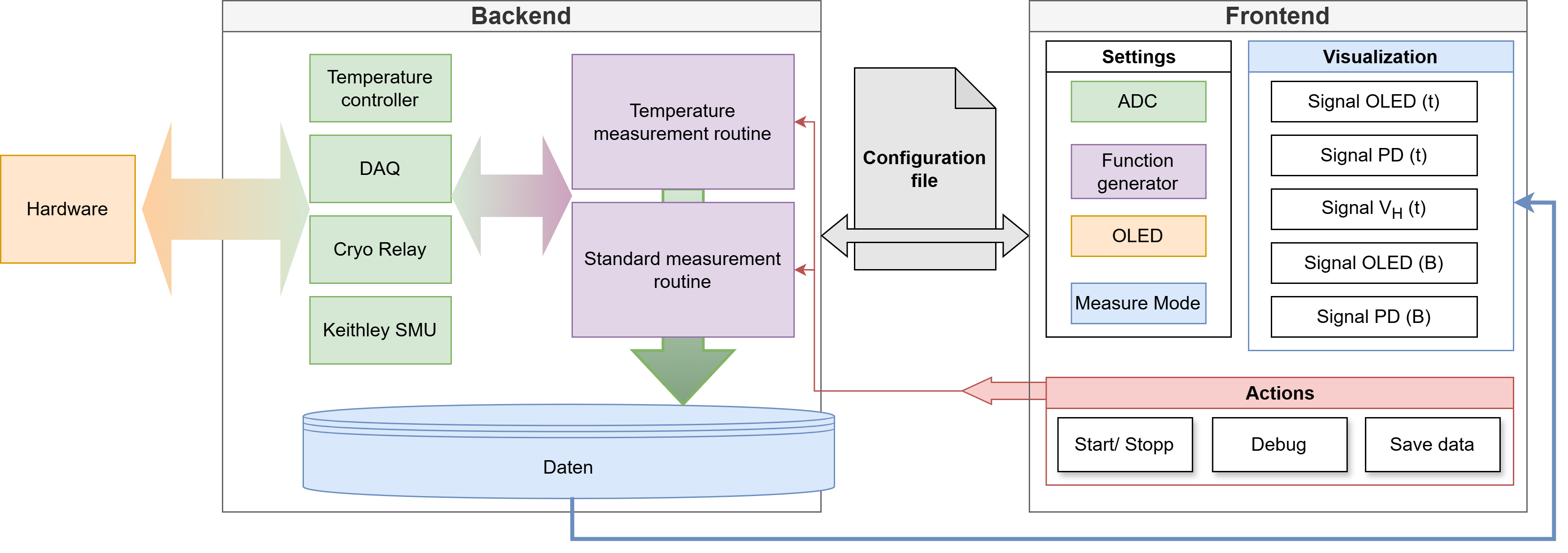}
    \caption{A scheme of the softwares' front- and backend components behind the measurement routine. The graphical user interface, shown in Figure \ref{fig:graphical user interface}, is governed by the frontend, while the backend manages the measurement routines and device control. The drivers (\textit{cf.} backend) enable reliable communication from the backend to the measurement hardware, which is described in detail. }
    \label{fig:front-backend}
\end{figure}

The measurement software is available on GitHub: \url{https://github.com/semiconductor-physics/Organic-MFE-Measurement}. A schematic representation of the measurement routine is shown in Figure \ref{fig:front-backend}. The software architecture is divided into three main components: the backend, the frontend, and the configuration file. The communication from the backend to the hardware is included in Figure \ref{fig:front-backend}, while the hardware components are described in detail above.

The \textbf{backend} is responsible for hardware control, data acquisition, and data management. It consists of four driver classes, each dedicated to controlling a specific hardware device. Importantly, the temperature controller and cryo relay classes are only used during temperature-dependent measurements. In contrast, for room temperature measurements, a simpler standard routine is employed, leaving the cryo relay and temperature controller unused.
The generated magnetic field is applied continuously via a triangular function from the function generator to the power supply. This process sweeps the magnetic field back and forth until the desired number of ramps, as specified by the user, is reached. Consequently, the collected data must be segmented appropriately during the post-processing.
\newline
During measurements, the Keithley SMU 2636 A applies either a constant voltage or current, depending on the users choice. The AD/DA converter records three critical signals:
\begin{itemize}
    \item the OLED response,
    \item the photodiode response, and
    \item the hall voltage from the Hall sensor.
\end{itemize}

The microcontroller continuously collects data from the ADC with the specified sample rate (here $866 \ \mathrm{Hz}$) and transmits it to the PC. The read-out of the serial buffer is independent of this process. This architecture enables time-constant sampling. 
Meanwhile, time-insensitive control tasks, such as sweeping the magnetic field, are also handled independently, ensuring they do not interfere with the real-time data acquisition process.\\

\begin{figure}
    \centering
    \includegraphics[width=0.8\linewidth]{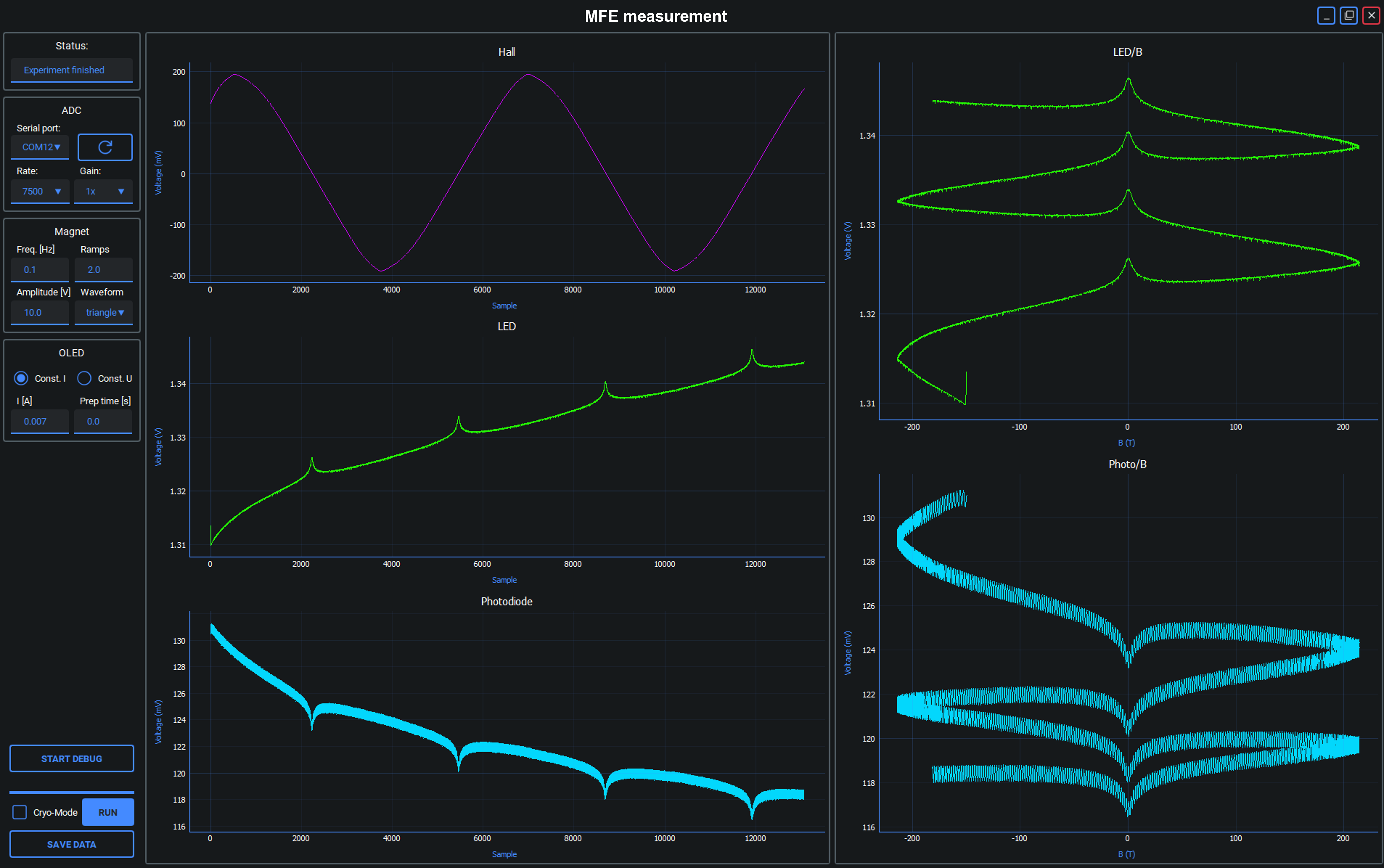}
    \caption{Graphical user interface displaying a measurement taken at a constant current of 7 mA. The left-hand side provides configuration options for the ADC, function generator (Magnet), OLED, and saving options.}
    \label{fig:graphical user interface}
\end{figure}

\begin{figure}
    \centering
    \includegraphics[width=0.5\linewidth]{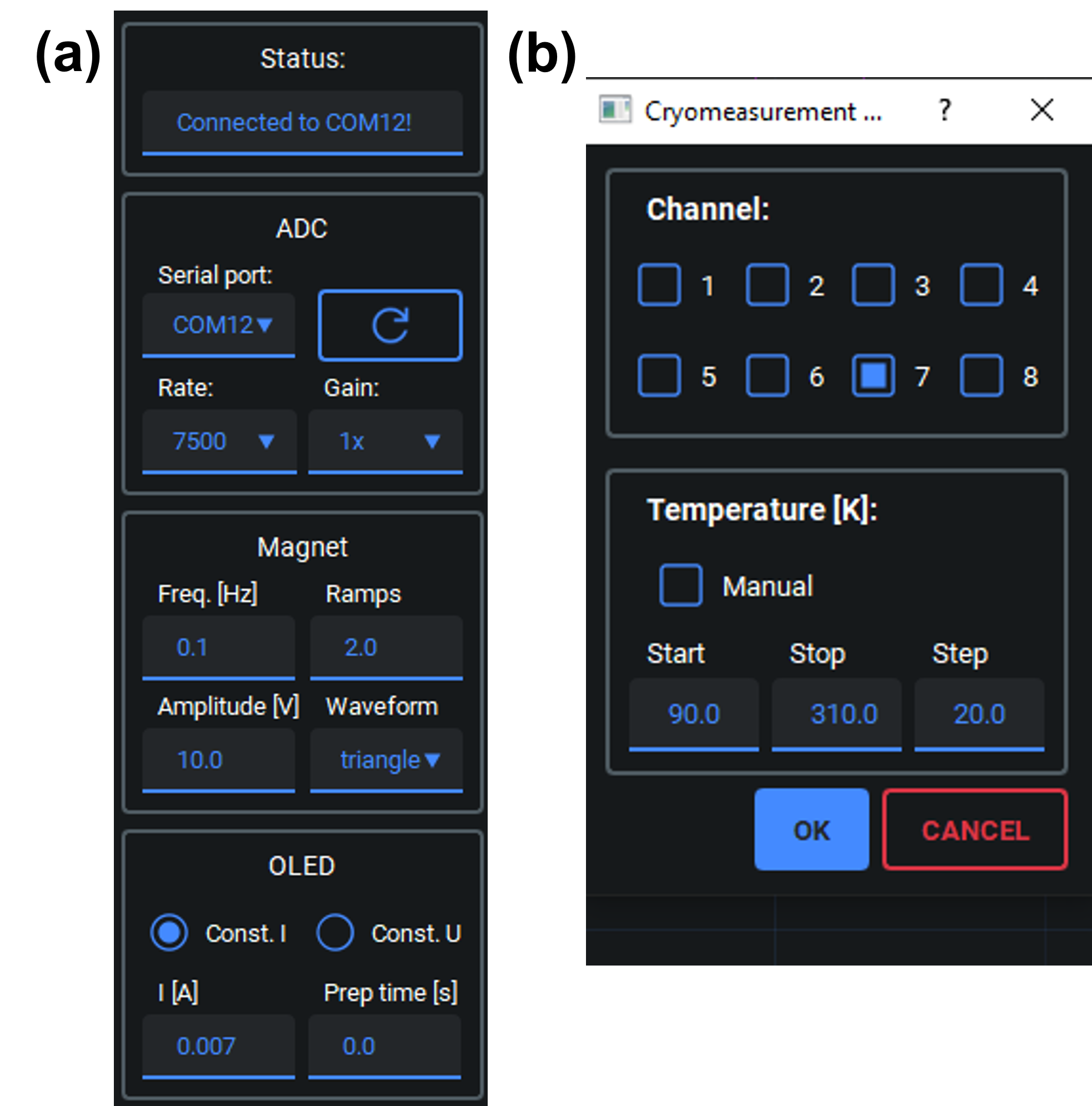}
    \caption{(a) Enlargement of the setting options shown on the left hand side in Figure \ref{fig:graphical user interface}. Additionally, selecting the \textit{Cryo mode} button opens the window shown in (b), allowing the user to select the pixel to operate and define a specific temperature range. The \textit{Manual mode} can be used to operate the OLED in the cryostat at a constant temperature.}
    \label{fig:graphical-userinterface-enlarged}
\end{figure}

This data is subsequently stored and visualized by the \textbf{frontend}.
The frontend provides a user-friendly interface that enables users to configure settings and visualize measured data, as shown in Figure \ref{fig:graphical user interface}. It effectively separates user interaction from data handling, ensuring clarity and ease of use. The available settings, displayed on the left side of Figure \ref{fig:graphical-userinterface-enlarged} (a), include the sampling rate for the ADC and the gain. The function generator settings allow users to adjust the frequency, the number of ramps corresponding to the period, the amplitude, and the waveform. The OLED can be operated in either constant current or voltage mode, with users able to select the current (I) and voltage (U) values. If the device needs to be powered on before measurement, a preparation time can be set before starting the measurement. The \textit{Debug mode} can be used to adjust the amplification for the photodiode or to check the OLED signal, as the ADC is read out, while bypassing the main measurement routine. Additionally, the \textit{Cryo Mode} (see Figure \ref{fig:graphical-userinterface-enlarged} (b)) allows users to select the pixel to be measured and define the appropriate temperature range. The \textit{Manual Mode} offers the option to turn on the pixels without sweeping the temperature, enabling measurements at a fixed temperature without looping.  
\newline
The measured data is displayed over time, in the first three diagrams in the middle of the graphical user interface. On top, the hall voltage is shown, while in the middle the OLED signal and on the bottom, the photodiode response are depicted. The right-hand side shows the signal from the OLED and photodiode in dependence on the external magnetic field. 

Figure \ref{fig:front-backend} illustrates the clear distinction between the backend and frontend, ensuring that data handling and user interaction remain well-separated. We break down components into independent modules to enhance reusability and maintainability.

\subsubsection{Data Processing Software}
\begin{figure}
    \centering
    \includegraphics[width=0.9\linewidth]{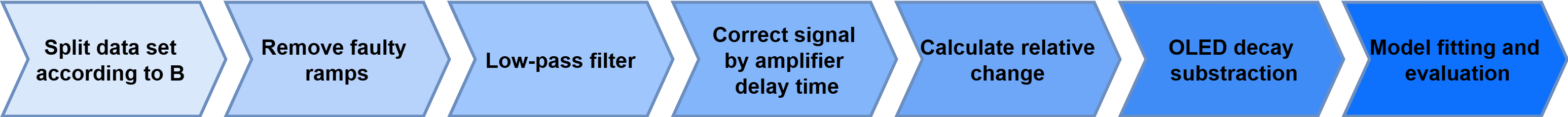}
    \caption{Schematic representation for the post-processing routine.}
    \label{fig:flow-chart-post-processing}
\end{figure}

The raw signals must undergo pre-processing and data analysis to generate the $\mathrm{MFE(\%)}$. For exact details, we refer to our GitHub link (see above). The simplified flow chart in Figure \ref{fig:flow-chart-post-processing} schematically depicts the workflow of the pre-processing routine and the analysis process. The first step involves separating the measurements based on the magnetic field (x-axis), while discarding faulty ramps. Faulty ramps are attributed to incomplete ramps because the magnetic field is continuously swept, and the starting point can randomly fall anywhere along the triangular function. So far, only the first and last ramps have been discarded. To further improve, an outlier detection is suggested. An infinite impulse response (IIR) low-pass filter of the type Butterworth is then applied to remove 50 Hz noise, with the cut-off frequency being user-defined. This kind of filter was chosen as they have a very consistent frequency response in the passband, allowing the signal of interest to be accurately represented.

Once the data is filtered, the signal is time-corrected to compensate for the delay introduced by the amplifiers. This delay was determined by applying a rectangular test signal both directly to the oscilloscope and via the amplifier. A time shift of $3$ ms was observed and subsequently corrected during post-processing.
Further, the relative change is calculated for the MFE response and subtracted from the OLED decay during the measurement. Afterwards, the data is fitted using models that are selected by the user in the post-processing configuration file. The least-squares method was consistently used to fit the data throughout the study. To ensure optimal fitting quality, users can select the best-fitting model based on various metrics, such as the Bayesian Information Criterion (BIC). Two primary classes of fitting models are available: \textit{Lorentzian} and \textit{Cole-Cole} models. Previous work has demonstrated that the Cole-Cole-Lorentzian fit accurately replicates the MFE response of exciplex-driven TADF materials (\textit{cf.} \cite{morgenstern2024analysis}). Nevertheless, throughout the present study, the Double-Lorentzian fitting function revealed the best fit.
\newline
For the Cole-Cole fitting function, a correlation exists between the mean polaron pair lifetime and the broadening of the magnetic field effect. As a result, the post-processing routine can also evaluate the distribution of the polaron pair lifetime, reflecting the dispersive dynamics of the polaron pair species. The distribution determination was implemented as discussed in ref. \cite{mondal2023degradation, mclaughlin2022study}. The fitted data, along with the raw data, is separated for each ramp and stored in individual folders. Additionally, a data file containing all fitted values from the fitting routine is generated. The distribution of the polaron pair lifetime is also stored separately.
\newline
For temperature measurements, the post-processing procedure also considers an extra column containing the pixel number and the measured temperature. In addition to the steps mentioned earlier, the data measured at each temperature step and pixel is averaged, and the standard deviation is calculated and stored in a separate file. This enables a direct observation of the temperature trend, with error bars representing the standard deviation. Please note that no safety mechanism has been implemented thus far. As a result, any outliers that occur during temperature-dependent measurements are included in the statistical analysis and may influence the results. In future work, a robust outlier detection method should be developed and incorporated accordingly.

\FloatBarrier

\subsubsection{Experimental Results and Analysis}
\begin{figure}
    \centering
    \includegraphics[width=0.9\linewidth]{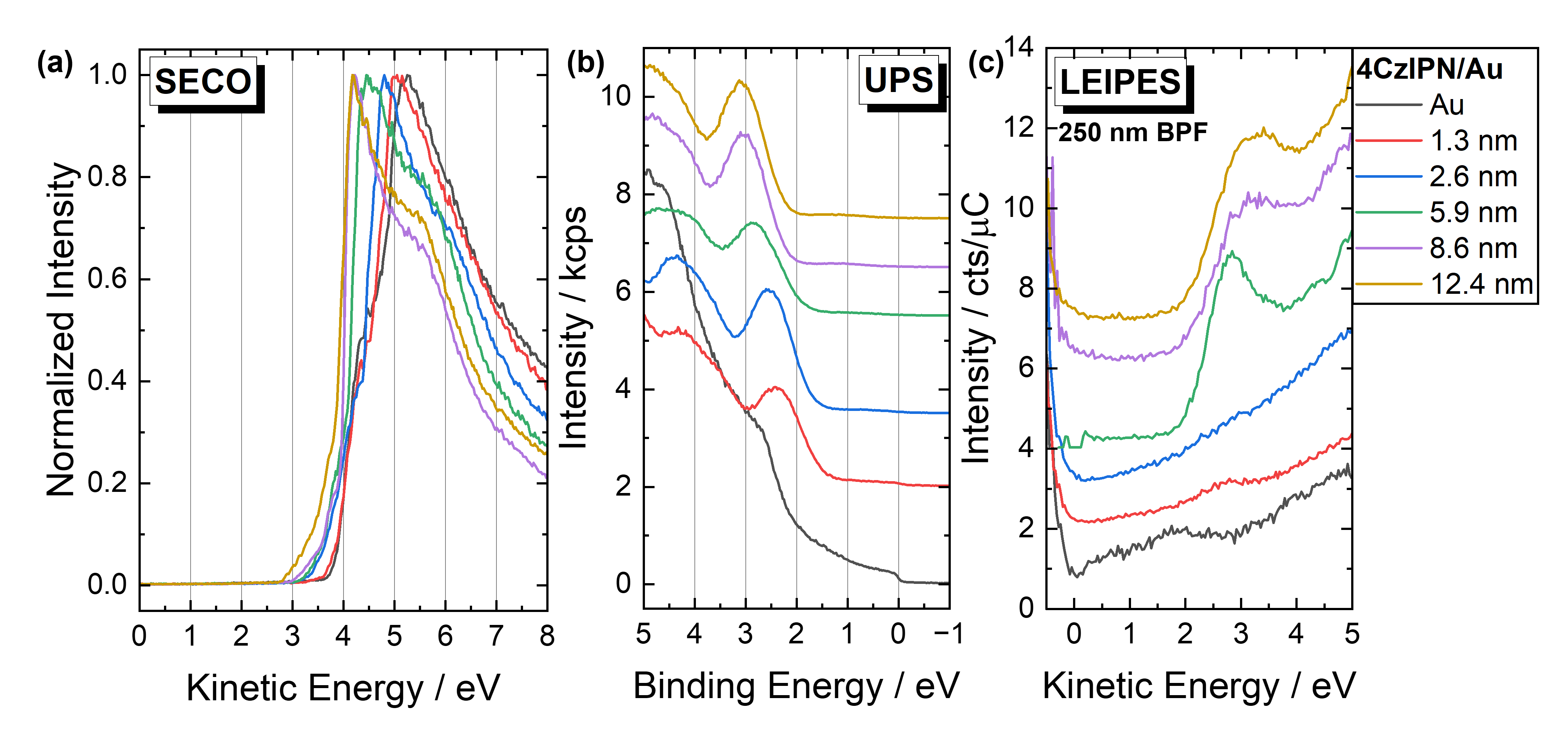}
    \caption{Combined UPS and LEIPES measurements of 4CzIPN thin films thermally evaporated on Au with progressive film thickness. (a) UPS of the SECO measured under $-10$ V sample bias, for which the spectra are corrected,(b) UPS of the valence region, (c) LEIPES measured using a $250$ nm BPF.}
    \label{fig:SECO-UPS-LEIPES}
\end{figure}

Figure \ref{fig:SECO-UPS-LEIPES} shows the UPS measurements of 4CzIPN, as well as sample LEIPES spectra measured using a $250$ nm bandpass filter (BPF). After preparation of the Au substrate by successive sonication in acetone, ethanol and DI water, it was loaded into the UHV chamber. An initial characterization of the Au substrate was performed by measurements of the valence region and the secondary-electron cut-off (SECO) using UPS, of the unoccupied density of states (UDOS) using LEIPES, and of the core-levels using XPS.  Sample XPS spectra are shown in Figure \ref{fig:XPS-Au-C} and are discussed later on. The base pressure in the analysis chamber was $1 \times 10^{-10}$ mbar. For UPS measurements, the He I emission line at $21.2182$ eV from a He discharge lamp was utilized by establishing a He gas pressure of $3 \times 10^{-8}$ mbar to suppress higher energy He emission lines. The SECO was measured under a bias voltage of $-10$ V to ensure electrons can surpass the potential barrier given by the analyzer work function. The spectra in Figure \ref{fig:SECO-UPS-LEIPES} (a) are corrected for that bias. The LEIPES measurement of Au were performed using an incident electron current of $2 \ \mu\mathrm{A}$ above the turn-on point of electron emission. For well-conductive samples the turn-on point is given by the work function of the BaO cathode and it constitutes the position of $0$ eV electron kinetic energy. As the isochromat mode is employed for LEIPES measurements, the electron energy is varied while using a constant probing energy of the emitted photons, achieved by the BPF. The electron energy is set by a fixed electron emission energy of $10$ eV from the electron source, while increasing the negative bias on the sample in steps of $40$ mV. The bandpass energy facilitates the position of the vacuum level on the kinetic energy scale. For the spectra in Figure \ref{fig:SECO-UPS-LEIPES} (c), it is at $4.96$ eV. These spectra are achieved by normalization of the raw signal from the photodetector by the incident electron current and the acquisition time per point. This normalization also causes the apparent increase in intensity below $0$ eV, because the constant background signal is divided by the electron current which is approaching $0$ in that range.
The initial SECO measurement of Au shows more than one cut-off due to the additional presence of adventitious carbon on the substrate surface. The higher energy cut-off was considered for the determination of the work function of Au. The valence region shows a clearly visible Fermi edge as well as the Au5d valence features superimposed with the C3s and C3p features originating from the adventitious carbon on the Au surface. The LEIPES measurement of Au also shows a Fermi edge near $0$ eV. However, it is likely to be partially cut off, due to the work function being close to the bandpass energy.
After the initial measurements of the Au substrate, it was moved to the preparation chamber for the successive deposition of 4CzIPN by thermal evaporation. The base pressure in the evaporation chamber was $2 \times 10^{-9}$ mbar at the beginning of the experiment. The material was introduced as powder into a water-cooled evaporation cell. The material temperature was controlled by a type-S thermocouple in direct contact with the crucible. Evaporation rates were controlled with a quartz crystal microbalance (QCM) and kept near $\sim 1$ nm/min. Before the first deposition the evaporation cell was degassed at approximately $200^\circ$C until the chamber pressure normalized at about $1 \times 10^{-8}$ mbar, which was also the typical pressure during deposition. After the first deposition the base pressure decreased to $5 \times 10^{-10}$ mbar. The typical evaporation temperature during deposition was approximately $230^\circ$ C.
The SECO measurements of the organic films show a gradual shift of the main cut-off at higher energy, which is also the cut-off considered for the determination of the work function. The lower energy features previously associated with adventitious carbon decrease in relative intensity which is expected as the carbon should be gradually buried below the 4CzIPN film. In the valence region, clearly visible HOMO peaks appear while the Fermi edge of Au vanishes. For increasing film thickness the absolute intensity of those peaks barely changes, as the information depth of UPS is in the range of only a few Angstroms, which is significantly lower than the deposited film thickness. A gradual movement of the peak positions with increasing film thickness is observable, saturating for the last two measurements. 
One of the main advantages of LEIPES, compared to conventional inverse photoemission spectroscopy, is that by employing electron kinetic energies below $5$ eV, much of the damage that the electrons can deal to organic materials is avoided. However, this is often not enough, as high electron currents can still damage organics even at low energies after sufficient time. As the typical acquisition time per point in our setup is in the range of $120\ – \ 150$ s, a single spectrum takes several hours which can definitely be sufficient to still observe degradation behavior. Therefore, we restricted the electron current to approximately $800$ nA for the measurements on the 4CzIPN films. In combination with the relatively large beam diameter of about $5$ mm, this helped to drastically decrease possible degradation effects, while still sustaining sufficient signal intensity.
For the LEIPES measurements of the first two evaporation steps only a very tiny signal potentially originating from the LUMO levels can be observed. There are two main reasons for this. Firstly, the interaction cross-section for the inverse photoemission process can be up to 5 orders of magnitude lower than for the direct photoemission process, making the sample characteristics harder to detect. Secondly, the information depth given by the electron inelastic mean free path (IMFP) is higher for LEIPES than for UPS. The minimum of the universal IMFP curve is near $40$ eV, amounting to an information depth of approximately one atomic layer. As mentioned above, for He-I-UPS, the information depth is slightly higher, more than about $5 \ \mathrm{\r{A}}$. At $5$ eV, which is the maximum kinetic energy used in our LEIPES experiments, the IMFP is at almost $6$ nm. Due to that, the largest part of the electrons would travel right through ultra-thin surface layers without even interacting with the film. The resulting low intensity also makes the analysis of the data more challenging and uncertain. However, after the third deposition step a strong LUMO peak can be observed, which moves to higher energy after the fourth step and remains at the same position thereafter.

\begin{figure}
    \centering
    \includegraphics[width=0.8\linewidth]{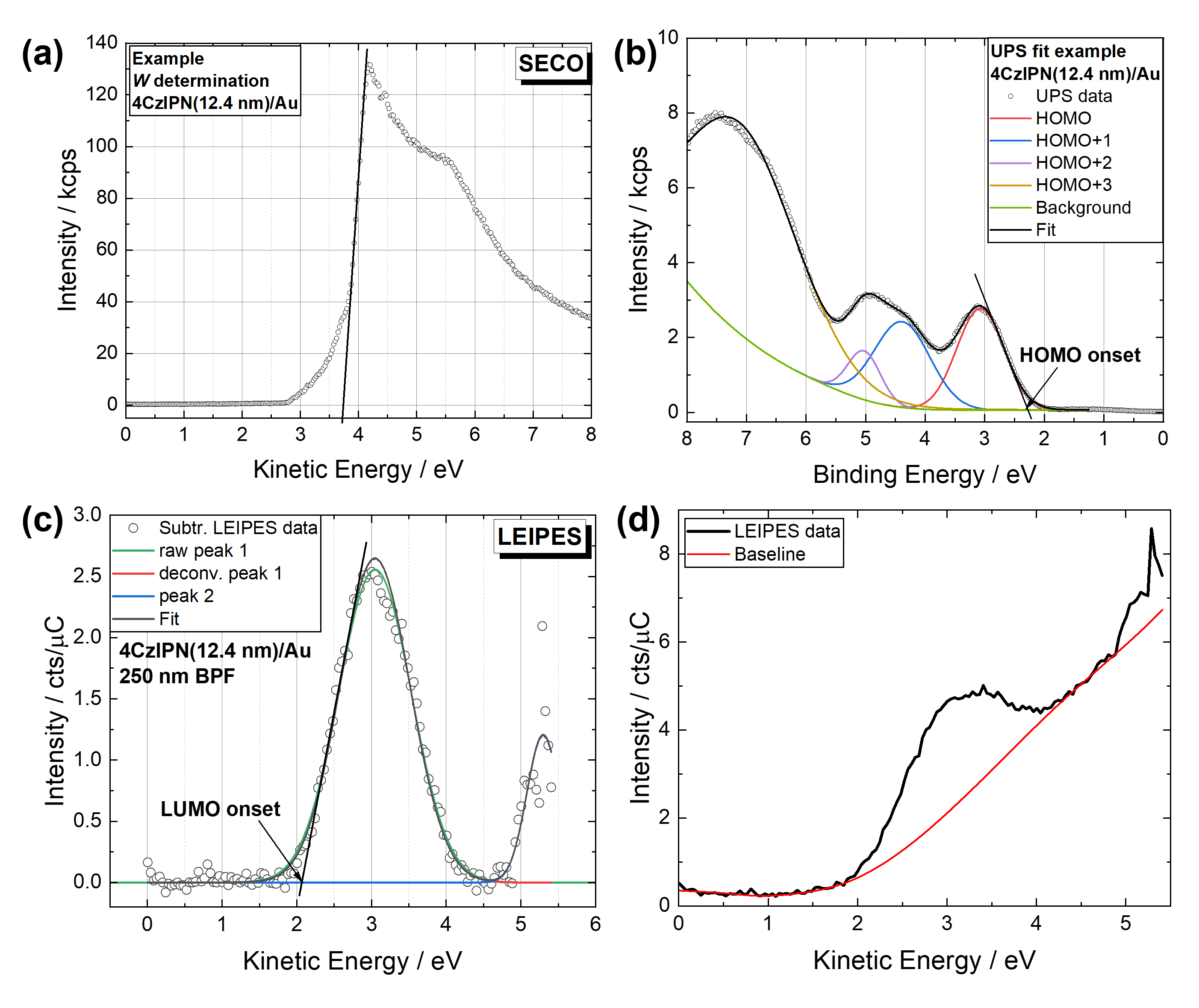}
    \caption{Sample analyses of the SECO, valence, and LEIPES spectra. (a) Determination of the work function $W$ of the fifth deposition step of 4CzIPN on Au using a linear extrapolation of the cut-off, (b) peak fitting of the HOMO levels for the same deposition step and determination of the HOMO onset using a linear extrapolation of the Voigt peak, (c) Peak fit of the LEIPES data for the same deposition step measured with a $250$ nm BPF, deconvolution of the experimental broadening and determination of the LUMO onset via linear extrapolation of the deconvoluted Gaussian peak, (d) Baseline subtraction of the polynomial background of the LEIPES measurement.}
    \label{fig:SECO-UPS-LEIPES-fit}
\end{figure}

\begin{figure}
    \centering
    \includegraphics[width=0.8\linewidth]{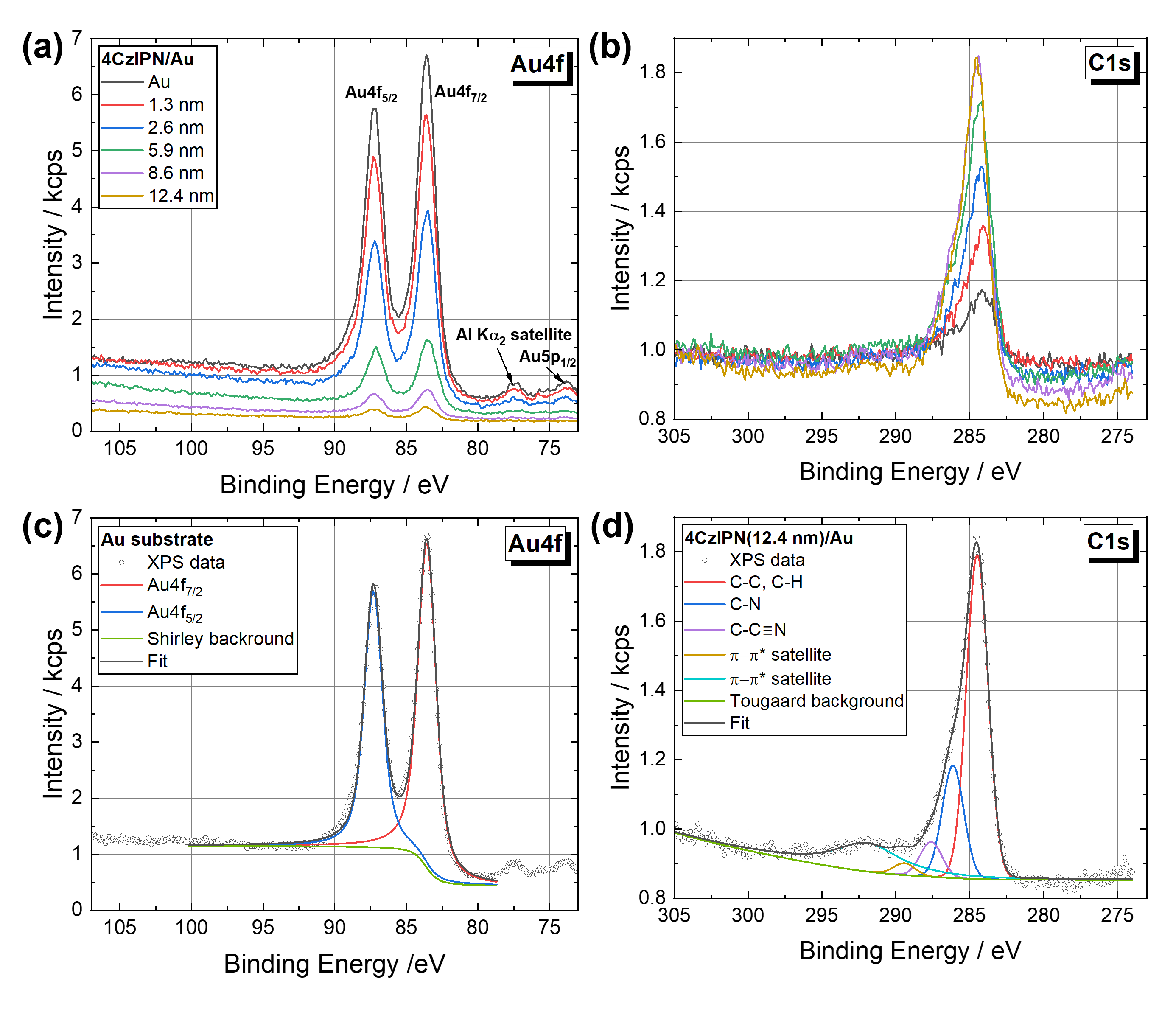}
    \caption{XPS core-levels and there analysis. (a) Au4f core-levels for the progressive deposition steps, (b) C1s core-levels for the progressive deposition steps, (c) sample fit of Au4f for the Au substrate, (d) sample fit of C1s for the fifth deposition step of 4CzIPN on Au.}
    \label{fig:XPS-Au-C}
\end{figure}

Figure \ref{fig:SECO-UPS-LEIPES-fit} depicts sample analyses of the resulting spectra. As mentioned above, for the SECO measurements we considered to higher energy cut-off for the determination of the work function. It is determined by simply taking the intersect of the linear extrapolation of the cut-off and the energy axis, as shown in Figure \ref{fig:SECO-UPS-LEIPES} (a).
For the analysis of the valence region, peak fitting as in Figure \ref{fig:SECO-UPS-LEIPES-fit} (b) is employed. For this analysis, we utilized the data analysis tool of the measurement software SPECSLab Prodigy. A Tougaard background was assumed, as well as a Voigt line shape for the HOMO peaks. The HOMO onset is taken as the intersect of the linear extrapolation of the approximately linear part of the first Voigt peak and the baseline.
For LEIPES, a similar approach was taken, as depicted in Figures \ref{fig:SECO-UPS-LEIPES-fit} (c) and (d). However, a polynomial background and a Gaussian peak shape were assumed. An additional consideration was taken into account, as it was shown that better agreement between calculation and experiment could be achieved in inverse photoemission spectroscopy, by deconvolution of the experimental broadening \cite{zahn2006transport}. To that end, we employed a pseudo-deconvolution as
\begin{equation}
    FWHM^{2}_\mathrm{LUMO} = FWHM^{2}_\mathrm{measured} - FWHM^{2}_\mathrm{setup}
\end{equation}

Here, $FWHM_\mathrm{LUMO}$ is the true broadening of the LUMO peak, $FWHM_\mathrm{measured}$ is the broadening determined by a direct Gaussian fit of the LUMO peak, and $FWHM_\mathrm{setup}$ is the broadening due to the spectral resolution. The latter was determined by a reference measurement of the Fermi edge of Ag to be $0.3$ eV. Figure \ref{fig:SECO-UPS-LEIPES-fit} (c) shows the difference between the directly fitted Gaussian, the “raw” peak, and the deconvoluted one. As the measured broadening is at about 1 eV more than three times as large as the experimental broadening the effect may seem rather small. But it can make a difference of several $10$ meV in the final value of the electron affinity.

The main purposes of XPS in our experiment were to additionally verify the deposition of the 4CzIPN film and its chemical integrity, as well as to determine the film thickness, as XPS can provide quite accurate and consistent thickness values in the range below $20$ nm where many other techniques may have difficulties. Also, since we can measure XPS without changing the setup, exposure to the ambient will have no influence on the results. Figure \ref{fig:XPS-Au-C} shows the analysis of the main core-levels which were considered for the thickness determination, \textit{i.e.} Au4f and C1s. All spectra were measured under the same incident X-ray power, same pass energy ($20$ eV), and the same position. It is clearly visible how the intensity of Au4f gradually decreases with each deposition step, while the C1s peak rises. It is also worth noting that C1s has a contribution for the “pure” Au substrate, which attributes to the aforementioned adventitious carbon on the Au surface. According to ref. \cite{jablonski2019evaluation}, the thickness of an adlayer can be determined from the measurement of a substrate core level and a core level of the adlayer via the numerical evaluation of the formula

\begin{equation}
    \frac{I_f}{I_s} = \frac{I^\infty_f}{I^\infty_s} \frac{[1-exp(-d/L_f (E_f)cos(\alpha)]}{exp(-d/L_f (E_s)cos(\alpha)]}
\end{equation}

where $I_f$ would be the measured intensity of the C1s peak from the 4CzIPN film, Is the measured intensity of Au4f from the substrate, both normalized to their respective interaction cross-section, $I^\infty_f$ and $I^\infty_s$ the respective intensities from the bulk material, d the thickness, $L(E_f)$ the electron attenuation length of C1s electrons in the 4CzIPN layer, $L(E_s)$  the electron attenuation length of Au4f electrons in the 4CzIPN layer, and $\alpha$ the detection angle. The difference between the electron attenuation length and the IMFP is that the former also takes elastic electron scattering effects into account. The ratio $I^\infty_f/I^\infty_s$, and the attenuation lengths can be calculated from considerations of electron energy, bandgap energy, and mass density. For this, we used the Avantage software by Thermo Fisher Scientific. To determine the measured core level intensities, peak fitting was carried out as depicted in Figures \ref{fig:XPS-Au-C} (c) and (d). For Au4f, a typical asymmetric line-shape has to be considered, which was provided by a convolution of Gaussian and the asymmetric Doniach-Sunjic peak functions. For C1s, Voigt peak functions were used. Figure \ref{fig:XPS-Au-C} (d) also shows assignments of which chemical bonds are likely to contribute to the C1s line shape. We would normally consider the two peaks labeled as $\pi - \pi^*$ satellites more likely to be just one broader feature, however accounting for the second smaller peak resulted in a more reasonable fit. Such satellites are typically observed in materials containing benzene-like structures. 

\begin{figure}
    \centering
    \includegraphics[width=0.7\linewidth]{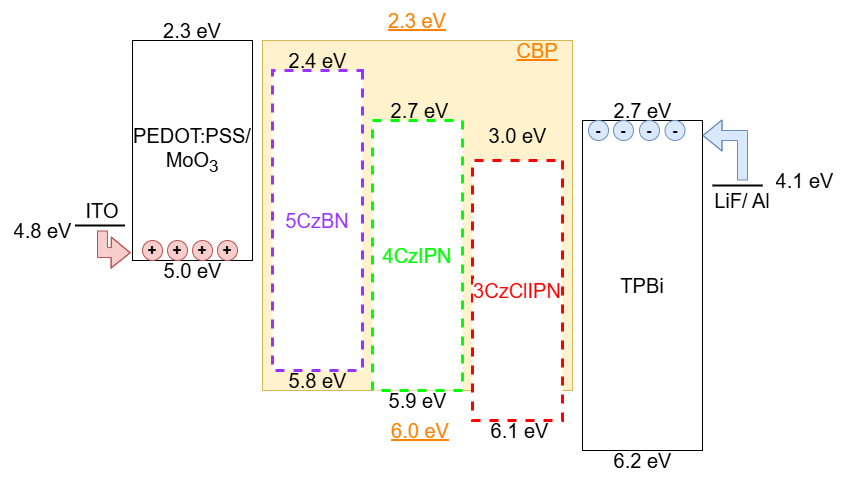}
    \caption{Band alignment scheme for the layer stack used throughout this study. In devices based on bare emitter molecules, the host matrix CBP was omitted, while all other layers remained unchanged (\textit{cf.} Figure 1 (a) and (b)).}
    \label{fig:band-alignment}
\end{figure}

\vspace{1em}
Figure \ref{fig:band-alignment} shows a scheme of the band alignment for the layer stack used in this study. The host matrix CBP was incorporated only in the host-guest systems and omitted in the bare molecular devices. The calculated absolute HOMO and LUMO energies of CBP ($–5.58$ eV and $–1.54$ eV, respectively) are significantly lower than the values reported in the literature \cite{lo2002green}. This discrepancy likely arises from the level of theory employed, which is specifically optimized for cyanoarene systems (\textit{vide infra}) and demonstrates excellent agreement with our experimental data for those molecules. However, deviations for structurally different compounds such as CBP would not be uncommon. Consequently, we used the literature values for the HOMO and LUMO position for CBP as well as for the injection materials (\textit{cf.} \cite{weber2024exciplex}). The emitter HOMO-LUMO positions were used from our calculations (\textit{cf.} Figure 2(a)). As illustrated, the energy offsets between the hole-injection layer (PEDOT:PSS/$\mathrm{MoO_3}$) and the emitter molecules are, in some cases, quite significant. A similar mismatch was observed between the electron-injection layer (TPBi) and the emitters. These energy barriers hinder charge injection, thereby limiting device performance and contributing to the relatively high turn-on voltage (see Table \ref{tab:deivce-performance}). Despite these limitations, the simplified device architecture allows for a clear assessment of the intrinsic charge transport properties of the emitter molecules and their organic magnetic field effects. Optimizing the band alignment would be essential for improving device efficiency.

\begin{figure}
\centering
  \includegraphics[width=0.8\linewidth]{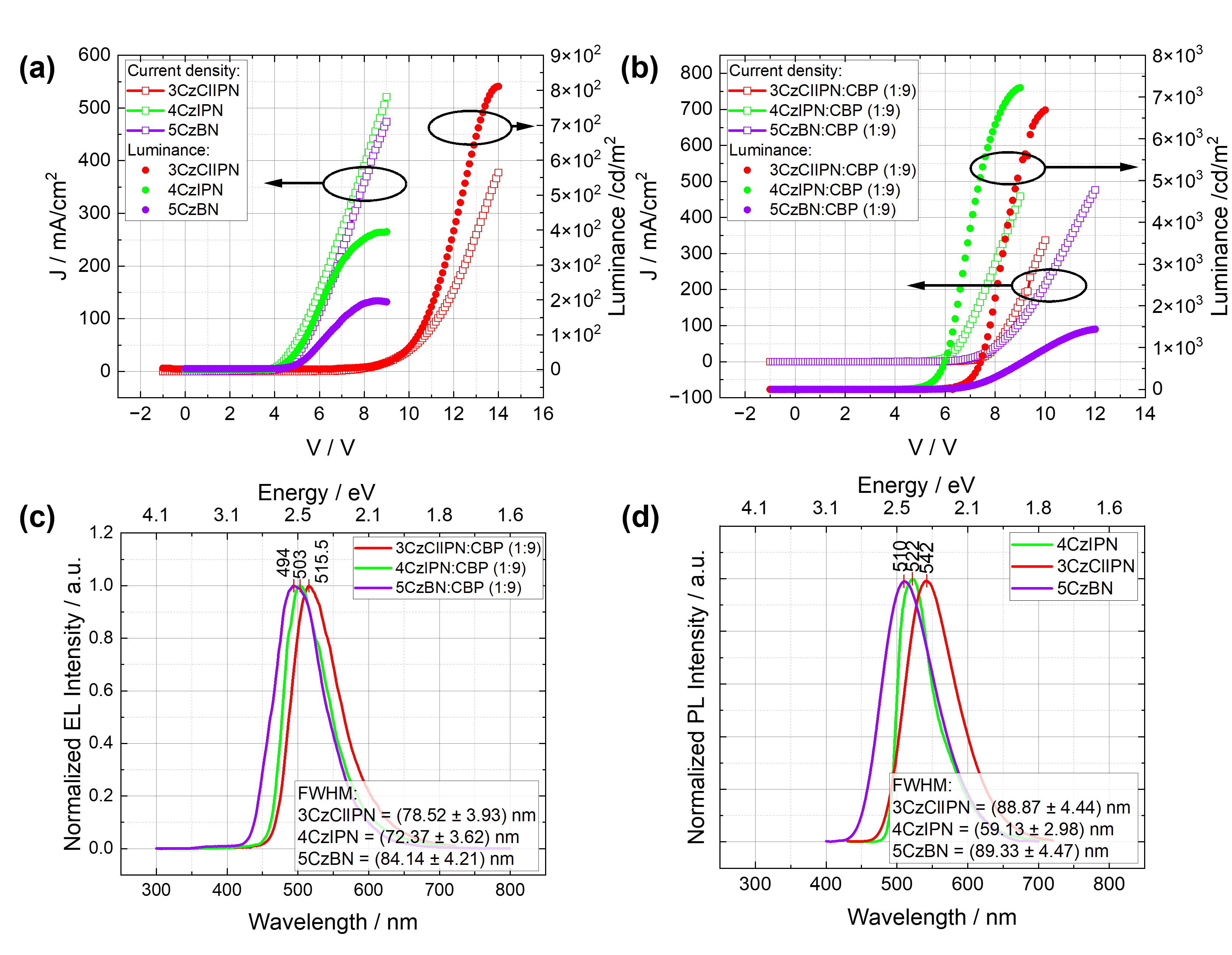}
  \caption{Current density and luminance as functions of the applied bias voltage analyzed for (a) the neat films of the materials 3CzClIPN, 4CzIPN, and 5CzBN as well as (b) the corresponding host-guest systems, where the same materials (3CzClIPN, 4CzIPN, and 5CzBN) were employed as guest emitters, embedded in a CBP host matrix. (c) EL spectra for the host-guest systems used at (b). (d) PL spectra for the three emitter molecules 3CzClIPN, 4CzIPN, and 5CzBN with the respective full width at half maxima (FWHM).}
  \label{fig:J-V-Luminance}
\end{figure}

\vspace{1em}
In Figure \ref{fig:J-V-Luminance} (a) and (b), the J-V and luminance-voltage characteristics of both bare-molecule OLEDs and host-guest systems are presented. Notably, at high bias voltages, a decrease in luminance is observed, which correlates with non-radiative recombination. Embedding the TADF emitters in a host-guest matrix significantly enhances luminance and, consequently, the EQE. This improvement is attributed to the reduced aggregation and interaction of the TADF emitter molecules \cite{mao2024interacting, stavrou2020photophysics}.  
\newline
The EL response of the host-guest systems and the PL response of the bare TADF emitters are shown in Figure \ref{fig:J-V-Luminance} (c) and (d), respectively. The host-guest system exhibits a much narrower EL response (\textit{cf.} Figure 3 (b) manuscript), indicating higher selectivity and improved color purity \cite{dos2016using, stavrou2020photophysics}. When comparing EL and PL spectra, no peak shift is observed, suggesting similar behavior under both optical and electrical excitation. The exact values for all optoelectronic devices throughout this study can be found in Table \ref{tab: Device performance}.

\begin{table}
 \caption{Device performance parameters for the devices discussed in this work. The corresponding spectra can be found in Figure 3 (b) and (c) and \ref{fig:J-V-Luminance}. The Turn-on voltage was determined for each device at a Luminance (L) of $\mathrm{10 \ cd/m^2}$.}
 \label{tab:deivce-performance}
  \begin{tabular}[htbp]{@{}lllll@{}}
    \hline
    Device & Turn-on V / V & Maximum EQE / $\%$ & Maximum L / $\mathrm{cd/m^2}$ & EL peak position / nm\\
    \hline
    3CzClIPN  & ($9.10 \pm 0.46$)  & ($0.11 \pm 0.02$) & ($723.89 \pm 36.68$) & ($545.20 \pm 0.12$)  \\
    4CzIPN  & ($4.22 \pm 0.19$)  & ($0.11 \pm 0.02$) & ($395.08 \pm 19.71$) & ($520.50 \pm 0.12$)  \\
    5CzBN  & ($4.78 \pm 0.24$)  & ($0.13 \pm 0.01$) & ($197.34 \pm 9.85$) & ($512.00 \pm 0.13$)  \\
    3CzClIPN:CBP (1:9)  & ($5.08 \pm 0.25$)  & ($2.76 \pm 0.14$) & ($6689.04 \pm 334.45$) & ($515.50 \pm 0.34$)  \\
    4CzIPN:CBP (1:9) & ($4.44 \pm 0.22$)  & ($2.61 \pm 0.13$) & ($7279.18 \pm 363.96$) & ($503.00 \pm 0.12$)  \\
    5CzBN:CBP (1:9)  & ($5.93 \pm 0.29$)  & ($0.54 \pm 0.03$) & ($1440.93 \pm 72.05$) & ($494.00 \pm 0.14$)  \\
    
    \hline
    \label{tab: Device performance}
  \end{tabular}
\end{table}

The CIE values extracted from the chromaticity diagram shown in Figure 3 (d) are summarized in Table \ref{tab:CIE values}. It can be observed that the emitted color of 5CzBN is significantly cooler compared to the other two TADF emitters, which is associated with its increased optical bandgap. Additionally, the host-guest systems also exhibit a blue-shifted emission color, likely due to the altered surrounding environment in these systems.

\begin{table}[htbp]
  \caption{CIE values for the pristine molecules 3CzClIPN, 4CzIPN, and 5CzBN, together with their respective host-guest systems}
  \label{tab:CIE values}
  \centering
  \begin{tabular}{lll}
    \hline
    Device & CIEx & CIEy \\ 
    \hline
    3CzClIPN & 0.401 & 0.564 \\ 
    4CzIPN & 0.288 & 0.583 \\ 
    5CzBN & 0.272 & 0.470 \\ 
    3CzClIPN:CBP (1:9) & 0.258 & 0.554 \\ 
    4CzIPN:CBP (1:9) & 0.230 & 0.539 \\ 
    5CzBN:CBP (1:9) & 0.166 & 0.032 \\ 
    \hline
  \end{tabular}
\end{table}

\begin{figure}
    \centering
    \includegraphics[width=0.8\linewidth]{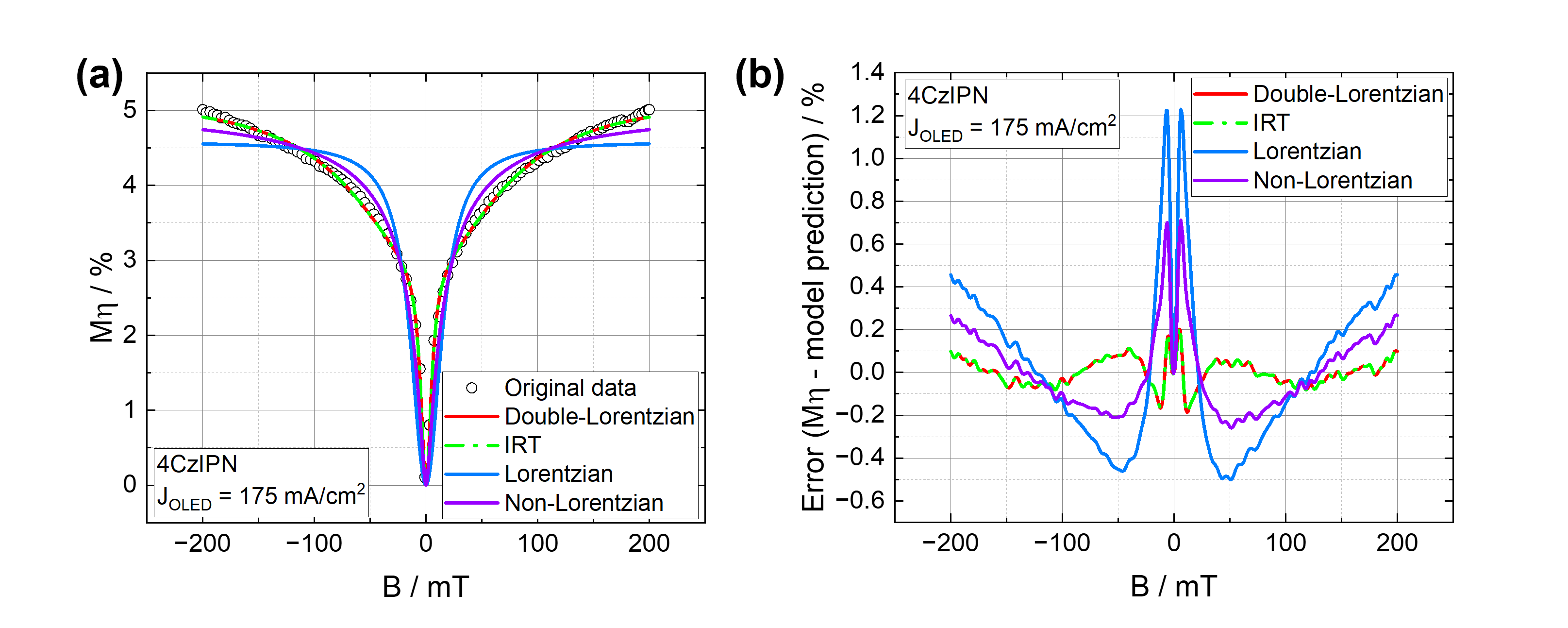}
    \caption{(a) The original $M_\mathrm{\eta}$ data for 4CzIPN at a constant current density of $J_\mathrm{OLED} = 175 \ \mathrm{mA/cm^2}$ is fitted by four different fitting functions as an example, namely Lorentzian, Non-Lorentzian, IRT (\textbf{I}SC-\textbf{R}ISC-\textbf{T}CA), and Double-Lorentzian, respectively. (b) demonstrates the error observed between the original and predicted data by the models used under (a).}
    \label{fig:model-evaluation}
\end{figure}

To evaluate model quality and identify a simple yet accurate representation of our data, we tested several models, whose mathematical expressions are listed in Table \ref{tab:model-formulas}. Additional models were examined but are not shown here. The selection criteria for the most suitable model were simplicity and the ability to accurately replicate the measured data. As an example, the fitted $M\eta$ response of a 4CzIPN OLED at a constant current density of $J_\mathrm{OLED} = 175 \ \mathrm{mA/cm^2}$ is presented in Figure \ref{fig:model-evaluation}. The extracted parameter values are summarized in Table \ref{tab:model-values}.

To account for different magnetic field effect regimes, the fitting was performed in three distinct regions: low-field (LF), mid-field (MF), and high-field (HF). The same terminology was used throughout the manuscript. As shown in Table \ref{tab:model-values}, the models vary in the number of fitting parameters used. The single-term models, namely the Lorentzian and Non-Lorentzian models, performed the worst, exhibiting the largest deviations between model predictions and the measured data (see Figure \ref{fig:model-evaluation} (b)). This implies that more than one charge transport mechanism is present. 

Among the multi-term models, the IRT (\textbf{I}SC-\textbf{R}ISC-\textbf{T}CA) model—proposed by Wang et al. \cite{wang2024understanding}—and the Double-Lorentzian model were compared. Notably, the amplitude of the third term in the IRT model is extremely small ($MFE_\mathrm{HF} = 2.35 \times 10^{-16}$), suggesting that this additional fitting term is unnecessary. Moreover, when examining the fitted data, the Double-Lorentzian and IRT models overlap almost perfectly. Therefore, the Double-Lorentzian model successfully replicates the data without overfitting and is thus recommended in this study. This fitting approach also reveals a strong correlation between the MF and HF terms in the IRT model. Consequently, reducing the number of parameters helps minimize parameter correlation. Furthermore, this analysis confirms that the TCA and RISC processes cannot be easily distinguished, as previously discussed in the manuscript. Since TCA decreases with the application of a magnetic field while RISC is enhanced, these two processes are inherently linked.

\begin{table}[h]
    \caption{Mathematical expressions for the models used to fit the $M\eta$ response.}
    \centering
    \begin{tabular}{l c}
        \hline
        model & formula \\
        \hline
        Lorentzian & $\mathrm{MFE(B)} = MFE_\mathrm{{LF}} \cdot \frac{B^2}{B^2 + B_0^2}$ \\
        Double-Lorentzian & $\mathrm{MFE(B)} = MFE_\mathrm{{LF}} \cdot \frac{B^2}{B^2 + B_\mathrm{{0-LF}}^2} + MFE_\mathrm{{HF}} \cdot \frac{B^2}{B^2 + B_\mathrm{{0-HF}}^2}$ \\
        Non-Lorentzian & $\mathrm{MFE(B)} = MFE_\mathrm{{LF}} \cdot \frac{B^2}{(B + B_0)^2}$ \\
        IRT & $\mathrm{MFE(B)} = MFE_\mathrm{{LF}} \cdot \frac{B^2}{B^2 + B_\mathrm{{0-LF}}^2} + MFE_\mathrm{{MF}} \cdot \frac{B^2}{B^2 + B_\mathrm{{0-MF}}^2} - MFE_\mathrm{{HF}} \cdot \frac{B^2}{(B + B_\mathrm{{0-HF}})^2}$ \\
        \hline
    \end{tabular}
    \label{tab:model-formulas}
\end{table}

\begin{table}[h]
    \caption{Characteristic values extracted by the models used in Figure \ref{fig:model-evaluation} to replicate the measured $M\eta$ response.}
    \centering
    \begin{tabular}{l c c c c c c}
        \hline
        Model & $B_\mathrm{{0-LF}}$ & $B_\mathrm{{0-MF}}$ & $B_\mathrm{{0-HF}}$ & $MFE_\mathrm{{LF}}$ & $MFE_\mathrm{{MF}}$ & $MFE_\mathrm{{HF}}$ \\
        \hline
        Double-Lorentzian & 5.68 & - & 77.32 & 3.12 & - & 2.27 \\
        IRT & 5.68 & 77.32 & 53.43 & 3.12 & 2.27 & $2.35 \times 10^{-16}$ \\
        Lorentzian & 16.30 & - & - & 4.75 & - & - \\
        Non-Lorentzian & 7.10 & - & - & 5.28 & - & - \\
        \hline
    \end{tabular}
    \label{tab:model-values}
\end{table}

\begin{figure}
    \centering
    \includegraphics[width=0.8\linewidth]{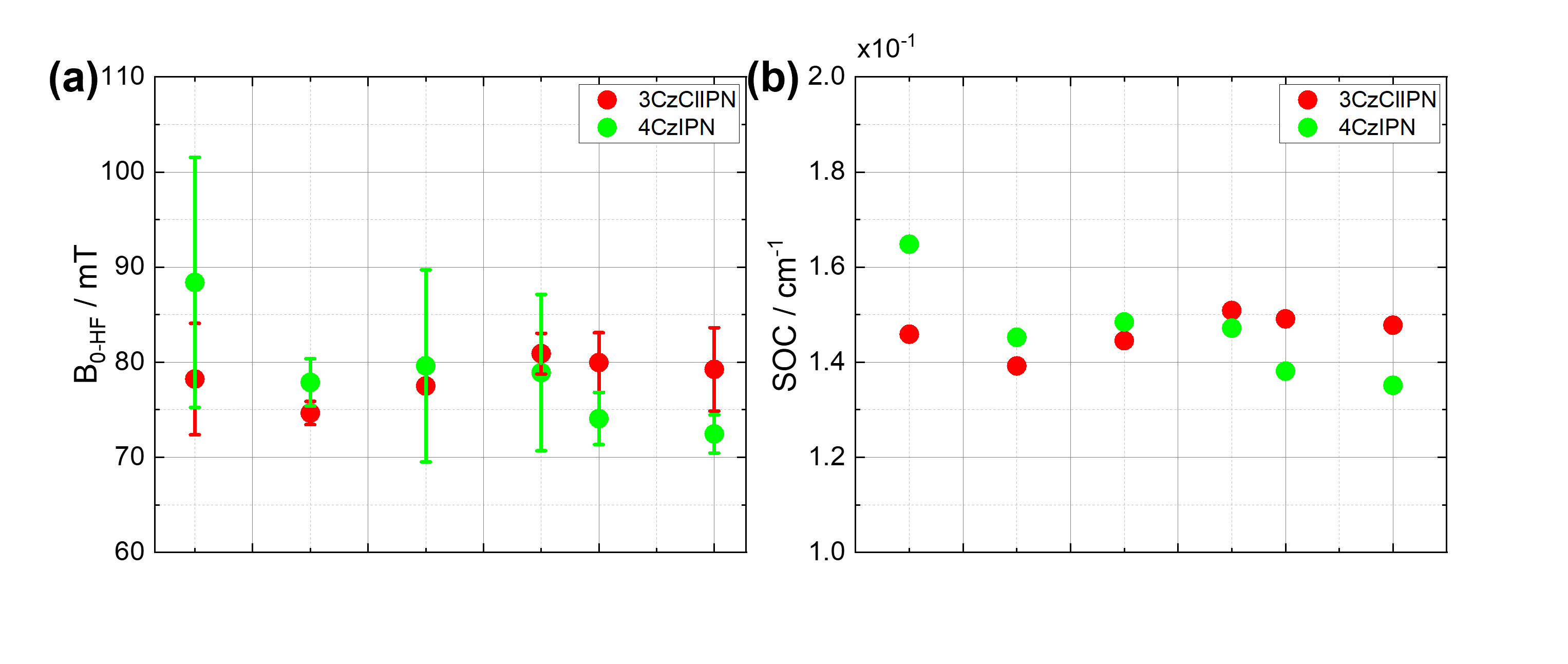}
    \caption{(a) $B_\mathrm{{0-HF}}$ for 3CzClIPN and 4CzIPN plotted as inset for the Figure  for a better visibility of the data points, (b) shows the corresponding SOC-values calculated according to Eq. (2)- manuscript.}
    \label{fig:Magnetoefficiency-B0-HF}
\end{figure}

For a better visibility, Figure \ref{fig:Magnetoefficiency-B0-HF} (a) shows the extracted $B_\mathrm{{0-HF}}$ values for 3CzClIPN and 4CzIPN, respectively. The values can also be found in the manuscript in Figure 8 (b). Figure \ref{fig:Magnetoefficiency-B0-HF} (b) shows the corresponding SOC values, which were calculated according to Eq. (2) and (3). The values match the theoretical values and are also implemented in Table 1 in the manuscript for a direct comparison. For comparison, we also listed the characteristic magnetic field values for the three TADF-emitter molecules in Table \ref{tab:SOC-CMF}, determined from the theoretical SOC values, which can be found in Table 1 in the manuscript.\newline 

\begin{table}[]
    \caption{characteristic magnetic field calculated according to Eq. (2) and (3) (manuscript) from the calculated theoretical SOC values which are listed in Table 1 of the manuscript.}
    \centering
    \begin{tabular}{c|c|c}
    \hline
        Molecule & SOC electronic state & characteristic magnetic field / mT \\
        \hline
         3CzClIPN &  $T_\mathrm{1}$ & 129.4 \\
          & $T_\mathrm{2}-S_\mathrm{1}$ MECP & 146.5\\
         \hline
         4CzIPN &  $T_\mathrm{1}$ & 161.5 \\
          &  $T_\mathrm{2}-S_\mathrm{1}$ MECP & 208.5 \\
         \hline
         5CzBN &  $T_\mathrm{1}$ & 737.9 \\
          &  $T_\mathrm{2}-S_\mathrm{1}$ $\mathrm{MECP^{\dagger}}$ & 422.4 \\
         \hline
    \end{tabular}
    \label{tab:SOC-CMF}
\end{table}

\begin{figure}
  \includegraphics[width=1\linewidth]{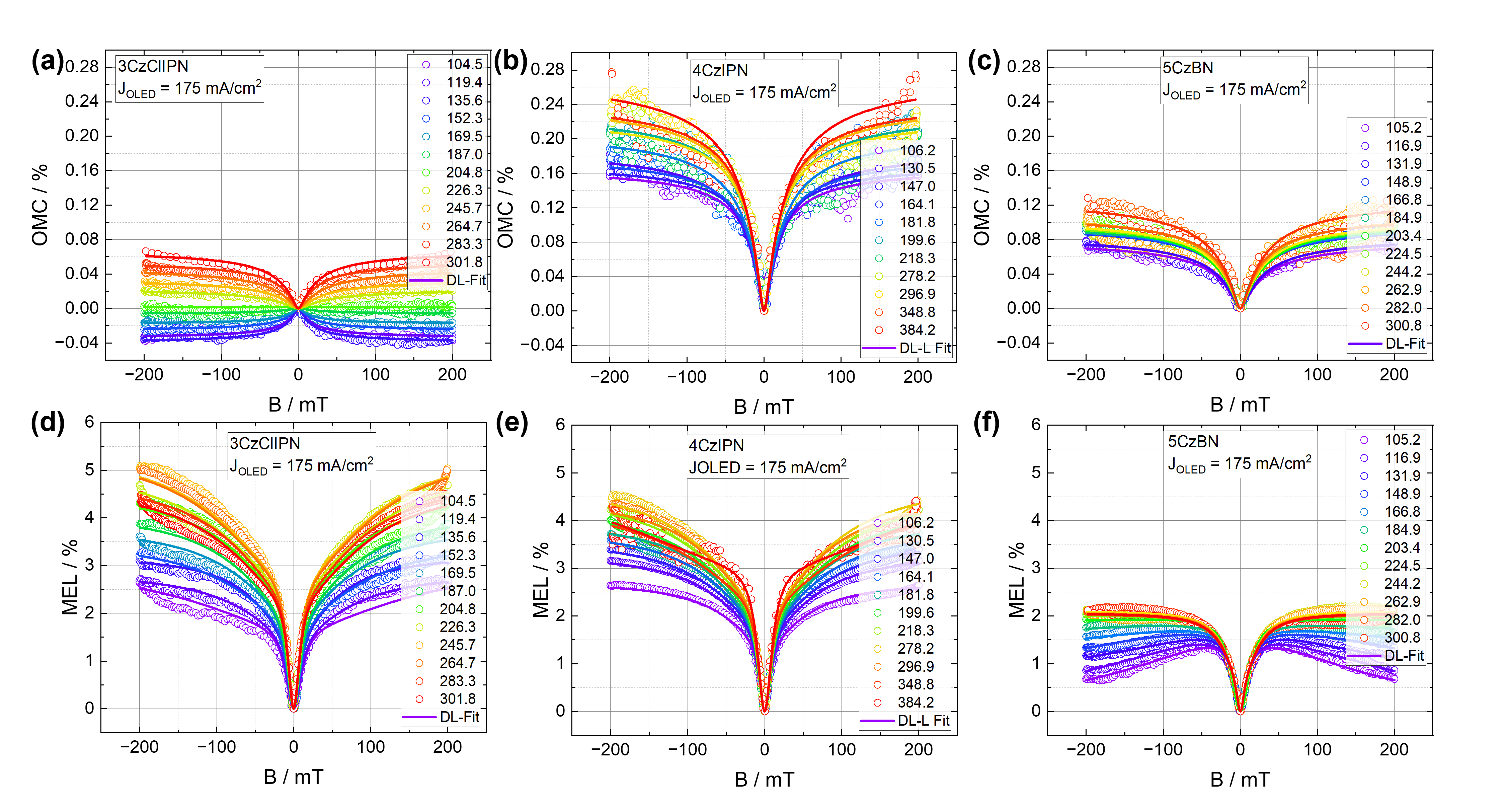}
  \caption{Temperature dependent OMC and MEL response for the three TADF-emitter molecules (a) and (d) 3CzClIPN, (b) and (e) 4CzIPN, and (c) and (f) 5CzBN.}
  \label{fig:Temp-OMC-MEL-response-full-data}
\end{figure}

\begin{figure}
    \centering
    \includegraphics[width=0.5\linewidth]{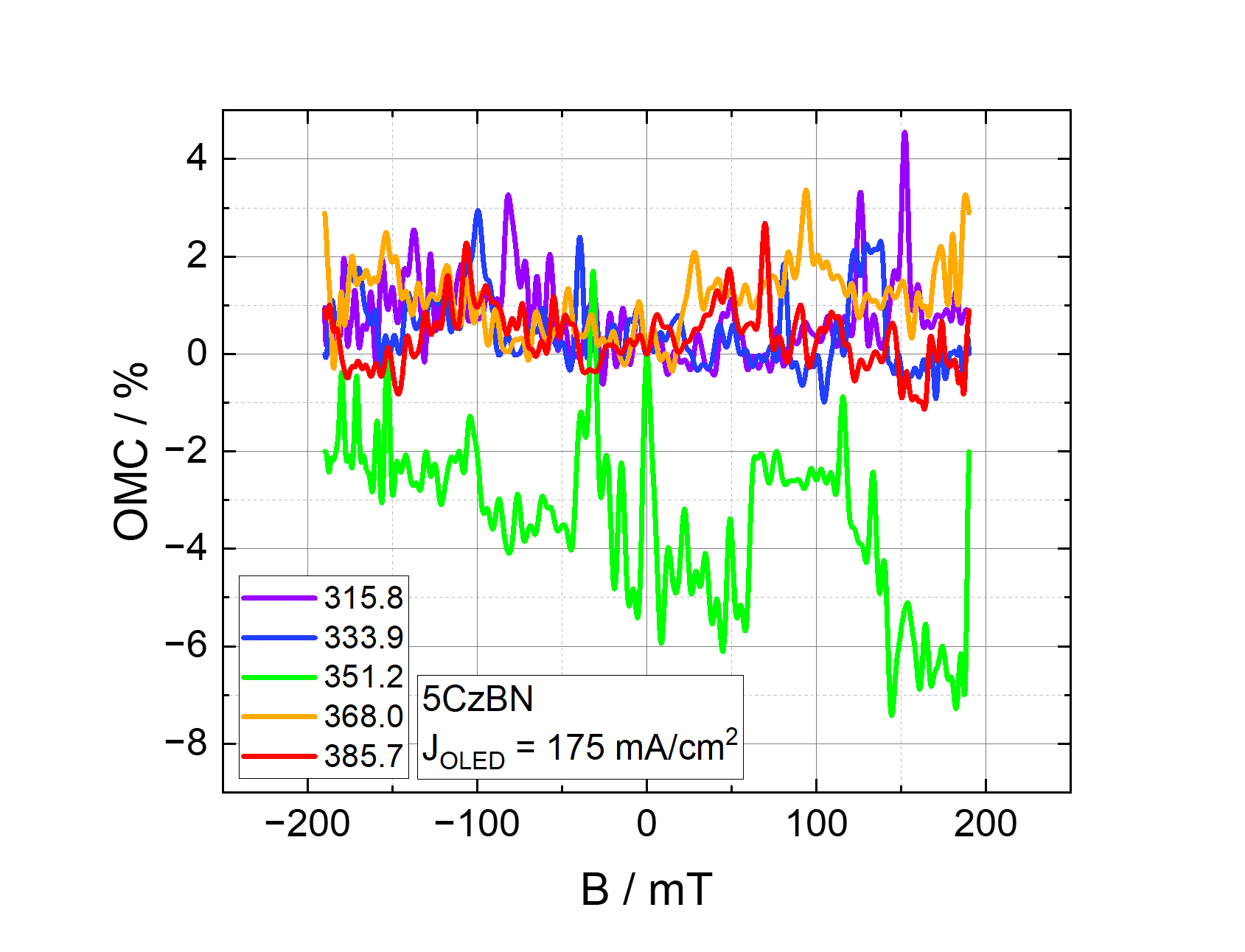}
    \caption{OMC response at temperatures exceeding room temperature observed for 5CzBN. The high-temperature regime causes significant structural changes, which reduce the OMC signal and lead to a high signal-to-noise ratio. Consequently, those measurements cannot be considered for activation energy evaluation.}
    \label{fig:5CzBN-high-temp}
\end{figure}

The determination of the activation energy was based on temperature-dependent OMC measurements, with the averaged data from several measurements presented in the manuscript. Additionally, the raw spectra for the OMC and MEL responses of 3CzClIPN, 4CzIPN, and 5CzBN are shown in Figure \ref{fig:Temp-OMC-MEL-response-full-data}. A clear temperature dependence can be observed for all three molecules. Notably, at temperatures exceeding room temperature (RT), the OMC response for 5CzBN became extremely noisy, which is likely due to structural changes that hinder charge transport (see Figure \ref{fig:5CzBN-high-temp}). 
For 3CzClIPN, the activation energy was determined at a lower temperature regime, consistent with its small $\Delta E_\mathrm{{ST}}$. In contrast, for 4CzIPN, no significant structural changes were observed even at temperatures above RT, allowing the activation energy to be determined at a higher temperature regime. This adjustment was necessary due to the increased singlet-triplet gap, which caused a shift of the linear onset in the ln(OMC) curve to higher temperatures, along with an increase in activation energy. For 5CzBN, by a theoretical approach (\textit{cf.} Table 2 in the manuscript) the activation energy was found to be much higher than for 3CzClIPN and 4CzIPN, making it impossible to determine it from the measurements conducted. It is anticipated that the onset would occur at significantly higher temperatures, which could not be achieved in our experiments due to the risk of molecular degradation.

\FloatBarrier

\subsection{Computational Details and Additional Information}
\subsubsection{Methodology}

Molecular geometries of 3CzClIPN, 4CzIPN, 5CzBN, and CBP were constructed using GaussView 6 \cite{gv6}. Subsequent ground state optimizations were carried out with Gaussian 16 \cite{g16}, using the B3LYP functional \cite{vosko1980accurate, lee1988development, becke1988density, becke1993density, stephens1994ab} augmented by Grimme's D3 dispersion correction \cite{grimme2010consistent} in conjunction with the 6-31+G* Pople basis set \cite{ditchfield1971a, hehre1972a, hariharan1973a, francl1982a, gordon1982a, clark1983a, spitznagel1987a}. This level of theory is based on our previous study of 3CzClIPN and 4CzIPN as thermally activated delayed fluorescence (TADF) emitter materials \cite{streiter2020impact}. All resulting optimized structures were confirmed as minimum energy geometries by vibrational frequency analysis. 
\newline
To explore potential alternative ground state minima of the cyanoarene systems, the global optimizer algorithm (GOAT) module \cite{de2025goat} implemented in Orca 6.0.1 \cite{ORCA, ORCA5} was used in combination with the semiempirical GFN2-XTB method \cite{bannwarth2019gfn2}. While no additional minima were identified for 3CzClIPN and 5CzBN, one further minimum was located for 4CzIPN. These geometries served as starting points for TD-DFT optimizations of the first singlet ($S_\mathrm{1}$) and the first two triplet ($T_\mathrm{1}$, $T_\mathrm{2}$) excited states using the same level of theory as detailed above. Due to the complex electronic structure of the higher excited triplet states of 5CzBN, convergence of the $T_\mathrm{2}$ state could not be achieved. Additionally, owing to the low Boltzmann population associated with the local minima of 4CzIPN in the ground state ($15.3$ \%), subsequent analyses have been focused on the respective global minima.
\newline
Linear interpolated pathways (LIPs) connecting the optimized excited-state geometries were generated in internal coordinates to avoid nonphysical distortions of the interpolated geometries. Intersections between the $S_\mathrm{1}$ and $T_\mathrm{2}$ states identified along these paths were used as initial guess structures for minimum energy crossing point (MECP) searches. MECPs were located using the KST48 program \cite{ma2022formal, KST48} interfaced with Orca 5.0.4, employing B3LYP-D3 with the larger def2-SVPD basis set \cite{weigend2005a, rappoport2010a} in TD-DFT.  Consistent with the previously noted difficulty in optimizing the $T_\mathrm{2}$ state of 5CzBN, no $T_\mathrm{2}-S_\mathrm{1}$ MECP was obtained for this molecule. Likewise, searches for $T_\mathrm{1}-S_\mathrm{1}$ MECPs across all cyanoarene systems were unsuccessful, in line with earlier reports on the non-intersecting nature of these states in 4CzIPN \cite{aizawa2020kinetic}.
\newline
To identify potential minimum energy conical intersections (MECIs) between the $T_\mathrm{1}$ and $T_\mathrm{2}$ states, a customized implementation of the BPUPD algorithm developed by Maeda \textit{et al.} \cite{maeda2010updated} was employed. Specifically, the \textit{pbupd} and \textit{bpupdforce} Python functions from the XMECP package \cite{xu2024xmecp} were adapted for compatibility with the KST48 code. The bpupdforce function was then integrated into the GDIIS procedure to compute the necessary gradients. This methodology yielded a MECI with the necessary accuracy exclusively for 4CzIPN. For further discussion on the necessity of MECIs for efficient RISC, please refer to the main text.
\newline
All electronic energies and spin-orbit coupling (SOC) constants were then calculated at the respective optimized geometries using Orca 6.0.1 and the B3LYP-D3 / def2-SVPD level of theory with TD-DFT. Based on previous work \cite{kurle2024red}, hyperfine coupling (HFC) constants were determined within the Tamm-Dancoff approximation (TDA) using the specialized pcH-2 basis set \cite{jakobsen2019a} obtained from Basis Set Exchange \cite{pritchard2019new}. Charge transfer (CT) character was analyzed with TheoDORE 3.2 \cite{plasser2020theodore, TheoDORE} by partitioning each molecule into a joint donor fragment and an acceptor fragment.
\newline
Molecular geometries and molecular orbitals were rendered using ChemCraft \cite{chemcraft}. Simulated UV-vis spectra were broadened with Gaussian functions using full width at half maximum (FWHM) values of $0.3$ eV for absorption and $0.2$ eV for emission to best match experimental results.
\newline
\bigskip
The following section contains additional data referenced in the main text.

\newpage
\subsubsection{Further Analysis}

\begin{figure}
\centering
  \includegraphics[width=0.5\linewidth]{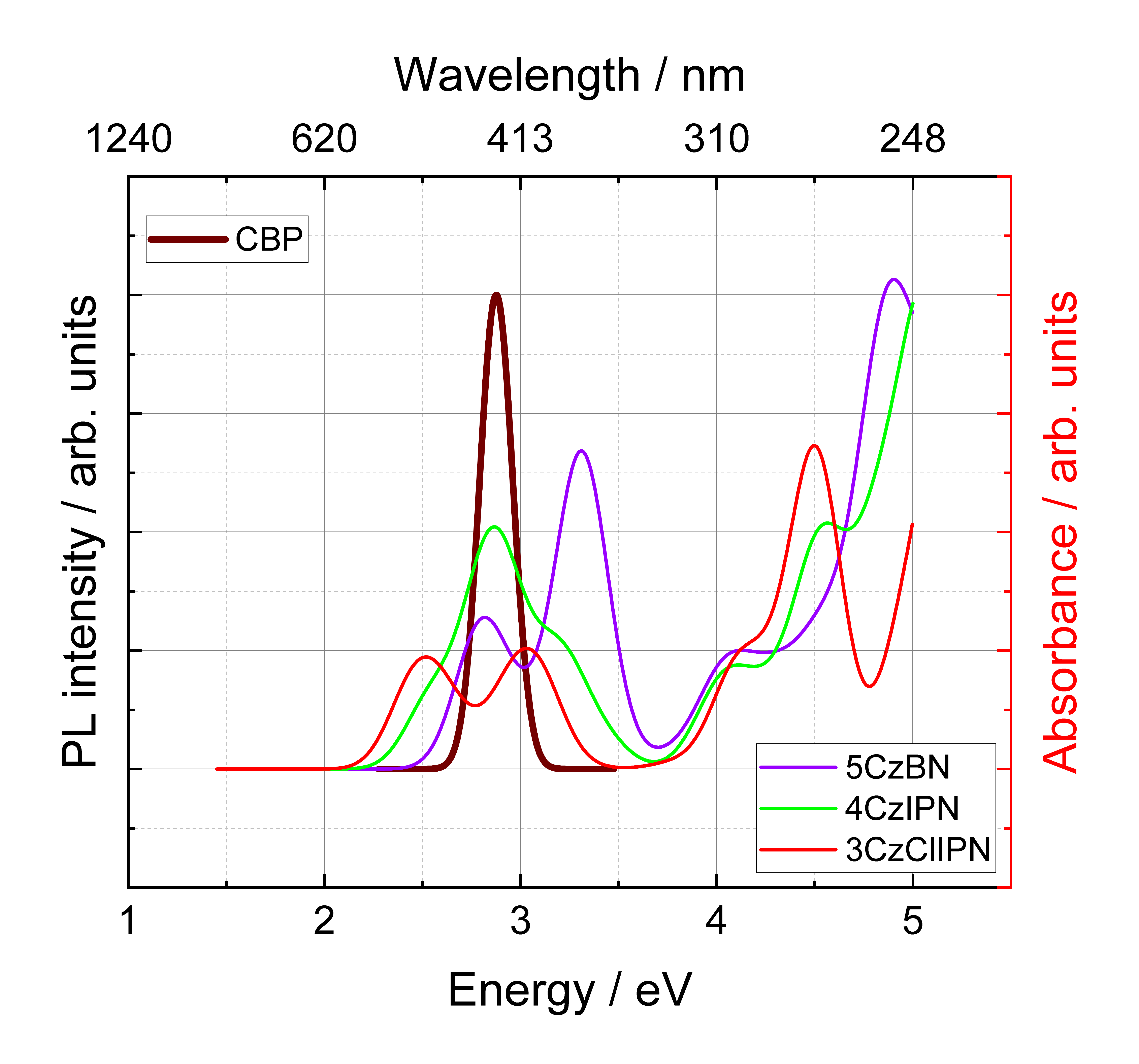}
  \caption{Calculated absorption spectra of the emitter molecules and PL spectrum of the CBP host. The visible shoulder observed in the experimental PL spectrum of CBP is likely due to vibrational progression and was not accounted for in the computations.}
  \label{fig:computational-PL-sepctra}
\end{figure}

\begin{figure}
\centering
  \includegraphics[width=0.8\textwidth]{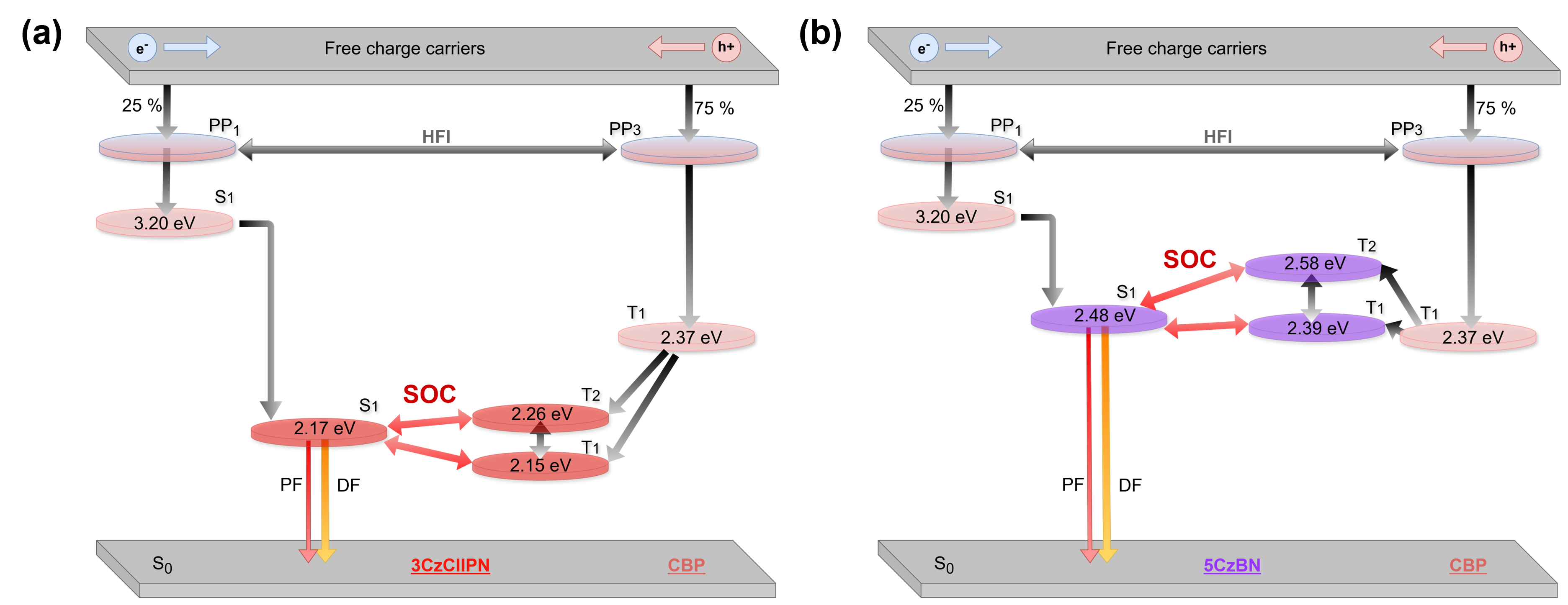}
  \caption{Jablonski diagrams for 3CzClIPN (a) and 5CzBN (b) based on TD-DFT calculations. As no equilibrium geometry for the $T_\mathrm{2}$ state of 5CzBN could be converged, the depicted $T_\mathrm{2}$ energy was calculated as a vertical excitation on the $T_\mathrm{1}$ equilibrium geometry as the closest approximation. This is indicated by a dagger.}
  \label{fig:Jablonski-3CzClIPN-5CzBN}
\end{figure}
\FloatBarrier

\begin{figure}
\centering
  \includegraphics[width=0.5\linewidth]{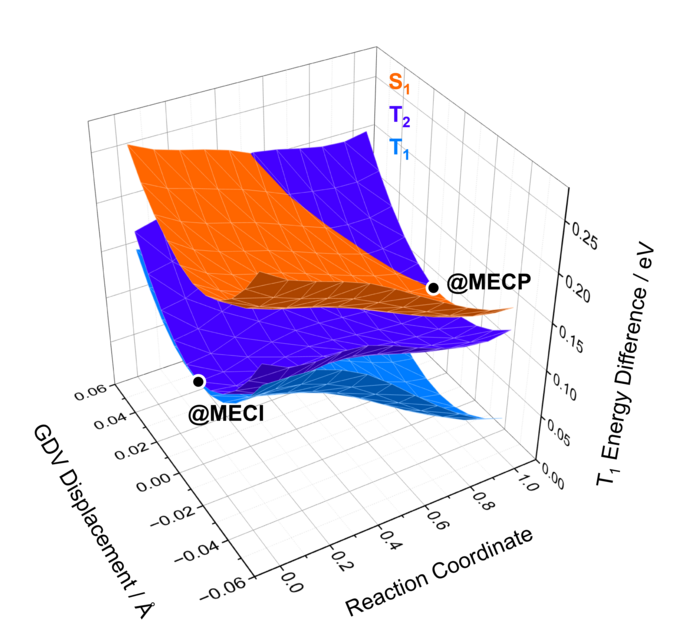}
\caption{Two-dimensional LIP between the $T_\mathrm{1}-T_\mathrm{2}$ MECI and the $T_\mathrm{2}-S_\mathrm{1}$ MECP. A second dimension corresponding to the displacement of the $GDV$ from the MECI and MECP is introduced to better visualize these points. The hyperline where the $T_\mathrm{2}$ and $S_\mathrm{1}$ states cross is clearly visible. All energies are in relation to the energy of the $T_\mathrm{1}$ state in its minimum.}
  \label{fig:2D-LIP-full-PES}
\end{figure}

\vspace{3\baselineskip}

\begin{table}
    \caption{CT character for the first excited singlet and triplet states of 3CzClIPN, 4CzIPN, and 5CzBN.}
    \centering
    \begin{tabular}{c|c|c}
    \hline
        Molecule & electronic state & CT character \\
        \hline
         3CzClIPN &  $S_\mathrm{1}$ & 0.86 \\
           & $T_\mathrm{1}$ & 0.85 \\
         \hline
         4CzIPN &  $S_\mathrm{1}$ & 0.84 \\
          &  $T_\mathrm{1}$ & 0.80 \\
         \hline
         5CzBN &  $S_\mathrm{1}$ & 0.81 \\
          &  $T_\mathrm{1}$ & 0.64 \\
         \hline
    \end{tabular}
    \label{tab:CT-character}
\end{table}
\FloatBarrier

\newpage

\begin{figure}
\centering
  \includegraphics[width=1\textwidth]{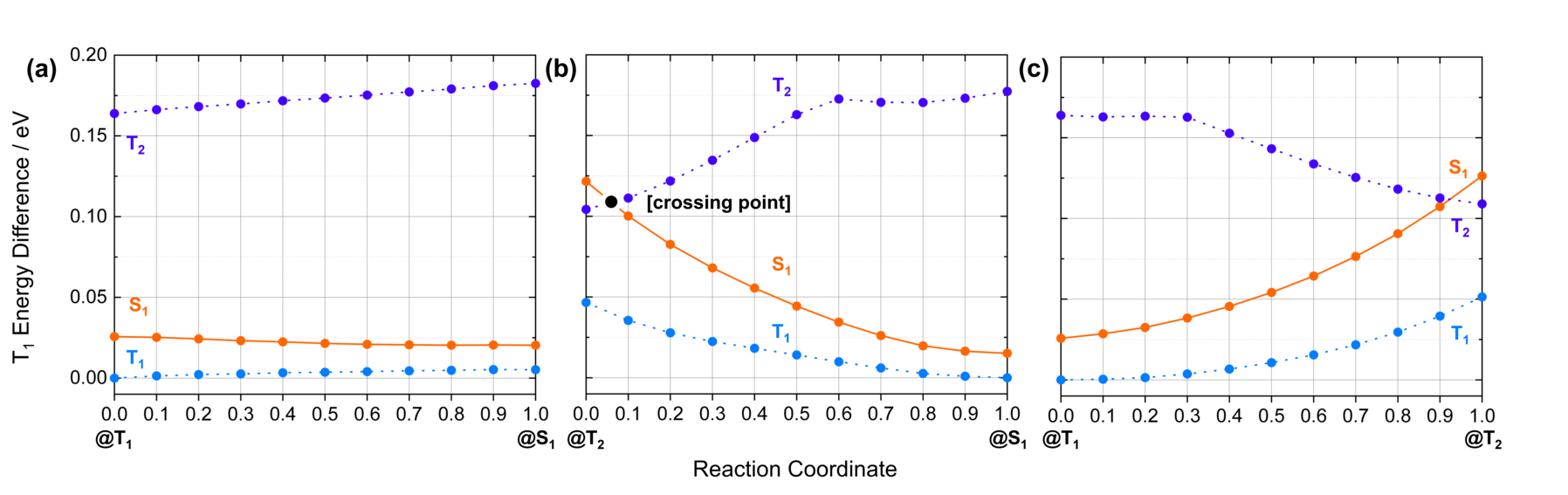}
  \caption{Excited state energies at LIPs between selected electronic state minima of 3CzClIPN. (a) shows the $T_\mathrm{1} \rightarrow S_\mathrm{1}$ LIP, (b) the $T_\mathrm{2} \rightarrow S_\mathrm{1}$ LIP, and (c) the $T_\mathrm{1} \rightarrow T_\mathrm{2}$ LIP. Dotted lines correspond to triplet PES, while continuous lines represent singlet state PES. All energies are in relation to the energy of the $T_\mathrm{1}$ state in its minimum.}
  \label{fig:1D-LIPs-3Cz}
\end{figure}

\begin{figure}
\centering
  \includegraphics[width=0.5\textwidth]{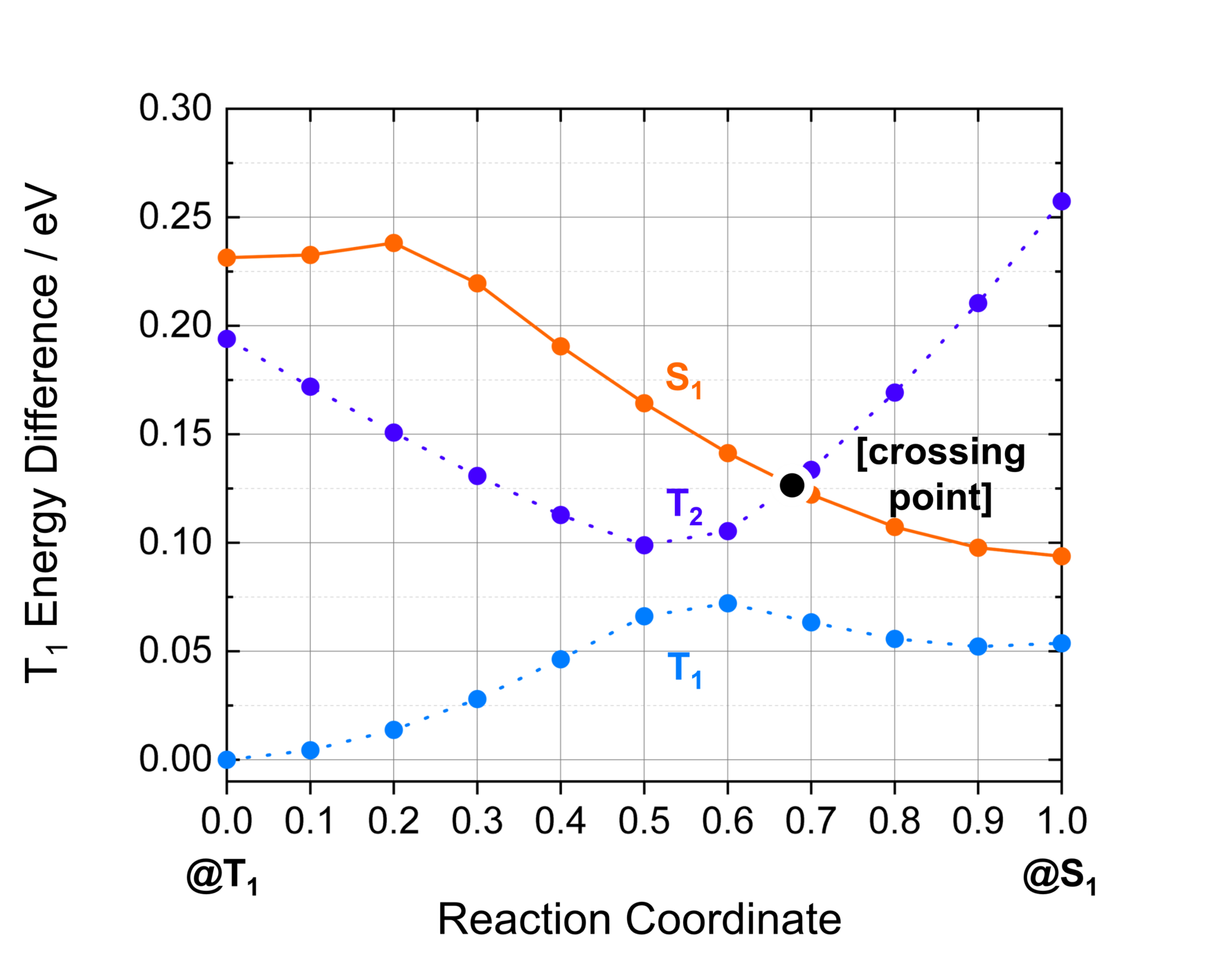}
  \caption{Excited state energies at LIP between the $T_\mathrm{1}$ and $S_\mathrm{1}$ electronic state minima of 5CzBN. Dotted lines correspond to triplet PES, while continuous lines represent singlet state PES. All energies are in relation to the energy of the $T_\mathrm{1}$ state in its minimum.}
  \label{fig:1D-LIP-5Cz}
\end{figure}

\clearpage

\newpage
\bibliographystyle{MSP}
\bibliography{bibliography}

\begin{thebibliography}{100}
\providecommand{\url}[1]{\texttt{#1}}
\providecommand{\urlprefix}{URL }

\bibitem{hong2021brief}
G.~Hong, X.~Gan, C.~Leonhardt, Z.~Zhang, J.~Seibert, J.~M. Busch, S.~Br{\"a}se,
\newblock \emph{Advanced Materials} \textbf{2021}, \emph{33}, 9 2005630.

\bibitem{dos2024golden}
J.~M. Dos~Santos, D.~Hall, B.~Basumatary, M.~Bryden, D.~Chen, P.~Choudhary, T.~Comerford, E.~Crovini, A.~Danos, J.~De, et~al.,
\newblock \emph{Chemical Reviews} \textbf{2024}, \emph{124}, 24 13736.

\bibitem{yin2016efficient}
D.~Yin, J.~Feng, R.~Ma, Y.-F. Liu, Y.-L. Zhang, X.-L. Zhang, Y.-G. Bi, Q.-D. Chen, H.-B. Sun,
\newblock \emph{Nature communications} \textbf{2016}, \emph{7}, 1 11573.

\bibitem{kim2013soft}
W.~Kim, S.~Kwon, S.-M. Lee, J.~Y. Kim, Y.~Han, E.~Kim, K.~C. Choi, S.~Park, B.-C. Park,
\newblock \emph{Organic Electronics} \textbf{2013}, \emph{14}, 11 3007.

\bibitem{kwon2018weavable}
S.~Kwon, H.~Kim, S.~Choi, E.~G. Jeong, D.~Kim, S.~Lee, H.~S. Lee, Y.~C. Seo, K.~C. Choi,
\newblock \emph{Nano letters} \textbf{2018}, \emph{18}, 1 347.

\bibitem{song2020organic}
J.~Song, H.~Lee, E.~G. Jeong, K.~C. Choi, S.~Yoo,
\newblock \emph{Advanced Materials} \textbf{2020}, \emph{32}, 35 1907539.

\bibitem{hauck2024perspective}
M.~Hauck, C.~Bickmann, A.~Morgenstern, N.~Nagel, C.~R. Meinecke, A.~Schade, R.~Tafat, L.~Viriato, H.~Kuhn, G.~Salvan, et~al.,
\newblock \emph{Energies (19961073)} \textbf{2024}, \emph{17}, 20.

\bibitem{cole1941dispersion}
K.~S. Cole, R.~H. Cole,
\newblock \emph{The Journal of chemical physics} \textbf{1941}, \emph{9}, 4 341.

\bibitem{wang2024boosting}
H.~Wang, Y.~Yuan, Z.~Wang, Y.~Wang,
\newblock \emph{ACS Applied Engineering Materials} \textbf{2024}, \emph{2}, 4 781.

\bibitem{adachi2001nearly}
C.~Adachi, M.~A. Baldo, M.~E. Thompson, S.~R. Forrest,
\newblock \emph{Journal of Applied Physics} \textbf{2001}, \emph{90}, 10 5048.

\bibitem{adachi2014third}
C.~Adachi,
\newblock \emph{Japanese Journal of Applied Physics} \textbf{2014}, \emph{53}, 6 060101.

\bibitem{nakanotani2021thermally}
H.~Nakanotani, Y.~Tsuchiya, C.~Adachi,
\newblock \emph{Chemistry Letters} \textbf{2021}, \emph{50}, 5 938.

\bibitem{uoyama2012highly}
H.~Uoyama, K.~Goushi, K.~Shizu, H.~Nomura, C.~Adachi,
\newblock \emph{Nature} \textbf{2012}, \emph{492}, 7428 234.

\bibitem{shizu2015enhanced}
K.~Shizu, M.~Uejima, H.~Nomura, T.~Sato, K.~Tanaka, H.~Kaji, C.~Adachi,
\newblock \emph{Physical Review Applied} \textbf{2015}, \emph{3}, 1 014001.

\bibitem{kotadiya2019efficient}
N.~B. Kotadiya, P.~W. Blom, G.-J.~A. Wetzelaer,
\newblock \emph{Nature Photonics} \textbf{2019}, \emph{13}, 11 765.

\bibitem{wang2024understanding}
Z.~Wang, X.~Jiang, J.~Xiong, B.~Xiao, Y.~Wang, X.~Zhou, R.~Pan, X.~Tang,
\newblock \emph{The Journal of Physical Chemistry Letters} \textbf{2024}, \emph{15}, 38 9630.

\bibitem{tanaka2020understanding}
M.~Tanaka, R.~Nagata, H.~Nakanotani, C.~Adachi,
\newblock \emph{Communications Materials} \textbf{2020}, \emph{1}, 1 18.

\bibitem{basel2016magnetic}
T.~Basel, D.~Sun, S.~Baniya, R.~McLaughlin, H.~Choi, O.~Kwon, Z.~V. Vardeny,
\newblock \emph{Advanced Electronic Materials} \textbf{2016}, \emph{2}, 2 1500248.

\bibitem{wang2016immense}
Y.~Wang, K.~Sahin-Tiras, N.~J. Harmon, M.~Wohlgenannt, M.~E. Flatt{\'e},
\newblock \emph{Physical Review X} \textbf{2016}, \emph{6}, 1 011011.

\bibitem{gibson2016importance}
J.~Gibson, A.~P. Monkman, T.~J. Penfold,
\newblock \emph{ChemPhysChem} \textbf{2016}, \emph{17}, 19 2956.

\bibitem{aizawa2020kinetic}
N.~Aizawa, Y.~Harabuchi, S.~Maeda, Y.-J. Pu,
\newblock \emph{Nature Communications} \textbf{2020}, \emph{11}, 1 3909.

\bibitem{pan2022disorder}
X.~Pan, O.~Kwon, D.~R. Khanal, B.~Choi, Z.~V. Vardeny,
\newblock \emph{Advanced Optical Materials} \textbf{2022}, \emph{10}, 4 2101334.

\bibitem{mondal2023degradation}
A.~K. Mondal, X.~Pan, O.~Kwon, Z.~V. Vardeny,
\newblock \emph{ACS Applied Materials \& Interfaces} \textbf{2023}, \emph{15}, 7 9697.

\bibitem{schellekens2010exploring}
A.~Schellekens, W.~Wagemans,
\newblock \emph{Master's thesis, Eindhoven University of Technology} \textbf{2010}.

\bibitem{morgenstern2024analysis}
A.~Morgenstern, D.~Weber, L.~Hertling, K.~Gabel, U.~T. Schwarz, D.~Schondelmaier, D.~R.~T. Zahn, G.~Salvan,
\newblock \emph{Scientific Reports} \textbf{2024}, \emph{14}, 1 30520.

\bibitem{weber2024exciplex}
D.~Weber, A.~Morgenstern, D.~Beer, D.~R.~T. Zahn, C.~Deibel, G.~Salvan, D.~Schondelmaier,
\newblock \emph{Applied Physics A} \textbf{2024}, \emph{130}, 6 1.

\bibitem{geng2018effect}
R.~Geng, R.~C. Subedi, H.~M. Luong, M.~T. Pham, W.~Huang, X.~Li, K.~Hong, M.~Shao, K.~Xiao, L.~A. Hornak, et~al.,
\newblock \emph{Physical review letters} \textbf{2018}, \emph{120}, 8 086602.

\bibitem{crooker2014spectrally}
S.~Crooker, F.~Liu, M.~Kelley, N.~Martinez, W.~Nie, A.~Mohite, I.~Nayyar, S.~Tretiak, D.~Smith, P.~Ruden,
\newblock \emph{Applied physics letters} \textbf{2014}, \emph{105}, 15.

\bibitem{ehrenfreund2012effects}
E.~Ehrenfreund, Z.~V. Vardeny,
\newblock \emph{Israel Journal of Chemistry} \textbf{2012}, \emph{52}, 6 552.

\bibitem{nguyen2010isotope}
T.~D. Nguyen, G.~Hukic-Markosian, F.~Wang, L.~Wojcik, X.-G. Li, E.~Ehrenfreund, Z.~V. Vardeny,
\newblock \emph{Nature materials} \textbf{2010}, \emph{9}, 4 345.

\bibitem{hu2019spin}
Y.~Hu, X.~Tang, R.~Pan, J.~Deng, H.~Zhu, Z.~Xiong,
\newblock \emph{Physical Chemistry Chemical Physics} \textbf{2019}, \emph{21}, 32 17673.

\bibitem{zhao2023abundant}
X.~Zhao, J.~Chen, L.~Cheng, S.~Yang, B.~Wang, T.~Peng, J.~Liu, H.~Lu, S.~Zhang, Z.~Xiong,
\newblock \emph{ACS Photonics} \textbf{2023}, \emph{11}, 1 230.

\bibitem{ogiwara2015mechanism}
T.~Ogiwara, Y.~Wakikawa, T.~Ikoma,
\newblock \emph{The Journal of Physical Chemistry A} \textbf{2015}, \emph{119}, 14 3415.

\bibitem{mukherjee2021modeling}
S.~Mukherjee, D.~A. Fedorov, S.~A. Varganov,
\newblock \emph{Annual Review of Physical Chemistry} \textbf{2021}, \emph{72}, 1 515.

\bibitem{bearpark1994direct}
M.~J. Bearpark, M.~A. Robb, H.~B. Schlegel,
\newblock \emph{Chemical physics letters} \textbf{1994}, \emph{223}, 3 269.

\bibitem{maeda2010updated}
S.~Maeda, K.~Ohno, K.~Morokuma,
\newblock \emph{Journal of Chemical Theory and Computation} \textbf{2010}, \emph{6}, 5 1538.

\bibitem{matsika2011nonadiabatic}
S.~Matsika, P.~Krause,
\newblock \emph{Annual review of physical chemistry} \textbf{2011}, \emph{62}, 1 621.

\bibitem{bassan2023visible}
E.~Bassan, R.~Inoue, D.~Fabry, F.~Calogero, S.~Potenti, A.~Gualandi, P.~G. Cozzi, K.~Kamogawa, P.~Ceroni, Y.~Tamaki, et~al.,
\newblock \emph{Sustainable Energy \& Fuels} \textbf{2023}, \emph{7}, 14 3454.

\bibitem{uda1998work}
M.~Uda, A.~Nakamura, T.~Yamamoto, Y.~Fujimoto,
\newblock \emph{Journal of electron spectroscopy and related phenomena} \textbf{1998}, \emph{88} 643.

\bibitem{flores2009modelling}
F.~Flores, J.~Ortega, H.~V{\'a}zquez,
\newblock \emph{Physical Chemistry Chemical Physics} \textbf{2009}, \emph{11}, 39 8658.

\bibitem{sugie2023dependence}
A.~Sugie, K.~Nakano, K.~Tajima, I.~Osaka, H.~Yoshida,
\newblock \emph{The Journal of Physical Chemistry Letters} \textbf{2023}, \emph{14}, 50 11412.

\bibitem{stavrou2020photophysics}
K.~Stavrou, L.~G. Franca, A.~P. Monkman,
\newblock \emph{ACS applied electronic materials} \textbf{2020}, \emph{2}, 9 2868.

\bibitem{karunathilaka2020suppression}
B.~S. Karunathilaka, U.~Balijapalli, C.~A. Senevirathne, S.~Yoshida, Y.~Esaki, K.~Goushi, T.~Matsushima, A.~S. Sandanayaka, C.~Adachi,
\newblock \emph{Nature communications} \textbf{2020}, \emph{11}, 1 4926.

\bibitem{viezbicke2015evaluation}
B.~D. Viezbicke, S.~Patel, B.~E. Davis, D.~P. Birnie~III,
\newblock \emph{physica status solidi (b)} \textbf{2015}, \emph{252}, 8 1700.

\bibitem{zhang2019suppressing}
Y.~Zhang, Z.~Li, C.~Li, Y.~Wang,
\newblock \emph{Frontiers in Chemistry} \textbf{2019}, \emph{7} 302.

\bibitem{nakanotani2013promising}
H.~Nakanotani, K.~Masui, J.~Nishide, T.~Shibata, C.~Adachi,
\newblock \emph{Scientific reports} \textbf{2013}, \emph{3}, 1 2127.

\bibitem{kim2014study}
J.~W. Kim, S.~I. You, N.~H. Kim, J.-A. Yoon, K.~W. Cheah, F.~R. Zhu, W.~Y. Kim,
\newblock \emph{Scientific Reports} \textbf{2014}, \emph{4}, 1 7009.

\bibitem{zhang2009low}
Y.~Zhang, R.~Liu, Y.~Lei, Z.~Xiong,
\newblock \emph{Applied Physics Letters} \textbf{2009}, \emph{94}, 8.

\bibitem{pan2019extraordinary}
R.~Pan, X.~Tang, Y.~Hu, H.~Zhu, J.~Deng, Z.~Xiong,
\newblock \emph{Journal of Materials Chemistry C} \textbf{2019}, \emph{7}, 8 2421.

\bibitem{wu2022identifying}
F.~Wu, X.~Zhao, H.~Zhu, X.~Tang, Y.~Ning, J.~Chen, X.~Chen, Z.~Xiong,
\newblock \emph{ACS Photonics} \textbf{2022}, \emph{9}, 8 2713.

\bibitem{niedermeier2008enhancement}
U.~Niedermeier, M.~Vieth, R.~P{\"a}tzold, W.~Sarfert, H.~Von~Seggern,
\newblock \emph{Applied Physics Letters} \textbf{2008}, \emph{92}, 19.

\bibitem{kim2017concentration}
H.~S. Kim, S.-R. Park, M.~C. Suh,
\newblock \emph{The Journal of Physical Chemistry C} \textbf{2017}, \emph{121}, 26 13986.

\bibitem{nguyen2012isotope}
T.~D. Nguyen, T.~Basel, Y.-J. Pu, X.~Li, E.~Ehrenfreund, Z.~Vardeny,
\newblock \emph{Physical Review B—Condensed Matter and Materials Physics} \textbf{2012}, \emph{85}, 24 245437.

\bibitem{kalinowski2003magnetic}
J.~Kalinowski, M.~Cocchi, D.~Virgili, P.~Di~Marco, V.~Fattori,
\newblock \emph{Chemical Physics Letters} \textbf{2003}, \emph{380}, 5-6 710.

\bibitem{lei2016ultralarge}
Y.~Lei, Q.~Zhang, L.~Chen, Y.~Ling, P.~Chen, Q.~Song, Z.~Xiong,
\newblock \emph{Adv. Opt. Mater} \textbf{2016}, \emph{4}, 5 694.

\bibitem{ling2015large}
Y.~Ling, Y.~Lei, Q.~Zhang, L.~Chen, Q.~Song, Z.~Xiong,
\newblock \emph{Applied Physics Letters} \textbf{2015}, \emph{107}, 21.

\bibitem{zhang2015magnetic}
C.~Zhang, D.~Sun, C.~Sheng, Y.~Zhai, K.~Mielczarek, A.~Zakhidov, Z.~Vardeny,
\newblock \emph{Nature Physics} \textbf{2015}, \emph{11}, 5 427.

\bibitem{liu2020isotope}
X.~Liu, H.~Popli, O.~Kwon, H.~Malissa, X.~Pan, B.~Park, B.~Choi, S.~Kim, E.~Ehrenfreund, C.~Boehme, et~al.,
\newblock \emph{Advanced Materials} \textbf{2020}, \emph{32}, 48 2004421.

\bibitem{hu2009magnetic}
B.~Hu, L.~Yan, M.~Shao,
\newblock \emph{Advanced Materials} \textbf{2009}, \emph{21}, 14-15 1500.

\bibitem{noda2019T2intermediate}
H.~Noda, X.-K. Chen, H.~Nakanotani, T.~Hosokai, M.~Miyajima, N.~Notsuka, Y.~Kashima, J.-L. Brédas, C.~Adachi,
\newblock \emph{Nature Materials} \textbf{2019}, \emph{18}, 10 1084.

\bibitem{monka2022understanding}
M.~Mo{\'n}ka, I.~E. Serdiuk, K.~Kozakiewicz, E.~Hoffman, J.~Szumilas, A.~Kubicki, S.~Y. Park, P.~Bojarski,
\newblock \emph{Journal of Materials Chemistry C} \textbf{2022}, \emph{10}, 20 7925.

\bibitem{kaminski2024balancing}
J.~M. Kaminski, T.~B{\"o}hmer, C.~M. Marian,
\newblock \emph{The Journal of Physical Chemistry C} \textbf{2024}, \emph{128}, 33 13711.

\bibitem{plasser2020theodore}
F.~Plasser,
\newblock \emph{The Journal of chemical physics} \textbf{2020}, \emph{152}, 8.

\bibitem{serdiuk2021vibrationally}
I.~E. Serdiuk, M.~Mo\'{n}ka, K.~Kozakiewicz, B.~Liberek, P.~Bojarski, S.~Y. Park,
\newblock \emph{The Journal of Physical Chemistry B} \textbf{2021}, \emph{125}, 10 2696.

\bibitem{bloom2008temperature}
F.~Bloom, W.~Wagemans, B.~Koopmans,
\newblock \emph{Journal of Applied Physics} \textbf{2008}, \emph{103}, 7.

\bibitem{speckmeier2018toolbox}
E.~Speckmeier, T.~G. Fischer, K.~Zeitler,
\newblock \emph{Journal of the American Chemical Society} \textbf{2018}, \emph{140}, 45 15353.

\bibitem{weber2023cost}
D.~Weber, R.~Heimburger, G.~Schondelmaier, T.~Junghans, A.~Zetzl, D.~R.~T. Zahn, D.~Schondelmaier,
\newblock \emph{SN Applied Sciences} \textbf{2023}, \emph{5}, 1 21.

\bibitem{yoshida2015principle}
H.~Yoshida,
\newblock \emph{Journal of Electron Spectroscopy and Related Phenomena} \textbf{2015}, \emph{204} 116.

\bibitem{erdman1982low}
P.~W. Erdman, E.~C. Zipf,
\newblock \emph{Review of Scientific Instruments} \textbf{1982}, \emph{53}, 2 225.

\bibitem{mclaughlin2022study}
R.~McLaughlin, X.~Pan, D.~Sun, O.~Kwon, Z.~V. Vardeny,
\newblock \emph{Advanced Optical Materials} \textbf{2022}, \emph{10}, 16 2200499.

\bibitem{zahn2006transport}
D.~R.~T. Zahn, G.~N. Gavrila, M.~Gorgoi,
\newblock \emph{Chemical Physics} \textbf{2006}, \emph{325}, 1 99.

\bibitem{jablonski2019evaluation}
A.~Jablonski,
\newblock \emph{Surface Science} \textbf{2019}, \emph{688} 14.

\bibitem{lo2002green}
S.-C. Lo, N.~A. Male, J.~P. Markham, S.~W. Magennis, P.~L. Burn, O.~V. Salata, I.~D. Samuel,
\newblock \emph{Advanced Materials} \textbf{2002}, \emph{14}, 13-14 975.

\bibitem{mao2024interacting}
Y.-H. Mao, M.-K. Hung, S.-T. Chung, S.~Sharma, K.-W. Tsai, S.-A. Chen,
\newblock \emph{ACS Applied Materials \& Interfaces} \textbf{2024}, \emph{16}, 44 60715.

\bibitem{dos2016using}
P.~L. Dos~Santos, J.~S. Ward, M.~R. Bryce, A.~P. Monkman,
\newblock \emph{The journal of physical chemistry letters} \textbf{2016}, \emph{7}, 17 3341.

\bibitem{gv6}
R.~Dennington, T.~A. Keith, J.~M. Millam,
\newblock Gaussview {V}ersion {6}, \textbf{2019},
\newblock Semichem Inc. Shawnee Mission KS.

\bibitem{g16}
M.~J. Frisch, G.~W. Trucks, H.~B. Schlegel, G.~E. Scuseria, M.~A. Robb, J.~R. Cheeseman, G.~Scalmani, V.~Barone, G.~A. Petersson, H.~Nakatsuji, X.~Li, M.~Caricato, A.~V. Marenich, J.~Bloino, B.~G. Janesko, R.~Gomperts, B.~Mennucci, H.~P. Hratchian, J.~V. Ortiz, A.~F. Izmaylov, J.~L. Sonnenberg, D.~Williams-Young, F.~Ding, F.~Lipparini, F.~Egidi, J.~Goings, B.~Peng, A.~Petrone, T.~Henderson, D.~Ranasinghe, V.~G. Zakrzewski, J.~Gao, N.~Rega, G.~Zheng, W.~Liang, M.~Hada, M.~Ehara, K.~Toyota, R.~Fukuda, J.~Hasegawa, M.~Ishida, T.~Nakajima, Y.~Honda, O.~Kitao, H.~Nakai, T.~Vreven, K.~Throssell, J.~A. Montgomery, {Jr.}, J.~E. Peralta, F.~Ogliaro, M.~J. Bearpark, J.~J. Heyd, E.~N. Brothers, K.~N. Kudin, V.~N. Staroverov, T.~A. Keith, R.~Kobayashi, J.~Normand, K.~Raghavachari, A.~P. Rendell, J.~C. Burant, S.~S. Iyengar, J.~Tomasi, M.~Cossi, J.~M. Millam, M.~Klene, C.~Adamo, R.~Cammi, J.~W. Ochterski, R.~L. Martin, K.~Morokuma, O.~Farkas, J.~B. Foresman, D.~J. Fox,
\newblock Gaussian˜16 {R}evision {C}.02, \textbf{2016},
\newblock Gaussian Inc. Wallingford CT.

\bibitem{vosko1980accurate}
S.~H. Vosko, L.~Wilk, M.~Nusair,
\newblock \emph{Canadian Journal of physics} \textbf{1980}, \emph{58}, 8 1200.

\bibitem{lee1988development}
C.~Lee, W.~Yang, R.~G. Parr,
\newblock \emph{Physical review B} \textbf{1988}, \emph{37}, 2 785.

\bibitem{becke1988density}
A.~D. Becke,
\newblock \emph{Physical review A} \textbf{1988}, \emph{38}, 6 3098.

\bibitem{becke1993density}
A.~D. Becke,
\newblock \emph{The Journal of chemical physics} \textbf{1993}, \emph{98}, 7 5648.

\bibitem{stephens1994ab}
P.~J. Stephens, F.~J. Devlin, C.~F. Chabalowski, M.~J. Frisch,
\newblock \emph{The Journal of physical chemistry} \textbf{1994}, \emph{98}, 45 11623.

\bibitem{grimme2010consistent}
S.~Grimme, J.~Antony, S.~Ehrlich, H.~Krieg,
\newblock \emph{The Journal of chemical physics} \textbf{2010}, \emph{132}, 15.

\bibitem{ditchfield1971a}
R.~Ditchfield, W.~J. Hehre, J.~A. Pople,
\newblock \emph{J. Chem. Phys.} \textbf{1971}, \emph{54} 724.

\bibitem{hehre1972a}
W.~J. Hehre, R.~Ditchfield, J.~A. Pople,
\newblock \emph{J. Chem. Phys.} \textbf{1972}, \emph{56} 2257.

\bibitem{hariharan1973a}
P.~C. Hariharan, J.~A. Pople,
\newblock \emph{Theor. Chim. Acta} \textbf{1973}, \emph{28} 213.

\bibitem{francl1982a}
M.~M. Francl, W.~J. Pietro, W.~J. Hehre, J.~S. Binkley, M.~S. Gordon, D.~J. DeFrees, J.~A. Pople,
\newblock \emph{J. Chem. Phys.} \textbf{1982}, \emph{77} 3654.

\bibitem{gordon1982a}
M.~S. Gordon, J.~S. Binkley, J.~A. Pople, W.~J. Pietro, W.~J. Hehre,
\newblock \emph{J. Am. Chem. Soc.} \textbf{1982}, \emph{104} 2797.

\bibitem{clark1983a}
T.~Clark, J.~Chandrasekhar, G.~W. Spitznagel, P.~V.~R. Schleyer,
\newblock \emph{J. Comput. Chem.} \textbf{1983}, \emph{4} 294.

\bibitem{spitznagel1987a}
G.~W. Spitznagel, T.~Clark, P.~v.~R. Schleyer, W.~J. Hehre,
\newblock \emph{J. Comput. Chem.} \textbf{1987}, \emph{8} 1109.

\bibitem{streiter2020impact}
M.~Streiter, T.~G. Fischer, C.~Wiebeler, S.~Reichert, J.~Langenickel, K.~Zeitler, C.~Deibel,
\newblock \emph{The Journal of Physical Chemistry C} \textbf{2020}, \emph{124}, 28 15007.

\bibitem{de2025goat}
B.~de~Souza,
\newblock \emph{Angewandte Chemie International Edition} \textbf{2025}, e202500393.

\bibitem{ORCA}
F.~Neese,
\newblock \emph{WIRES Comput. Molec. Sci.} \textbf{2012}, \emph{2}, 1 73.

\bibitem{ORCA5}
F.~Neese,
\newblock \emph{WIRES Comput. Molec. Sci.} \textbf{2022}, \emph{12}, 1 e1606.

\bibitem{bannwarth2019gfn2}
C.~Bannwarth, S.~Ehlert, S.~Grimme,
\newblock \emph{Journal of chemical theory and computation} \textbf{2019}, \emph{15}, 3 1652.

\bibitem{ma2022formal}
Y.~Ma, A.~A. Hussein,
\newblock \emph{ChemistrySelect} \textbf{2022}, \emph{7}, 37 e202202354.

\bibitem{KST48}
{Y. Ma, KST48: A Powerful Tool for MECP locating},
\newblock available from \url{https://github.com/RimoAccelerator/KST48},
\newblock Accessed: 2024-12-02.

\bibitem{weigend2005a}
F.~Weigend, R.~Ahlrichs,
\newblock \emph{Phys. Chem. Chem. Phys.} \textbf{2005}, \emph{7} 3297.

\bibitem{rappoport2010a}
D.~Rappoport, F.~Furche,
\newblock \emph{J. Chem. Phys.} \textbf{2010}, \emph{133} 134105.

\bibitem{xu2024xmecp}
J.~Xu, J.~Hao, C.~Bu, Y.~Meng, H.~Xiao, M.~Zhang, C.~Li,
\newblock \emph{Journal of Chemical Theory and Computation} \textbf{2024}, \emph{20}, 9 3590.

\bibitem{kurle2024red}
P.~Kurle-Tucholski, C.~Wiebeler, L.~K{\"o}hler, R.~Qin, Z.~Zhao, M.~{\v{S}}im{\'e}nas, A.~P{\"o}ppl, J.~Matysik,
\newblock \emph{The Journal of Physical Chemistry B} \textbf{2024}, \emph{128}, 18 4344.

\bibitem{jakobsen2019a}
P.~Jakobsen, F.~Jensen,
\newblock \emph{J. Chem. Phys.} \textbf{2019}, \emph{151} 174107.

\bibitem{pritchard2019new}
B.~P. Pritchard, D.~Altarawy, B.~Didier, T.~D. Gibson, T.~L. Windus,
\newblock \emph{Journal of chemical information and modeling} \textbf{2019}, \emph{59}, 11 4814.

\bibitem{TheoDORE}
{F. Plasser, TheoDORE: A package for theoretical density, orbital relaxation, and exciton analysis},
\newblock available from \url{https://theodore-qc.sourceforge.io/},
\newblock Accessed: 2025-04-10.

\bibitem{chemcraft}
{Chemcraft: A graphical software for visualization of quantum chemistry computations, Version 1.8, build 682},
\newblock available from \url{https://www.chemcraftprog.com}.

\end{thebibliography}

\end{document}